\documentclass[preprintnumbers,amsmath,11pt,amssymb,floatfix,superscriptaddress,nofootinbib]{revtex4}
\usepackage{graphicx}
\usepackage{epsfig}
\usepackage{bm}
\usepackage{amsfonts}

\input epsf

\begin{document}

\title{{ Power-Law Entropy-Corrected Holographic Dark Energy in Ho\v{r}ava-Lifshitz Cosmology with Granda-Oliveros Cut-off}}
\author{Antonio Pasqua}
\email{toto.pasqua@gmail.com} \affiliation{Department of Physics,
University of Trieste, Trieste, Italy.}

\author{Surajit Chattopadhyay}
\email{surajit_2008@yahoo.co.in, surajcha@iucaa.ernet.in}
\affiliation{ Pailan College of Management and Technology, Bengal
Pailan Park, Kolkata-700 104, India.}

\author{Ratbay Myrzakulov}
\email{rmyrzakulov@gmail.com}
\affiliation{Eurasian International Center for Theoretical Physics, Eurasian National University, Astana 010008, Kazakhstan.}

\date{\today}
\newpage

%\author[a]{Antonio Pasqua}
%\affiliation[a]{Department of Physics,
%University of Trieste, Via Valerio, 2 34127 Trieste, Italy.}
% \author[b]{Surajit Chattopadhyay}
%\affiliation[b]{Pailan College of Management and Technology, Bengal Pailan Park, Kolkata-700 104, India.}
% \author[c]{Ratbay Myrzakulov}
%\affiliation[c]{Eurasian International Center for Theoretical Physics, Eurasian National University, Astana 010008, Kazakhstan}

%\emailAdd{toto.pasqua@gmail.com}
%\emailAdd{surajcha@iucaa.ernet.in}
%\emailAdd{rmyrzakulov@gmail.com}

\begin{abstract}
\textbf{Abstract}: In this paper, we study the Power Law Entropy Corrected Holographic Dark Energy (PLECHDE) model in the framework of a non-flat Universe and of Ho\v{r}ava-Lifshitz cosmology with infrared cut-off given by  recently proposed  Granda-Oliveros cut-off, which contains one term proportional to the Hubble parameter squared $H^2$ and one term proportional to the first time derivative of the Hubble parameter $\dot{H}$. Moreover, this cut-off is characterized by two constant parameters, $\alpha$ and $\beta$. For the two  cases corresponding to non-interacting and interacting  DE and Dark Matter (DM), we derive the evolutionary form of the energy density of DE $\Omega_D'$, the Equation of State (EoS) parameter of DE $\omega_D$  and the deceleration parameter $q$. Using the parametrization of the EoS parameter  $\omega_D\left(z\right)=\omega_0+\omega_1 z$, we obtain the expressions of the two parameters $\omega_0$ and $\omega_1$.  We also study the statefinder parameters   $\left\{  r,s \right\}$, the snap and lerk cosmographic parameters  and the squared speed of the sound $v_s^2$. We also calculate the values of the quantities we study for different values of the running parameter $\lambda$ and for different set of values of $\alpha$ and $\beta$.
\end{abstract}

%\arxivnumber{arXiv:***}
\flushbottom
%\pacs{98.80.Jk, 98.80.-k}
\maketitle

\section{Introduction}
Recent  cosmological and astrophysical data obtained with observations done thanks to the Supernova Cosmology Project, the  Wilkinson Microwave Anisotropy Probe (WMAP) and the Planck satellites, the Sloan Digital Sky Survey (SDSS) and X-ray experiments \citep{sn1,sn3,sn4,cmb1,cmb2,planck,sds1,xray, sds2,1c} give clear indications that the observable present day Universe is experiencing a phase of  expansion with accelerated rate, which is practically the expansion with accelerated rate which the Universe undergoes, with the first happened during the inflationary period.  The present day cosmic acceleration is one of the biggest challenges in the understanding of the standard models of gravity and particle physics.\\
Three main different classes of models have been suggested and well studied till now with the aim to give a proper  explanation to the accelerated expansion of the present day observable Universe:
\begin{enumerate}
  \item the Cosmological Constant (CC) $\Lambda_{CC}$ model;
  \item Dark Energy (DE) models;
  \item Theories of Modified Gravity models.
\end{enumerate}
The first and also the simplest candidate introduced with the aim to explain the present day observed accelerated expansion of the Universe is the Cosmological Constant $\Lambda_{CC}$, which can be considered as an extra term added to Einstein's equations. One of the main features of the Cosmological Constant $\Lambda_{CC}$ is that it has an Equation of State (EoS) parameter $\omega$ exactly equal to $ -1$, i.e. $\omega_{\Lambda_{CC}} \equiv -1$.  According to what we know thanks to the Quantum Field Theory (QFT), a cut-off at the Planck (or at the electro-weak scale) leads to the production of a Cosmological Constant $\Lambda_{CC}$ which is of the order of $10^{123}$ (or $10^{55}$), respectively, times bigger than the value we are able to observe. The fact that we still do not have a fundamental symmetry which is able to put the precise value of the Cosmological Constant $\Lambda_{CC}$ to exactly zero (i.e., $\Lambda_{CC}\equiv 0$) or, instead, to  a very small value (i.e., $\Lambda_{CC} \approx 0$) produces the so-called Cosmological Constant problem, also known as fine tuning problem.
Moreover, it is also well-known that the Cosmological Constat $\Lambda_{CC}$ model is affected by another problem, which is the Cosmic Coincidence problem \citep{16}.  The Cosmic Coincidence problem states that the DM and the vacuum energy are almost equal at the present epoch of the Universe even if they had an independent evolution and they had an evolution starting from different mass scales. Many proposals have been suggested till now with the purpose and the hope to obtain an explanation to the Cosmic Coincidence problem \citep{delcampoa,delcampoc,delcampob,delcampo,delcampoe,delcampod,delcampof,delcampol}.\\
The second class of models which are suggested and widely studied with the aim to give a plausible explanation to the present day accelerated expansion of the Universe considers Dark Energy (DE) models. \\
The observational evidences of the cosmic accelerated expansion imply that, if, on cosmological scale, the theory of Einstein's General Relativity (GR) is valid, we must have that the present day observable Universe must have as dominant component an unknown missing energy component which has  some particular features, in particular:  1) its pressure $p$ must be sufficiently negative if it wants to be able to produce the rate of accelerated expansion of the Universe we are able to observe and 2) it must not be clustered on cosmological scales (i.e. on large scale length). The present day observed cosmic accelerated expansion of the Universe can be described, in relativistic cosmology, introducing a perfect fluid with energy density $\rho $ and  pressure $p$ satisfying the following condition: $\rho + 3p < 0$ (which implies that the pressure $p$ must be negative in order the condition is satisfied). This kind of fluid with a sufficient negative pressure $p$ in order to satisfy the condition $\rho+3p<0$ is referred as Dark Energy (DE). The fact that the relation $\rho + 3p < 0$ must be satisfied leads to the fact that the EoS parameter $\omega $ (defined as the ratio of the pressure $p$ and the energy density $\rho$)  satisfies the following condition: $\omega \equiv \frac{p}{\rho} <-1/3$. Instead, from an observational point of view, it is still a challenging task to constrain its exact value. The fundamental theory  which can explain the microscopic physics of DE is still not known up to now, for this reason scientists continue to reconstruct and suggest different possible models which are mainly based on its macroscopic behavior.\\
Furthermore, recent cosmological experiments and observations have clearly indicated that the largest part of the total energy density $\rho_{tot}$ of the present Universe is contained in the two Dark Sectors \citep{twothirds}, i.e. DE and DM, which represent, respectively, the 68.3$\%$ and the 26.8$\%$ of the total energy density of the present day observable Universe. We also know that the Baryonic Matter (BM) we are able to observe with our scientific  instruments contributes only for approximately the $ 4.9 \%$ of the total energy density $\rho_{tot}$ of the present day Universe. Moreover, we have that the contribution produced by the radiation term to the total cosmic energy density $\rho_{tot}$ can be safely considered practically negligible, i.e we have that $\rho_{rad} \approx 0$.\\
Other different candidates introduced and suggested for the  DE problem are given by the dynamical DE scenarios with a time dependent EoS parameter $\omega$, then not anymore constant. According to analysis of the  available SNe Ia observational data, it has been derived that time-varying DE models lead to a better fit compared with a model with Cosmological Constant $\Lambda_{CC}$. There are two main different categories suggested for dynamical DE scenarios: (i) scalar fields models, which include k-essence \citep{kess2,kess3,kess4}, quintessence \citep{quint1,quint2,quint3}, tachyon \citep{tac1,tac2,tac3}, phantom \citep{pha1,pha2,pha5,pha6}, dilaton \citep{dil1,dil2,dil3} and quintom field \citep{qui1,qui2,qui6,qui8}, (ii) interacting DE models, which include for example Chaplygin gas \citep{cgas1,cgas2,cgas3} and Agegraphic DE (ADE) models \citep{ade1,ade2}. \\
The complete description of the DE features and nature must come from a consistent theory of Quantum Gravity (QG).
Unfortunately, we still do not have a complete and widely accepted theory of Quantum Gravity (QG) and then some approximations for this theory can be made: some examples are given by the Loop Quantum Gravity (LQG) and String Theory.\\
The third and last class of models proposed in order to give an explanation to the present day accelerated expansion of Universe involves extended theories of gravity, which correspond to a modification of the action of the gravitational fields. Some of the most famous and studied models of Modified Gravity are the $f\left(T\right)$ modified gravity model (with $T$ being the torsion scalar), braneworld models, the $f \left(G\right)$ modified gravity model (with $G$ being the Gauss-Bonnet invariant defined as $G=R^2-4R_{\mu \nu}R^{\mu \nu} + R_{\mu \nu \lambda \sigma}R^{\mu \nu \lambda \sigma}$, $R_{\mu \nu}$ being the Ricci curvature tensor and $R_{\mu \nu \lambda \sigma}$ being the Riemann curvature tensor), the $f \left(R\right)$ modified gravity model (with $R$ being the Ricci scalar curvature),  the Dvali-Gabadadze-Porrati (DGP) model,  the $f \left(R,T\right)$ modified gravity model, the Dirac-Born-Infeld (DBI) model and the Brans-Dicke  model \citep{miofg1,frt2,frt5,dgp1,fr1,fr3,miofr,fr6,fr8,fr10,fr11,fr12,fr13,fr14,fr15,mioft1,ft1,ft3,ft6,ft8,bra1,bra2,miors}.\\
Using the holographic principle which was recently introduced by Fischler $\&$ Susskind \citep{holo1,holo2,holo4,holo5}, a model dubbed as Holographic DE (HDE) model has been recently proposed in the paper of Li \citep{li}. The HDE model is one of the most famous and studied candidate of DE \citep{fis3,fis5,fis7,fis8,hde1,hde5,hde10,hde12,hde13,hde20,hde26,hde33,hde35,saridakis11,33,30,15,saridakis2,saridakis3}.
It is well-known that the holographic principle assumes a fundamental role in both black hole and string
theories. It was recently demonstrated in the work of Cohen et al. \citep{15cohen} that, in the framework of the QFT, the UV cut-off, which is indicated with $\Lambda_{UV}$, is related to the IR cut-off, which is given by $L$, due to the limitations produced by the formation of a black hole. If the vacuum energy $\rho_D$ density produced by the UV cut-off is given by the relation $\rho_D = \Lambda_{UV}^4$, then we have that the total energy density of a given size $L$ must be less or at least equal to the mass corresponding to the system-size black hole, i.e. we must have that:
\begin{eqnarray}
E_D \leq E_{BH} , \label{1}
\end{eqnarray}
which implies that:
\begin{eqnarray}
 L^3 \rho_D \leq M_p^2 L, \label{1cinzia}
\end{eqnarray}
where $M_p = \left( 8\pi G_N  \right)^{-1/2} \approx 10^{18}\, GeV$ represents the reduced Planck mass and $G_N=6.67\cdot10^{-11}Nm^2kg^{-2}$ represents the Newton's gravitational constant. If the largest possible cut-off $L$ of the system is that one which is able to saturate the inequality given in Eq. (\ref{1cinzia}), we derive the following expression for the energy density $\rho_D$ of the HDE model:
\begin{eqnarray}
\rho_D = 3n^2 M_p^2 L^{-2}, \label{2}
\end{eqnarray}
where $n$ represents a dimensionless constant parameter. It has been obtained that, in the case of a Universe that is not flat (i.e. for a value of the curvature parameter which is different from zero) the value of such constant is given by $n = 0.815^{+0.179}_{-0.139}$ while for  a flat Universe (i.e. when the curvature parameter is equal to zero), we have that the value of $n$ is given by $n=0.818_{-0.097}^{+0.113}$ \citep{n2primo}.
The expression of the energy density $\rho$ of the HDE model can be also obtain using a different approach  \citep{8!!!}. It must be here underlined that the black hole entropy $S$ has an important role in the derivation of the HDE energy density $\rho_D$. In fact, we know that the derivation of the HDE energy density strongly depends on the entropy-area relation given, in Einstein's gravity, by the relation $S \approx A \approx L^2$ (where $A$ gives the area of the black hole horizon). According to the laws of the thermodynamics of black holes \citep{9!!!,9a!!!}, a maximum value of the entropy in a box with a dimension of $L$ (which is also referred as Bekenstein-Hawking entropy bound),  is given by the relation $S_{BH} \approx M_p^2 L^2$, which goes as the area $A$ of the box (given approximatively by the expression $A \approx L^2$) rather than the volume $V$ of the box (which is given by $V \approx L^3$). Moreover, for  macroscopic systems having some self-gravitation effects which cannot be ignored, we have that the expression of the Bekenstein entropy bound (which is indicated with $S_B$) can be obtained multiplying the energy $E$, given by the relation $E \approx \rho_DL^3$, and the linear size $L$ of the system. If we impose that the Bekenstein entropy bound must be smaller than the Bekenstein-Hawking entropy (i.e., if we impose that $S_B \leq S_{BH}$, which implies that $E\cdot L \leq M_p^2 L^2$), we obtain the same result obtained from energy bound arguments, i.e. we obtain that $\rho_D \leq M_p^2L^{-2}$.\\
Using the holographic principle, Cohen et al. \citep{15cohen} recently proposed that the vacuum energy density must be proportional to the Hubble parameter $H$. In this particular model, both the fine-tuning and coincidence problems can be solved, but it is still not possible to give a reasonable explanation to the present day cosmic accelerated expansion of the Universe since the effective Equation of State (EoS) parameter $\omega_{eff}$ for such vacuum energy is equal to zero, then it is different from what it is requested for the HDE model. In a recent paper, Li \citep{li} suggested that the future event horizon of the Universe can be used as possible IR cut-off. This DE model not only has a reasonable value for the DE energy density but it also leads to an accelerated solution for the cosmological expansion. \\
Jamil et al. \citep{11} studied the EoS parameter $\omega_D$ of the HDE model choosing a  Newton's gravitational constant $G_N$ which is not constant but it is time dependent, i.e. we have $G_N\left(t\right)$; furthermore, they obtained that the EoS parameter $\omega_D$ can be significantly modified when the low-redshift $z$ limit is considered.
Chen et al. \citep{A73} studied the HDE model in order to obtain an inflationary epoch in the early evolutionary stages of our Universe. The HDE model was recently considered in other works with different IR cut-offs, for example the Hubble horizon, the particle horizon and the future event horizon \citep{12a,12c,12e,12f,12g,13a}. Moreover, correspondences between some scalar field models and the HDE model have been recently proposed \citep{14,14b}, while in other works, the HDE model was accurately studied in different modified gravity theories, like for example scalar-tensor gravity, $f\left(R\right)$, DGP model, braneworld and Brans-Dicke cosmology \citep{15a,15b,15c,15d,15e,15g,15h,15i}.\\
Different HDE models have also been constrained and tested by using different astronomical and cosmological observations \citep{cons1,cons4,cons6,cons7,cons8,cons9} and also thanks to the anthropic principle \citep{ant1}.  It is also known that the HDE model fits well cosmological data obtained using the data obtained from observations of SNeIa and CMB radiation anisotropies \citep{16a,16b,16d,16e}.\\
The definition of the entropy-area relation can be  modified considering quantum effects which are motivated from the Loop Quantum Gravity (LQG). The relation entropy-area  $S\left(A\right)$ has an interesting modification (correction), i.e. the power-law correction \citep{das18,das18a} which arises in dealing with the entanglement of quantum fields in and out the horizon. \\
The power-law corrected entropy-area  relation $S\left( A \right)$  has the following specific form \citep{das18,das18a}:
\begin{eqnarray}
    S\left( A \right)=c_0 \left( \frac{A}{a_1^2}  \right)\left[ 1+c_1f\left( A \right)  \right],\label{powerlawentropyold}
\end{eqnarray}
where the term $ f\left( A \right) $ is given by the following power-law  relation:
\begin{eqnarray}
 f\left( A \right)  = \left( \frac{A}{a_1^2}  \right)^{-\nu},\label{powerlawentropyold}
\end{eqnarray}
with $c_0$ and $c_1$ indicating two constant parameters, $a_1$ being the UV cut-off at the horizon and $\nu$ being a fractional power which depends on the amount of mixing of ground and excited states. For a large horizon area (i.e. for $A>>a_1^2$), the contribution given by the term $f \left(A\right)$ to the entropy $S\left( A \right)$ can be considered practically negligible and, therefore, the mixed state entanglement entropy asymptotically approaches the ground state (Bekenstein-Hawking) entropy.\\
Another useful way to write the expression of the entropy area relation  $S\left( A \right)$ for the power-law corrected entropy is given by the following relation:
\begin{eqnarray}
  S\left( A \right)= \frac{A}{4G}\left( 1-K_{\alpha}A^{1-\alpha /2}   \right), \label{1--}
\end{eqnarray}
with $\alpha$ representing a dimensionless constant parameter and the term $K_{\alpha}$ is a constant which is defined as follows:
\begin{eqnarray}
  K_{\alpha} = \frac{\alpha \left( 4\pi  \right)^{\alpha /2 -1}}{\left( 4-\alpha  \right)r_c^{2-\alpha}}, \label{2}
\end{eqnarray}
where the term $r_c$ indicates the cross-over scale. Moreover, we have that the quantity $A=4\pi R_h^2$ gives the area of the horizon (with the term $R_h$ indicating the radius of the horizon). The second term present in Eq. (\ref{1--}) gives the power-law correction to the entropy-area law. In order the entropy is a well-defined quantity, we need to have that the parameter $\alpha$ is positive defined, i.e. we must have that the condition $\alpha>0$ must be satisfied. Motivated by the relation defined in Eq. (\ref{1--}), a new version of HDE (also known with the name of Power-Law Entropy-Corrected HDE (PLECHDE) model) was recently introduced as follows:
\begin{eqnarray}
\rho_D = 3n^2M_p^2L^{-2} - \varepsilon M_p^2 L^{-\delta}, \label{lgo2}
\end{eqnarray}
with $\varepsilon$ being a positive dimensionless parameter and $\delta$ begin a positive exponent. \\
In the limiting case of $ \varepsilon =0$ (or, equivalently, for $\delta \rightarrow \infty$), Eq. (\ref{lgo2}) reduces to the well-known expression of the HDE energy density. The correction term present in Eq. (\ref{lgo2}) is of the same order of to the first one only when $L$ assumes a very small value. Then, at the very early evolutionary phases and stages of our Universe history (i.e., when the Universe underwent the inflationary phase), the contribution of the correction term in the PLECHDE energy density can be safely  considered relevant but, when the Universe became larger, the PLECHDE energy density reduced to the ordinary HDE energy density. Therefore, PLECHDE model can be also considered as a model of entropic cosmology which is able to unify the early-time inflation and late-time cosmic acceleration of the Universe.\\
In some recent works, Ho\v{r}ava \citep{23,A22,A176} recently introduced a new theory of gravity which is renormalizable with higher spatial derivatives in four dimensions. This theory  leads to the Einstein's gravity (i.e.
 to GR) with non-vanishing value of the Cosmological Constant $\Lambda_{CC}$ in the infrared (IR) limit and it also have  some improved behaviors and features in the ultraviolet (UV) regime. Ho\v{r}ava gravity can be also considered similar to a scalar field theory previously proposed by Lifshitz \citep{lif1} in which we have that the temporal dimension $t$ has a weight equal to three if the space dimension has a weight of one. For this reason, the gravity theory proposed by Ho\v{r}ava is also known with the name of Ho\v{r}ava-Lifshitz gravity. The Ho\v{r}ava-Lifshitz gravity has been extended and studied  in detail in some papers, like for example \citep{A148,A171,A192,A140,A146,A202,A141,A145}, and it has been applied as a possible cosmological framework of our Universe  \citep{46,76,A126,A156,A143,A201,A151,A170,A188,A163,A172,A190,A36,A7}.
Furthermore, Ho\v{r}ava-Lifshitz theory is not Lorentz invariant (with the exception of the IR limit), test particles do not follow geodesics, it is non-relativistic and we also have that the speed of light $c$ diverges in the UV limit. We have four different versions of Ho\v{r}ava-Lifshitz theory of gravity:
(i) with projectability condition, (ii) without projectability condition, (iii) with detailed balance and (iv) without detailed balance. Having a first look, it seems that this model of Quantum Gravity (QG) has a well defined IR limit and it also reduces to General Relativity, but as it was obtained by Mukohyama \citep{57,58}, Ho\v{r}ava-Lifshitz gravity behaves like General Relativity plus DM. For some relevant works on the scenario where the cosmological evolution is ruled by Ho\v{r}ava-Lifshitz gravity see \citep{59,63,53,74}.\\
Because of these characteristics, a great effort in extending,  examining and improving the physical features and properties of the theory itself have been done \citep{A135,A112,A113,73}. Furthermore, applications of Ho\v{r}ava-Lifshitz gravity as a cosmological context produces the Ho\v{r}ava-Lifshitz cosmology, which has some interesting features. For example, it is possible to examine  the perturbation spectrum \citep{A221}, some particular solution subclasses \citep{mina}, the matter bounce \citep{A69}, the production of gravitational waves \citep{A131},   the phenomenology of DE \citep{app}, the properties of black hole \citep{A78,A110,A65} and the astrophysical phenomenology \citep{kai}. Ho\v{r}ava-Lifhsitz cosmology has been recently investigated raking into account and choosing different infrared cut (IR)-offs and different approaches.
Setare $\&$ Jamil \citep{suracon5} considered the HDE model with a varying Newton's gravitational constant $G$ in the framework of Ho\v{r}ava-Lifshitz cosmology.
Jamil et al. \citep{suracon3} studied the behavior of the Generalized Second Law of Thermodynamics (GSLT) in the context of Ho\v{r}ava-Lifshitz cosmology using  the dynamical apparent horizon as infrared (IR) cut-off of the system.
Karami et al. \citep{suracon1} studied the Logarithmic Entropy Corrected New Agegraphic DE (LECNADE) model in the context of Ho\v{r}ava-Lifshitz cosmology.
Jamil et al. \citep{suracon4} considered the NADE model in the context of the Ho\v{r}ava-Lifshitz cosmology.
Karami et al. \citep{suracon2} studied the Power-Law Entropy Corrected NADE (PLECNADE) model in the framework of Ho\v{r}ava-Lifshitz cosmology.
Pasqua et al. \citep{miohl1} studied the Power Law and the Logarithmic Ricci Dark Energy Models in the framework of  Ho\v{r}ava-Lifshitz Cosmology.
Jawad et al. \citep{miohl2} studied the power-law solution of the new agegraphic modified $f\left(R\right)$ Ho\v{r}ava-Lifshitz gravity.
Chattopadhyay and  Pasqua \citep{miohl3}  studied the  modified holographic Ricci DE (RDE) model in the framework of  modified $f\left( R\right)$ Ho\v{r}ava-Lifshitz gravity.
Jawad et al. \citep{miohl4} obtained a holographic reconstruction of the modified $f\left( R\right)$ Ho\v{r}ava-Lifshitz gravity with the scale factor $a\left(t\right)$ given in the in power-law form.
Anyway, even if this extended research is available, a lot of ambiguities are still presents about the fact that Ho\v{r}ava-Lifshitz gravity can be considered a reliable theory and it is able to accurately describe the cosmological behavior of our Universe.\\
This work differs from the ones cited above and other available in literature since we are considering an IR cut-off, known as Granda-Oliveros cut-off, based on purely dimensional grounds which Granda $\&$ Oliveros recently proposed. We must also underline here that DE models with Granda-Oliveros cut-off belong to generalized Nojiri-Odintsov HDEs classes \citep{valerossi}.  It also differs from the work of Pasqua $\&$ Chattopadhyay \citep{suracon6} since the Authors  considered the Logarithmic Entropy-Corrected Holographic Dark Energy (LECHDE) model in the framework of Ho\v{r}ava-Lifshitz cosmology with Granda-Oliveros cut-off while we are considering here the power law correction to the entropy. This new cut-off contains a term proportional to the time derivative of the Hubble parameter $\dot{H}$ and one term proportional to the squared Hubble parameter $H^2$ and it is indicated with $L_{GO}$. The final expression of $L_{GO}$ is given by  \citep{grandaoliveros,grandaoliverosa}:
\begin{equation}
L_{GO}=\left( \alpha H^{2}+\beta \dot{H}\right) ^{-1/2}.  \label{lgo5}
\end{equation}%
$\alpha $ and $\beta $ indicate two constant dimensionless parameters. In the limiting case corresponding to $ \alpha = 2$ and $\beta = 1$, we obtain that the expression of $L_{GO}$ defined in Eq. (\ref{lgo5}) becomes proportional to the average radius of the Ricci scalar curvature $R^{-1/2}$ when the curvature parameter $k$ assumes the values of zero. In a recent paper, Wang $\&$ Xu \citep{wangalfa} have constrained the HDE model with GO cut-off for a non-flat Universe using observational data. The best fit values of the pair $\left(\alpha, \beta   \right)$ with their confidence level they found are given by $\alpha  = 0.8824^{+0.2180}_{-0.1163}(1\sigma)\,^{+0.2213}_{-0.1378}(2\sigma)$ and $\beta = 0.5016^{+0.0973}_{-0.0871}(1\sigma)\,^{+0.1247}_{-0.1102}(2\sigma)$ for non flat Universe (i.e. for $k\neq 0$), while for a flat Universe (i.e. for $k=0$) they found that are $\alpha  = 0.8502^{+0.0984}_{-0.0875}(1\sigma)\,^{+0.1299}_{-0.1064}(2\sigma)$ and $\beta = 0.4817^{+0.0842}_{-0.0773}(1\sigma)\,^{+0.1176}_{-0.0955}(2\sigma)$.  \\
We decided to consider the GO scale $L_{GO}$ defined in Eq. (\ref{lgo5}) as IR cut-off for some specific reasons. If the IR cut-off chosen is given by the particle horizon, the HDE model cannot produce an accelerated expansion of the Universe \citep{hsunuovo}. If we consider as cut-off of the system the future event horizon, the HDE model has a causality problem.  The DE models which consider the GO scale depend only on local quantities, then it is possible to avoid the causality problem, moreover it is also possible to obtain the accelerated phase of the Universe.   \\
Replacing $L$ with $L_{GO}$ in the expression of the energy density of DE $\rho_D$ given in Eq. (\ref{lgo2}), we get the energy density of the  PLECHDE model $\rho_D$ as follows:
\begin{eqnarray}
 \rho_D=\frac{3n^2 M_p^2}{L_{GO}^2} - \frac{\varepsilon M_p^2}{L_{GO}^{\delta}}. \label{8}
\end{eqnarray}
In the following Sections we will study the main properties and features of the cosmological parameters obtained using the energy density $\rho_D$ given in Eq. (\ref{8}) of the PELCHDE model with GO cut-off we are studying.\\
This paper is organized in the following way. In Section 2, we describe the most important features of Ho\v{r}ava-Lifshitz cosmology. In Section 3, we study the PLECHDE model with Granda-Oliveros cut-off in the context of Ho\v{r}ava-Lifshitz cosmology. Moreover, we derive the evolutionary form of the energy density of DE, the Equation of State (EoS) parameter, the evolutionary form of the fractional energy density and the deceleration parameter for both non interacting and interacting Dark Sectors. In Section 4, we study the low redshift limit of the EoS parameter, which is parametrized as $\omega_D = \omega_0 + \omega_1 z$, obtaining the expressions of $\omega_0$ and $\omega_1$ for both cases corresponding to non interacting and interacting Dark Sectors. In Section 5, we study the statefinder pair $\left\{ r,s  \right\}$ for the model we are studying. In Section 6, we derive and study the expressions of the snap and of the lerk  cosmographic parameters for the model taken into account in this paper. In Section 7, we study the squared speed of the sound $v_s^2$ for the model considered in order to check its stability. Finally, in Section 8, we write the Conclusions of this work.

\section{HO\v{R}AVA-LIFSHITZ GRAVITY}
In this Section, we introduce the main physical and cosmological characteristics of Ho\v{r}ava-Lifshitz gravity. These information will be useful in order to obtain the cosmological information we want to derive for the model we are studying. \\
Considering the projectability condition, we have that the metric in the (3+1)-dimensional Arnowitt-Deser-Misner formalism can be written as follows \citep{arno}:
\begin{eqnarray}
ds^2 = -N^2dt^2+g_{ij}\left(dx^i+N^i dt \right)\left(dx^j+N^jdt \right), \label{9}
\end{eqnarray}
where $t$ indicates the cosmic time while the dynamical variables $g_{ij}$, $N$ and  $N^i$  indicate, respectively, the 3-dimensional metric tensor, the lapse function and the shift vector. The projectability condition leads to the fact that the lapse function $N$ is space-independent, instead  the 3-dimensional metric $g_{ij}$ and the shift vector $N^i$  still depend on both space and time. Moreover, we have that the indices are raised and lowered thanks to the metric tensor $g_{ij}$. The scaling transformations of the coordinates $x^i$ and $t$ are given by the following relations:
\begin{eqnarray}
x^i &&\rightarrow l x^i, \\
t&&\rightarrow l^z t, \label{tra}
\end{eqnarray}
where the quantities $t$, $z$, $\l$ and $x^i$   represent, respectively,  the temporal coordinate, the dynamical critical exponent, the scaling factor and the spatial coordinates.\\
In this paper, we have that $z=3$, then the scaling transformation of the temporal coordinate defined in Eq. (\ref{tra}) can be rewritten as follows:
\begin{eqnarray}
t\rightarrow l^3 t.
\end{eqnarray}

The gravitational action of Ho\v{r}ava-Lifshitz cosmology, indicated with $S_g$, can be decomposed into two different parts, i.e. a kinetic part, given by $L_K$, and a potential part, given by $L_V$, and it is given by the following relation:
\begin{eqnarray}
S_g = \int dt \, d^3x \sqrt{g} N \left( L_K + L_V   \right),
\end{eqnarray}
with $g$ indicating the determinant of the metric tensor $g^{\mu \nu}$.\\
The assumption of detailed balance \citep{A177} allows to reduce the number of possible terms in
the expression of the gravitational action $S_g$. Moreover, it also permit a quantum inheritance principle, because the $(D + 1)$-dimensional
theory takes the renormalization properties of the $D$-dimensional theory. Considering the detailed balance condition, the gravitational action of the Ho\v{r}ava-Lifshitz gravity $S_g$  is given by the following expression \citep{A177}:
\begin{eqnarray}
S_g &=& \displaystyle \int{d^3x \, dt \, N \sqrt{g}} \left\{\frac{2 \left(K_{ij}K^{ij}-\lambda K^2\right) }{\kappa^2} \right.\nonumber\\
&& + \left(\frac{\kappa^2}{2\omega^4}\right)C_{ij}C^{ij} -\left(\frac{\kappa^2\mu}{2\omega^2}\right)\left( \frac{\eta^{ijk}}{\sqrt{g}}\right)R_{il} \bigtriangledown_j R^l_k \nonumber\\
&& +\left(\frac{\kappa^2\mu^2}{8}\right)R_{ij}R^{ij}  \nonumber\\
&& + \left. \frac{\kappa^2\mu^2}{8\left(3\lambda - 1\right)}\left[\frac{\left(1-4\lambda\right)R^2}{4}+\Lambda R - 3 \Lambda^2\right]\right\}, \label{10}
\end{eqnarray}
where the terms $C_{ij}$ and $K_{ij}$  represent, respectively, the Cotton tensor and the extrinsic curvature which are defined in the following way:
\begin{eqnarray}
C_{ij} &=& \frac{e^{ijk}}{\sqrt{g}} \bigtriangledown_k \left(R^j_i - \frac{R \delta^j_i}{4}\right), \label{12}\\
K_{ij} &=& \frac{1}{2N}\left(\dot{g}_{ij}- \bigtriangledown_iN_j - \bigtriangledown_jN_i\right). \label{11}
\end{eqnarray}
Furthermore, the quantity $\Lambda$ represents a positive dimensionless constant which is related to the cosmological constant in the infrared (IR) limit, the quantity $\eta^{ijk}$ indicates the totally antisymmetric unit tensor and $\lambda$ represents a dimensionless constant (also known as running parameter). More information about the running parameter $\lambda$ will be given later on.\\
The three parameters $\mu$, $\kappa$ and $\omega$ represents three constants which has mass dimension, respectively, of 1, -1 and 0. \\
If we want to include the matter component in a Universe ruled by Ho\v{r}ava-Lifshitz gravity, we have that there are two options which can be taken into account. In the first option, we include a scalar field $\phi$ with action $S_{\phi}$ which is given by the following relation \citep{12calcagni}:
\begin{eqnarray}
S_m \equiv S_{\phi} &=& \int dtd^3x N\sqrt{g} \left[ \left(\frac{3\lambda -1}{4}\right)\frac{\dot{\phi}^2}{N^2} +m_1m_2 \phi \nabla ^2 \phi - \frac{1}{2}m_2^2\phi \nabla ^4 \phi \right. \nonumber \\
&&\left. + \frac{m_3^2\phi \nabla ^6 \phi}{2} - V\left( \phi  \right)  \right],
\end{eqnarray}
where the three quantities $m_1$, $m_2$ and $m_3$ represent three constant parameters while the term indicated with $V\left( \phi  \right) $ represents the potential term. Furthermore, we have that the equation of motion for the field $\phi$ can be written as follows:
\begin{eqnarray}
\ddot{\phi} + 3H\dot{\phi}+ \left(\frac{2}{3\lambda -1}\right) \frac{dV\left( \phi  \right)}{d \phi} = 0,
\end{eqnarray}
with the condition $3\lambda - 1 \neq 0$, i.e. $\lambda \neq 1/3$ in order to avoid singularities. Moreover, an overdot indicates a derivative with respect to the cosmic time $t$. \\
The second option we have in order to insert the matter component is obtained taking into account a hydrodynamical approximation adding a cosmological stress-energy tensor to the gravitational field equations; we must also consider  the condition that the formalism of the General Relativity must be obtained when the low-energy limit is considered \citep{14cha}. In this case, the energy density $\rho_m$ and the pressure $p_m$  of DM satisfy the following continuity equation:
\begin{eqnarray}
\dot{\rho}_m +3H\left(\rho_m+p_m\right) = 0. \label{13}
\end{eqnarray}
In this paper, we have decided to consider the hydrodynamical approximation.\\
Eq. (\ref{10}), as it is well-known, has several problems, like strong coupling problems,  instability and inconsistency \citep{57}. It is possible overcome these problems invoking the Vainshtein mechanism, as it was already done in the paper of Mukohyama in the case of spherical space-times \citep{57} and in the paper of Wang $\&$ Wu in the cosmological setting \citep{75}. These considerations were also carried out by considering the  gradient expansion method \citep{29}. Another possible approach which can be taken into account is given by the introduction of an extra $U\left(1\right)$ symmetry: this kind of approach  was considered for the first time by Ho\v{r}ava $\&$ Melby-Thompson \citep{24} in the limiting case corresponding to $\lambda =1$, and subsequently  generalized to the case with any possible value of $\lambda$ in the paper of da Silva \citep{17}. These works were also extended to the case with absence of the projectability condition \citep{89}. In both cases, i.e. with and without the projectability condition, the spin-0 gravitons are eliminated because of the $U\left(1\right)$ symmetry, therefore all the problems related to them  are then solved.\\
In the cosmological framework, we consider a FLRW metric which is recovered for the following values of $N$, $g_{ij}$ and $N^i$:
\begin{eqnarray}
N &=& 1, \label{14} \\
 g_{ij}&=&a^2\left(t\right)\gamma_{ij},\label{15} \\
N^i&=&0,  \label{16}
\end{eqnarray}
where $\gamma_{ij}$ is given be the following relation:
\begin{eqnarray}
\gamma_{ij}dx^idx^j=\frac{dr^2}{1-kr^2}+r^2d\Omega^2_2, \label{17}
\end{eqnarray}
with the term $d\Omega^2_2$ representing the angular part of the metric.\\
Taking the variation of the action $S_g$ obtained in Eq. (\ref{10}) with respect to the metric components $N$ and $g_{ij}$, we obtain the modified Friedmann equations in the framework of Ho\v{r}ava-Lifshitz cosmology as follows:
\begin{eqnarray}
H^2 &=& \left[\frac{\kappa^2}{6\left(3\lambda-1\right)}\right]\rho_m + \frac{\kappa^2}{6\left(3\lambda-1\right)}\left[\frac{3\kappa^2\mu^2k^2}{8\left(3\lambda-1\right)a^4} \right. \nonumber\\
&& + \left. \frac{3\kappa^2\mu^2\Lambda^2}{8\left(3\lambda-1\right)}\right] - \frac{\kappa^4\mu^2\Lambda k}{8\left(3\lambda - 1\right)^2a^2}\label{18} , \\
\dot{H}+\frac{3}{2}H^2 &=& - \left[\frac{\kappa^2}{4\left(3\lambda - 1\right)}\right]p_m \nonumber\\
&& -  \frac{\kappa^2}{4\left(3\lambda - 1\right)}\left[\frac{\kappa^2\mu^2k^2}{8\left(3\lambda-1\right)a^4} - \frac{3\kappa^2\mu^2\Lambda^2}{8\left(3\lambda-1\right)} \right] \nonumber\\
&& - \frac{\kappa^4\mu^2\Lambda k}{16\left(3\lambda - 1\right)^2a^2}. \label{19}
\end{eqnarray}
In the limiting case of a flat Universe, i.e. for $k = 0$, the higher order derivative terms do not produce contributions to the action. Instead, for a non flat universe, i.e. for $k \neq 0$, the higher derivative terms give a relevant contribution for small volumes, i.e. for small values of $a$,
while this contribution becomes practically negligible when $a$ assumes large values (in this case we recover a good agreement with the results of General Relativity). \\
Considering the Friedmann equations given in Eqs. (\ref{18}) and (\ref{19}), we define the energy density $\rho_D$ and the pressure $p_D$ of DE as follows:
\begin{eqnarray}
\rho_D &\equiv& \frac{3\kappa^2\mu^2k^2}{8\left(3\lambda-1\right)a^4} +  \frac{3\kappa^2\mu^2\Lambda^2}{8\left(3\lambda-1\right)}, \label{20} \\
p_D &\equiv& \frac{\kappa^2\mu^2k^2}{8\left(3\lambda-1\right)a^4} -  \frac{3\kappa^2\mu^2\Lambda^2}{8\left(3\lambda-1\right)}. \label{21}
\end{eqnarray}
The first term of the right hand side of both Eqs. (\ref{20}) and (\ref{21}) (which scales as $a^{-4}$) indicates effectively the dark radiation term which is present in Ho\v{r}ava-Lifshitz cosmology, instead the second term (which is constant) has a cosmological constant term-like behavior.
Furthermore, Eqs. (\ref{20}) and (\ref{21}) obey the following continuity equation:
\begin{eqnarray}
\dot{\rho}_D +3H\left(\rho_D+p_D\right) = 0. \label{22}
\end{eqnarray}
We also have that Eqs. (\ref{18}) and (\ref{19}) lead to the standard Friedmann equations if we take into account the following considerations:
\begin{eqnarray}
G_{cosmo} = \frac{\kappa^2}{16\pi\left(3\lambda-1\right)}, \label{23} \\
\frac{\kappa^4\mu^2\Lambda}{8\left(3\lambda-1\right)^2} = 1. \label{24}
\end{eqnarray}
The term $G_{cosmo}$ indicates the Newton's cosmological constant. We must underline that, in gravitational theories which lead to a violation of the Lorentz invariance (which happens in theories like Ho\v{r}ava-Lifshitz gravity), the Newton's gravitational constant $G_{grav}$ (which is one of the terms present in the gravitational action) is different from the Newton's cosmological constant $G_{cosmo}$ (which is one of the terms present in Friedmann equations). We have that $G_{cosmo}$ and $G_{grav}$ are equal if Lorentz invariance is recovered.\\
For completeness, we give the definition of $G_{grav}$, which can we written as follows:
\begin{eqnarray}
G_{grav} = \frac{\kappa^2}{32\pi}, \label{25}
\end{eqnarray}
as we easily derive using the results of  Eq. (\ref{10}). Moreover, we observe that, in the IR limit (corresponding to the limiting case of $\lambda = 1$), which also implies that the Lorentz invariance is restored, $G_{cosmo}$ and $G_{grav}$ assume an equivalent form. Then, we can also state that the running parameter $\lambda$ gives information about
possibility of breaking the Lorentz invariance. In fact, a value of $\lambda = 1$ indicates validity of
Lorentz invariance, while $\lambda \neq 1$ indicates that Lorentz invariance has been broken.\\
In a recent work of Dutta $\&$ Saridakis \citep{duttasari}, authors concluded that $|\lambda - 1 | \leq 0.02$ with $1\sigma$ confidence level  while its best fit value is $|\lambda_{b.f.} - 1 | \approx 0.02$.\\
Moreover, using Eqs. (\ref{20}), (\ref{21}), (\ref{23})  and (\ref{24}), it is possible to rewrite the modified Friedmann equations given in Eqs. (\ref{18}) and (\ref{19}) in the usual forms as follows:
\begin{eqnarray}
H^2+\frac{k}{a^2} &=&  \frac{8\pi G_{cosmo}}{3}\left(\rho_m+\rho_D \right), \label{26}\\
\dot{H}+ \frac{3}{2}H^2+\frac{k}{2a^2} &=& -4\pi G_{cosmo}  \left(p_m+p_D \right).  \label{27}
\end{eqnarray}
In the following Section, we will derive some important cosmological quantities for the model considered, in particular the Equation of State (EoS) parameter $\omega_D$, the evolutionary form of the energy density of DE $\Omega_D'$ and the deceleration parameter $q$.

\section{PLECHDE Model with GO cut-off IN Ho\v{r}ava-Lifshitz Cosmology}
We now discuss the main features and properties of the PLECHDE model with Granda-Oliveros cut-off in the context of Ho\v{r}ava-Lifhsitz cosmology. We must underline that we consider a spatially non-flat FLRW Universe which is filled by both Dark Sectors, i.e. DE and DM.\\
We start remembering that the DE energy density $\rho_D$ with GO cut-off can be written as follows:
\begin{eqnarray}
 \rho_D=\frac{3n^2 M_p^2}{L_{GO}^2} - \frac{\varepsilon M_p^2}{L_{GO}^{\delta}}. \label{8-1-1}
\end{eqnarray}
Since we have that the Planck mass can be expressed as $M_p^2 = \left( 8\pi G_{grav} \right)^{-1}$, we can rewrite the expression of the energy density $\rho_D$ defined in Eq. (\ref{8-1-1}) as follows:
\begin{eqnarray}
 \rho_D=\frac{3n^2 }{8\pi G_{grav} L_{GO}^2} - \frac{\varepsilon }{8\pi G_{grav} L_{GO}^{\delta}}. \label{8-1-2}
\end{eqnarray}
We now introduce the expressions of the dimensionless fractional energy densities for DM, DE and also for the curvature parameter $k$ which are defined, respectively, in the following way:
\begin{eqnarray}
\Omega_m &=& \frac{\rho_m}{\rho_{cr}}=\left(\frac{8\pi G_{cosmo}}{3H^2}\right)\rho_m, \label{28} \\
\Omega_D &=& \frac{\rho_D}{\rho_{cr}}=\left(\frac{8\pi G_{cosmo}}{3H^2}\right)\rho_D\nonumber \\
&=&\left(\frac{n^2}{H^2L_{GO}^2}\right)\gamma_n, \label{lgo3}  \\
\Omega_k &=& -\frac{k}{a^2H^2}, \label{30}
\end{eqnarray}
where $\rho_{cr}$ indicates the critical energy density (i.e. the energy density necessary to obtain the flatness of the Universe) which is defined as follows:
\begin{eqnarray}
\rho_{cr} = \frac{3H^2}{8\pi G_{cosmo}}, \label{31}
\end{eqnarray}
and the term $\gamma_n$ is given by the following relation:
\begin{eqnarray}
\gamma_n = \frac{G_{cosmo}}{G_{grav}}\left(1 - \frac{\varepsilon}{3n^2L_{GO}^{\delta-2}} \right). \label{lgo4}
\end{eqnarray}
Using the expressions of the fractional energy densities defined in Eqs. (\ref{28}), (\ref{lgo3}) and (\ref{30}), the first Friedmann equation defined in Eq. (\ref{26}) can be written in an equivalent way as follows:
\begin{eqnarray}
1 - \Omega_k = \Omega_D + \Omega_m . \label{32}
\end{eqnarray}
Eq. (\ref{32}) has the property that it relates all the fractional energy densities considered in this work.

\subsection{Non Interacting Case}
We start considering the case corresponding to a FLRW Universe containing DE and  pressureless DM  (i.e., we have $p_m =0$) and in absence of interaction between DE and DM. Moreover, we consider that the Dark Sectors evolve according to conservation laws which  expressions are given by the following continuity equations:
\begin{eqnarray}
\dot{\rho}_D +3H\left(1+\omega_D\right)\rho_D = 0, \label{38} \\
\dot{\rho}_m +3H\rho_m = 0, \label{39}
\end{eqnarray}
which are equivalent to the expressions:
\begin{eqnarray}
\rho'_D +3\left(1+\omega_D\right)\rho_D = 0, \label{38prime} \\
\rho'_m +3\rho_m = 0. \label{39prime}
\end{eqnarray}
We must also underline that the prime $'$ and the overdot are related by the following relation:
\begin{eqnarray}
\frac{d}{dt} = H \frac{d}{dx},
\end{eqnarray}
where we have that the parameter $x$ is defined as $x=\ln a$.\\
As described in the Introduction, we decided to choose as IR cut-off the GO scale $L_{GO}$, which has been previously defined in Eq. (\ref{lgo5}).
The first time derivative of the expression of $L_{GO}$ given in Eq. (\ref{lgo5}) is given by the following expression:
\begin{eqnarray}
\dot{L}_{GO}=-H^{3}L_{GO}^{3}\left[ \alpha \left(\frac{\dot{H}}{H^{2}}\right)+\beta \left(\frac{\ddot{H}}{2H^{3}}\right)\right].   \label{lgo7}
\end{eqnarray}
Instead, the  first derivative with respect to the cosmic time $t$ of the energy density of DE $\rho_D$ given in Eq. (\ref{lgo2}) is given by:
\begin{eqnarray}
\dot{\rho}_D = 6H^3 \left[ \alpha \left(\frac{\dot{H}}{H^{2}}\right)+\beta \left(\frac{\ddot{H}}{2H^{3}}\right)\right] \left( \frac{n^2}{8\pi G_{grav}} -\frac{\varepsilon \delta}{48\pi G_{grav}L_{GO}^{\delta -2}}      \right).  \label{lgo8}
\end{eqnarray}
where we have used the result of Eq. (\ref{lgo7}).\\
Differentiating the Friedmann equation given in Eq. (\ref{26}) with respect to the cosmic time $t$ and using Eq. (\ref{lgo8}), we obtain that the term $\left[ \alpha \left(\frac{\dot{H}}{H^{2}}\right)+\beta \left(\frac{\ddot{H}}{2H^{3}}\right)\right]$ can be rewritten as follows:
\begin{eqnarray}
\left[ \alpha \left(\frac{\dot{H}}{H^{2}}\right)+\beta \left(\frac{\ddot{H}}{2H^{3}}\right)\right]= \frac{1+\frac{\dot{H}}{H^2}+\Omega_D \left(\frac{u}{2} -1  \right) }{8\pi G_{cosmo}  \left( \frac{n^2}{8\pi G_{grav}}  -\frac{\varepsilon \delta}{48\pi G_{grav}L_{GO}^{\delta -2}}     \right)}, \label{lgo9}
\end{eqnarray}
where the dimensionless parameter $u$ is defined as follows:
\begin{eqnarray}
u= \frac{\Omega_m}{\Omega_D} = \frac{1-\Omega_k}{\Omega_D}-1. \label{lgu}
\end{eqnarray}
In Eq. (\ref{lgu}), we used the result of Eq. (\ref{32}) in order to find a result for $\Omega_m$.\\
Using the definition of $L_{GO}$ given in Eq. (\ref{lgo5}) and the expression of the fractional energy density of DE $\Omega_D$ given in Eq. (\ref{27}), after some algebraic calculations, we obtain that the term $\frac{\dot{H}}{H^2}$ can be written as follows:
\begin{eqnarray}
\frac{\dot{H}}{H^2} =\frac{1}{\beta} \left(\frac{\Omega_D}{n^2\gamma_n}  -\alpha \right) . \label{lgo6}
\end{eqnarray}
Inserting the result of Eq. (\ref{lgo6}) in Eq. (\ref{lgo9}) and later on  Eq. (\ref{lgo9}) into Eq. (\ref{lgo8}), we obtain the following expression for $\dot{\rho}_D$:
\begin{eqnarray}
\dot{\rho}_D = \frac{6H^3}{8\pi G_{cosmo} \beta}\left[\left(\frac{\Omega_D}{n^2\gamma_n}\right)  -\alpha + \beta  + \beta \Omega_D \left( \frac{u-2}{2} \right) \right],  \label{lgo10prim}
\end{eqnarray}
which leads to the following expression of the evolutionary form of the DE energy density $\rho'_D$:
\begin{eqnarray}
\rho'_D = \frac{\dot{\rho}_D}{H}  =\frac{6H^2}{8\pi G_{cosmo} \beta}\left[\left(\frac{\Omega_D}{n^2\gamma_n}\right) -\alpha + \beta  + \beta \Omega_D \left( \frac{u-2}{2} \right) \right].  \label{lgo10evo}
\end{eqnarray}
Using the definition of the fractional energy density of DE given in Eq. (\ref{lgo3}), we can rewrite the result of Eq. (\ref{lgo10evo}) as follows:
\begin{eqnarray}
\rho'_D = \frac{2\rho}{\Omega_D \beta}\left[\left(\frac{\Omega_D}{n^2\gamma_n}\right) -\alpha + \beta  + \beta \Omega_D \left( \frac{u-2}{2} \right) \right].  \label{lgo10evoother}
\end{eqnarray}
From the continuity equations for DE given in Eqs. (\ref{38}) and (\ref{38prime}), we can derive the following expression for the EoS parameter of DE $\omega_D$:
\begin{eqnarray}
\omega_D &=& -1 -\frac{\dot{\rho}_D}{3H\rho_D} \nonumber \\
&=& -1 -\frac{\rho'_D}{3\rho_D}. \label{omeganino}
\end{eqnarray}
Using in Eq. (\ref{omeganino}) the expression of $\dot{\rho}_D$ obtained in Eq. (\ref{lgo10prim}) or equivalently the expression of $\rho'_D$ obtained in Eq. (\ref{lgo10evo}), we can write $\omega_D$ as follows:
\begin{eqnarray}
\omega_D = -1 - \frac{2}{3\beta \Omega_D}\left[ \left(\frac{\Omega_D}{n^2 \gamma_n}\right) -\alpha + \beta + \beta \Omega_D \left(\frac{u}{2}  -1 \right)   \right].  \label{}
\end{eqnarray}
We now derive an expression for the evolutionary form of the fractional energy density of DE $\Omega_D'$.\\
Differentiating the expression of $\Omega_D$ given in Eq. (\ref{lgo3}) with respect to the variable $x$, we obtain the following expression:
\begin{eqnarray}
\Omega_D' = \Omega_D \left[  \frac{\rho_D'}{\rho_D} - 2\left(\frac{\dot{H}}{H^2}\right)  \right]  \label{maybritt}.
\end{eqnarray}
Using in Eq. (\ref{maybritt}) the expressions of  $\left(\frac{\dot{H}}{H^2}\right)$ and $\rho'_D$ given, respectively, in Eqs. (\ref{lgo6}) and (\ref{lgo10prim}), we obtain the evolutionary form of the fractional energy density of DE as follows:
\begin{eqnarray}
\Omega_D'= \frac{2}{\beta}\left[\left(\frac{\Omega_D}{n^2\gamma_n} - \alpha + \beta\right) \left( 1-\Omega_D  \right)  + \frac{\Omega_D \beta u}{2} \right].   \label{lgo11}
\end{eqnarray}

\subsection{Interacting Case}
We now extend the calculations we have made in the previous subsection obtaining the same cosmological quantities but in the case of presence of a kind of interaction between the two Dark Sectors, i.e. DM and DE.\\
The presence of an interaction between the two Dark Sectors implies that the energy conservation laws are not held separately, therefore we have the following expressions:
\begin{eqnarray}
\dot{\rho}_D +3H\left(1+\omega_D\right)\rho_D &=& -Q, \label{72} \\
\dot{\rho}_m +3H\rho_m &=& Q, \label{73}
\end{eqnarray}
which are equivalent to the following relations:
\begin{eqnarray}
\rho'_D +3\left(1+\omega_D\right)\rho_D &=& -\frac{Q}{H}, \label{72-1} \\
\rho'_m +3\rho_m &=& \frac{Q}{H}. \label{73-1}
\end{eqnarray}
The quantity $Q$ in the continuity equations given in Eqs. (\ref{72}), (\ref{73}), (\ref{72-1}) and (\ref{73-1}) indicates the interaction term which is in general a function of other cosmological parameters, like for example energy densities of DE and DM $\rho_D$ and $\rho_m$, the Hubble parameter $H$ and the deceleration parameter $q$. Many candidates have been suggested in order to describe the behavior of the interaction term $Q$, we decided to consider in this paper one interaction term which is proportional to the energy density of DE $\rho_D$ and to the Hubble parameter $H$ as follows \citep{A209}:
\begin{eqnarray}
    Q = 3b^2H \rho_D,\label{74}
\end{eqnarray}
where the quantity $b^2$ indicates a coupling parameter (also known with the name of transfer strength) between the Dark Sectors \citep{q1,q1-3,q1-5,q1-7,q1-10}. The limiting case corresponding to $b^2 = 0$ leads to the non-interacting FLRW Universe, which has been studied in the previous subsection.\\
The presence of interaction between DE and DM can be detected during the formation of the Large Scale Structures (LSS). It is suggested that the dynamical equilibrium of some collapsed structures like for example the clusters of galaxies (the cluster  Abell A586 is one good example) results to be modified because of the coupling between the two Dark Sectors \citep{1,10,9}. The main concept is that the virial theorem is affected by the exchange of energy between the two Dark Sectors which leads to a bias in the estimation made for the virial masses of clusters when the usual virial conditions are taken into account. This fact gives a probe in the near Universe of the coupling of the Dark components. Some other observational signatures on the dark sectors mutual interaction can be also found in the probes of the cosmic expansion history by using the results we know about Supernovae Ia (SNe Ia), Baryonic Acoustic Oscillations (BAO) and CMBR shift data \citep{A173,22}. Thanks to observational data of Gold SNe Ia samples, CMBR anisotropies and the Baryonic Acoustic Oscillations (BAO), it was possible to estimate that the coupling parameter between DM and DE must assume a small positive value. This condition satisfies the requirement for solving the cosmic coincidence problem and also constraints given by the second law of thermodynamics \citep{feng08}.
Observations of the CMBR and of clusters of galaxies indicate a value of the coupling parameter $b^2$ which is $b^2 < 0.025$. \citep{q4}. Negative values of  $b^2$ are not taken into account since they would lead to the violation of laws of thermodynamics.   We also need  to emphasize that other interaction terms can be also taken into account \citep{jamil-08-2008}.\\
Following the same procedure made for the non interacting case, we obtain the following expression for $\dot{\rho}_D$:
\begin{eqnarray}
\dot{\rho}_D &=& \frac{6H^3}{8\pi G_{cosmo} \beta}\left[\left(\frac{\Omega_D}{n^2\gamma_n}\right)  -\alpha + \beta  + \beta \Omega_D \left( \frac{u-2}{2} \right) - \left(\frac{8\pi G_{cosmo}}{6H^3}\right)\beta Q \right],  \label{lgo12!}
\end{eqnarray}
which leads to the following evolutionary form of the DE energy density:
\begin{eqnarray}
\rho'_D &=& \frac{6H^2}{8\pi G_{cosmo} \beta}\left[\left(\frac{\Omega_D}{n^2\gamma_n}\right)  -\alpha + \beta  + \beta \Omega_D \left( \frac{u-2}{2} \right)  -\left( \frac{8\pi G_{cosmo}}{6H^3}\right)\beta Q \right].  \label{lgo12}
\end{eqnarray}
Using in Eq. (\ref{lgo12}) the expression of $Q$ we have considered in Eq. (\ref{74}), we can write $\rho'_D$ as follows:
\begin{eqnarray}
\rho'_D &=& \frac{6H^2}{8\pi G_{cosmo} \beta}\left[\left(\frac{\Omega_D}{n^2\gamma_n}\right)  -\alpha + \beta  + \beta \Omega_D \left( \frac{u-2-3b^2}{2} \right) \right].  \label{lgo12evoint}
\end{eqnarray}
Using the definition of fractional energy density of DE given in Eq. (\ref{lgo3}), we can rewrite Eq. (\ref{lgo12evoint}) as follows:
\begin{eqnarray}
\rho'_D &=& \frac{2\rho_D}{\Omega_D \beta}\left[\left(\frac{\Omega_D}{n^2\gamma_n}\right)  -\alpha + \beta  + \beta \Omega_D \left( \frac{u-2-3b^2}{2} \right) \right].  \label{lgo12evoint2}
\end{eqnarray}
From the continuity equations for DE given in Eqs. (\ref{72}) and (\ref{72-1}), we can derive the following expression for the EoS parameter $\omega_D$:
\begin{eqnarray}
\omega_D &=& -1 -\frac{\dot{\rho}_D}{3H\rho_D} - \frac{Q}{3H\rho_D} \nonumber \\
&=& -1 -\frac{\rho'_D}{3\rho_D} - \frac{Q}{3H\rho_D}.\label{omeganinoint}
\end{eqnarray}
Using in Eq. (\ref{omeganinoint}) the expression of $\dot{\rho}_D$ obtained in Eq. (\ref{lgo12!}) or equivalently the expression of $\rho'_D$ obtained in Eq. (\ref{lgo12}) along with the definition of $Q$ given in Eq. (\ref{74}), we can write $\omega_D$ as follows:
\begin{eqnarray}
\omega_D = -1 - \frac{2}{3\beta \Omega_D}\left[ \frac{\Omega_D}{n^2 \gamma_n} -\alpha + \beta + \beta \Omega_D \left(\frac{u}{2}  -1 \right)   \right],\label{mona1}
\end{eqnarray}
which is the same result of the non interacting case.\\
We now want to find the expression for $\Omega_D'$ for the interacting case.\\
Using the general expression of $\Omega_D'$ given in Eq. (\ref{maybritt}) along with the expressions of $\frac{\dot{H}}{H^2}$ and $\rho_D'$ given, respectively, in Eqs. (\ref{lgo6}) and (\ref{lgo12}), we obtain the evolutionary form of the fractional energy density of DE as follows:
\begin{eqnarray}
\Omega_D' &=& \frac{2}{\beta}\left[\left(\frac{\Omega_D}{n^2\gamma_n} - \alpha + \beta\right)\left(   1-\Omega_D  \right)  + \frac{\Omega_D \beta u}{2} - \frac{3}{2}\Omega_D \beta b^2 \right]. \label{lgo13-1}
\end{eqnarray}

For completeness, we here obtain also the expression of the deceleration parameter $q$, which is generally defined as follows:
\begin{eqnarray}
	q&=&-\frac{\ddot{a}a}{\dot{a}^2} \nonumber \\
&=&  -\frac{\ddot{a}}{aH^2}  \nonumber \\
&=&-1-\frac{\dot{H}}{H^2}. \label{89}
\end{eqnarray}
The deceleration parameter $q$ can be used in order to quantify the status of the acceleration of the Universe \citep{dabro}. In particular, a negative value of the present day value of $q$ indicates an accelerating Universe, whereas a positive value of the present day value of $q$ indicates a Universe which is either decelerating or expanding at the coasting  \citep{alam}. In order to have a negative value of the deceleration parameter, we must have $\ddot{a}>0$.\\
Using the expression of $\frac{\dot{H}}{H^2}$ given in Eq. (\ref{lgo6}), we can write the deceleration parameter $q$ as follows:
\begin{eqnarray}
	q&=&-1-   \frac{1}{\beta}\left(  \frac{\Omega_D}{n^2\gamma_n}  -\alpha  \right)\nonumber \\
&=&-1 + \frac{\alpha}{\beta}  -\frac{1}{\beta}\left(  \frac{\Omega_D}{n^2\gamma_n}   \right)\nonumber \\
 &=& \frac{\left( \alpha - \beta  \right)n^2 \gamma_n - \Omega_D}{\beta n^2 \gamma_n} . \label{89-finale}
\end{eqnarray}
We can easily derive that the expression of $\omega_D$ given in Eq. (\ref{mona1}) and the expression of $q$ given in Eq. (\ref{89-finale}) are related through the following relation:
\begin{eqnarray}
	\omega_D = \frac{2q}{3\Omega_D} - \frac{1+u}{3}.
\end{eqnarray}
We must also underline that, in the limiting case of $b^2=0$ (i.e. in absence of interaction), we recover the same results of the non interacting case obtained in the previous subsection.

\section{Low Redshift Expansion}
In the previous Section, we have derived the general expression of the EoS parameter of DE $\omega_D$ as function of the other cosmological parameters.\\
We now consider a particular parametrization of the EoS parameter of DE $\omega_D$ as function of the redshift $z$, which is given by $\omega_D\left(z\right) = \omega_0 + \omega_1 z$, and we will calculate the expressions of the two parameters $\omega_0$ and $\omega_1$ for both non interacting and interacting Dark Sectors.

\subsection{Non Interacting Case}
\begin{figure}[ht]
\centering\includegraphics[width=8cm]{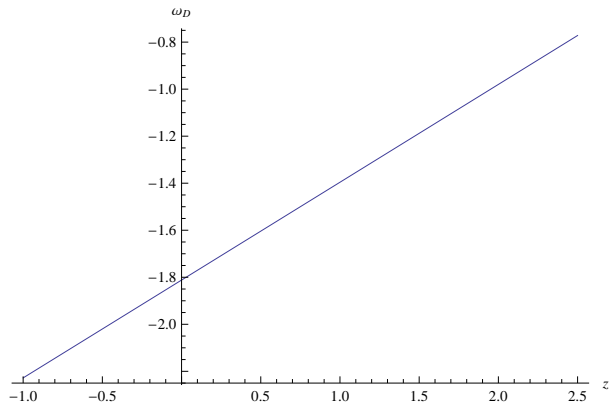}
\caption{Plot of  the EoS parameter of DE $\omega_D$ given in Eq. (\ref{ceizze2}) for $\lambda = 1.02$. } \label{omegad1}
\end{figure}

\begin{figure}[ht]
\centering\includegraphics[width=8cm]{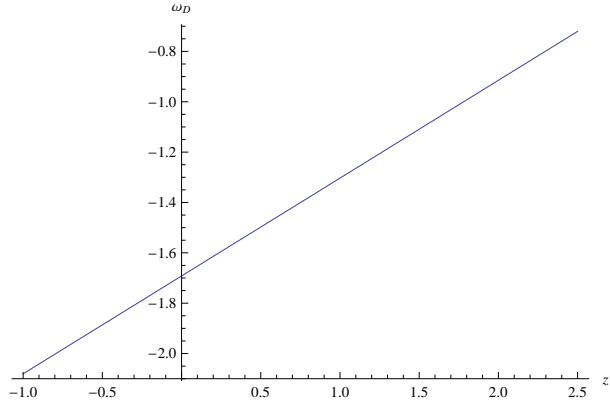}
\caption{Plot of  the EoS parameter of DE $\omega_D$ given in Eq. (\ref{ceizze1}) for $\lambda = 0.98$. } \label{omegad2}
\end{figure}

\begin{figure}[ht]
\centering\includegraphics[width=8cm]{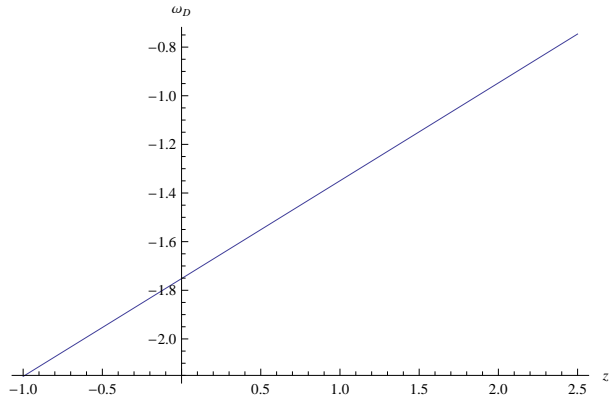}
\caption{Plot of  the EoS parameter of DE $\omega_D$ given in Eq. (\ref{ceizze3}) for $\lambda = 1.00$. } \label{omegad3}
\end{figure}

We start studying the case corresponding to absence of interaction between the two Dark Sectors. \\
As previously stated, the EoS parameter of DE $\omega_D$ written in a parameterized way as function of the redshift $z$ is given by the following relation \citep{28}:
\begin{eqnarray}
\omega_D\left(z\right) = \omega_0 + \omega_1 z, \label{49}
\end{eqnarray}
therefore we obtain an expression of the EoS parameter of DE $\omega_D$ which is function of the redshift $z$ too while the dependence on the other cosmological parameters will appear in the final expressions of the two parameters $\omega_0$ and $\omega_1$.\\
We must also remember here that the relation between the scale factor $a$ and the redshift $z$ is given by the following expression:
\begin{eqnarray}
a = \frac{1}{1+z} =\left( 1+z  \right)^{-1} ,     \label{50}
\end{eqnarray}
which leads to the following expression for the redshift $z$:
\begin{eqnarray}
 z= a^{-1}-1.     \label{50a}
\end{eqnarray}
Using the continuity equation for DE obtained in Eq. (\ref{38}) in the definition of $\omega_D$ given in (\ref{49}), we derive that the energy density of DE $\rho_D$ evolves according to the following relation \citep{84}:
\begin{eqnarray}
\frac{\rho_D}{\rho_{D_0}} = a^{-3 \left( 1+ \omega_0 - \omega_1 \right)} e^{3\omega_1z}, \label{51}
\end{eqnarray}
where $\rho_{D_0}$ indicates the present day value of the energy density of DE $\rho_D$.\\
The Taylor expansion of the energy density of DE $\rho_D$ around the point $a_0 = 1$ yields:
\begin{eqnarray}
\ln{\rho_D} = \ln{\rho_{D_0}} + \left. \frac{d\ln{ \rho_D}}{d\ln{ a}}\right|_0\ln{a} + \left. \frac{1}{2} \frac{d^2\ln{ \rho_D}}{d\left(\ln{ a}\right)^2}\right|_0\left(\ln{a}\right)^2 + ... , \label{52}
\end{eqnarray}
where $a_0$ indicates the present value of the scale factor $a$. Using the expression of $a$ given in Eq. (\ref{50}) in the Taylor expansion given in Eq. (\ref{52}), we obtain, for small redshifts, the following relation:
\begin{eqnarray}
\ln a = -\ln \left( 1+z  \right) \simeq -z + \frac{z^2}{2}, \label{53}
\end{eqnarray}
where we used the property of logarithm $\ln x^n = n \ln x$ (with $n$ real number) and the Taylor expansion of the logarithm  given by $\ln \left(1 + x \right) = \sum _{n=1}^{\infty} \left( -1  \right)^{n+1}\frac{x^n}{n} $.\\
Therefore, Eqs. (\ref{51}) and (\ref{52}) lead, respectively, to the following two relations:
\begin{eqnarray}
\frac{\ln{\left(\rho_D/\rho_{D_0}\right)}}{\ln{a}} &=& -3\left(1+\omega_0\right) - \frac{3}{2}\omega_1z, \label{54} \\
\frac{\ln{\left(\rho_D/\rho_{D_0}\right)}}{\ln{a}} &=& \left. \frac{d\ln{ \rho_D}}{d\ln{ a}}\right|_0 - \left.  \frac{1}{2} \frac{d^2\ln{ \rho_D}}{d\left(\ln{ a}\right)^2}\right|_0 z. \label{55}
\end{eqnarray}
Making a comparison of the results obtained in Eqs. (\ref{54}) and (\ref{55}), we easily derive the following relations for the two parameters $\omega_0$ and $\omega_1$:
\begin{eqnarray}
\omega_0  &=& \left. -\frac{1}{3}\frac{d\ln{ \rho_D}}{d\ln{ a}}\right|_0 - 1, \label{56} \\
\omega_1  &=& \left. \frac{1}{3}\frac{d^2\ln{ \rho_D}}{d\left(\ln{ a}\right)^2}\right|_0. \label{57}
\end{eqnarray}
Using the definition of DM and DE given, respectively, in Eqs. (\ref{28}) and (\ref{lgo3}), we can obtain the following relation between the critical energy density $\rho_{cr}$ and the energy densities of DM and DE $\rho_m$ and $\rho_D$:
\begin{eqnarray}
\rho_{cr} = \frac{\rho_m}{\Omega_m} = \frac{\rho_D}{\Omega_D},  \label{58a}
\end{eqnarray}
which leads to the following result for $\rho_D$:
\begin{eqnarray}
\rho_D = \left(\frac{\rho_m}{\Omega_m}\right)\Omega_D. \label{58b}
\end{eqnarray}
From the continuity equation for DM obtained in Eq. (\ref{39}), we derive that the energy density $\rho_m$ of DM evolves according to the relation $\rho_m = \rho_{m_0}a^{-3}$, where the constant $\rho_{m_0}$ indicates the present day value of $\rho_m$. Using the expression of $\rho_m$ we have above obtained along with the relation between all the fractional energy densities given in Eq. (\ref{32}), we can rewrite Eq. (\ref{58b}) as follows:
\begin{eqnarray}
\rho_D = \left[\frac{\rho_{m_0}a^{-3}}{\left(1 - \Omega_k - \Omega_D\right)}\right]\Omega_D. \label{58}
\end{eqnarray}
Substituting Eq. (\ref{58})  into Eq. (\ref{56}), we derive the following general expression for the parameter $\omega_0$:
\begin{eqnarray}
\omega_0 = -\frac{1}{3} \left[\frac{\Omega'_D}{\Omega_D} + \frac{\Omega'_D + \Omega'_k}{\left(1 - \Omega_k - \Omega_D\right)} \right]_0. \label{59}
\end{eqnarray}
Moreover, inserting the result of Eq. (\ref{58}) in Eq. (\ref{57}), we easily derive  the following general expression for the parameter $\omega_1$:
\begin{eqnarray}
\omega_1= \frac{1}{3} \left[\frac{\Omega''_D}{\Omega_D} - \frac{\Omega'^2_D}{\Omega^2_D} + \frac{\Omega''_D + \Omega''_k}{\left(1 - \Omega_k - \Omega_D\right)} + \frac{ \left( \Omega'_D + \Omega'_k \right)^2}{\left(1 - \Omega_k - \Omega_D\right)^2} \right]_0.
 \label{60}
\end{eqnarray}
We now want to obtain the final expression for both $\omega_0$ and $\omega_1$, therefore we must obtain the quantities involved in the relations we have obtained in Eqs. (\ref{59}) and (\ref{60}).\\
We have already obtained the evolutionary form of the fractional energy density of DE in Eq. (\ref{lgo11}), which is given by:
\begin{eqnarray}
\Omega_D'= \frac{2}{\beta}\left[\left(\frac{\Omega_D}{n^2\gamma_n} - \alpha + \beta\right) \left( 1-\Omega_D  \right)  + \frac{\Omega_D \beta u}{2} \right].   \label{lgo11new}
\end{eqnarray}
We now need to find the expressions of the quantities $\Omega_D''$, $\Omega_k'$ and $\Omega_k''$ in order to have the expressions of all the quantities necessary in order to calculate the final expressions of both $\omega_0$ and $\omega_1$.\\
Differentiating the expression of the fractional energy density of the curvature parameter $\Omega_k$ given in Eq. (\ref{30}) with respect to the variable $x$, we obtain the following relation:
\begin{eqnarray}
\Omega'_k = -2 \Omega_k \left(  1+ \frac{\dot{H}}{H^2}  \right). \label{lgo13}
\end{eqnarray}
Inserting in Eq. (\ref{lgo13}) the expression of $ \left(\frac{\dot{H}}{H^2}\right) $ derived in Eq. (\ref{lgo6}), we obtain the following final expression for the evolutionary form of the fractional energy density of curvature $\Omega_k'$:
\begin{eqnarray}
\Omega_k'=-\frac{2\Omega_k}{\beta}\left( \frac{\Omega_D}{n^2\gamma_n}  -\alpha + \beta  \right). \label{lgo15}
\end{eqnarray}
Differentiating the expression of $\Omega_D'$ given in Eq. (\ref{lgo11new}) with respect to the variable $x$, we derive that the second derivative of $\Omega_D$ with respect to the variable $x$ is given by:
\begin{eqnarray}
\Omega_D'' = \frac{2}{\beta}\left\{ \Omega_D'\left[\frac{1-2\Omega_D}{n^2\gamma_n} +\alpha + \beta\left(\frac{ u}{2}-1\right)  \right] +\Omega_D \left[ \frac{\beta u'}{2}
- \frac{\gamma_n'\left( 1-\Omega_D  \right)}{n^2 \gamma_n^2} \right]  \right\}. \label{lgo19}
\end{eqnarray}
Moreover, differentiating the expression of $\Omega_k'$ given in Eq. (\ref{lgo15}) with respect to the variable $x$, we derive that the second derivative of $\Omega_k$ with respect to the variable $x$ is given by:
\begin{eqnarray}
\Omega_k'' = -\frac{2}{\beta}\left[ \Omega'_k \left(\frac{\Omega_D}{n^2 \gamma_n}  +\beta - \alpha  \right)
+\frac{\Omega_k}{n^2 \gamma_n}\left( \Omega'_D - \frac{\Omega_D \gamma'_n}{\gamma_n}    \right)\right]. \label{lgo16}
\end{eqnarray}
We can now make some consideration about the quantities $\gamma_n$ and $u$ and also their derivatives and their present day values.\\
Using the definitions of $\gamma_n$ and $u$ given, respectively, in Eqs. (\ref{lgo4}) and (\ref{lgu}), we can easily derive their derivatives with respect to the variable $x$, which are given by the following relations:
\begin{eqnarray}
u'&=&\frac{\dot{u}}{H} = -\frac{\Omega_k'}{\Omega_D} - \left( 1 - \Omega_k\right)\frac{\Omega_D'}{\Omega_D^2}\nonumber \\
 &=&-\frac{\Omega_k'}{\Omega_D} - \left( u +1 \right) \frac{\Omega_D'}{\Omega_D}, \label{lgo17old} \\
\gamma_n' &=& \frac{H^2 \varepsilon \left( 2-\delta  \right)}{3 \beta n^2 L^{\delta-4}\left\{n^2   -\frac{\varepsilon \delta}{8L^{\delta -2}}   \right\}} \left\{ \beta +\frac{\Omega_D}{n^2 \gamma_n} -\alpha +\beta \Omega_D \left(\frac{u}{2} -1  \right) \right\}. \label{lgo18}
\end{eqnarray}
Using in Eq. (\ref{lgo17old}) the expressions of $\Omega_D'$ and $\Omega_k'$ derived, respectively, in Eqs. (\ref{lgo11}) and (\ref{lgo13}), we obtain the following expression for $u'$:
\begin{eqnarray}
u'&=&\frac{2}{\beta}\left( \frac{1}{n^2 \gamma_n} + \frac{\beta - \alpha}{\Omega_D}   \right) \left[\Omega_k - \left(u+1   \right)  \left(1-\Omega_D  \right)   \right] -u\left( u+1  \right). \label{lgo17}
\end{eqnarray}
In order to study the behavior of the quantities $\gamma_{n}$ and $\gamma_{n}'$, we need to make some preliminary considerations. We know that the power law correction to the entropy area law gives a reasonable contribution only at early stages of the Universe while with the passing of the time its contribution becomes less important. For this reason, we can tell that, for $\gamma_{n}$ and $\gamma_{n}'$, the power law corrections can be considered practically negligible. We then have:
\begin{eqnarray}
\gamma_{n} &=& \frac{G_{cosmo}}{G_{grav}} = \frac{2}{3\lambda -1}, \label{lim1}\\
\gamma_{n}' &=& 0. \label{lim2}
\end{eqnarray}
Taking into account the above reasons, in the following calculations, we will neglect the derivatives of $\gamma_{n}$ and we will consider as expression for $\gamma_{n}$ the one obtained in Eq. (\ref{lim1}).\\
Using the general definition of $u$ given in Eq. (\ref{lgu}), we have that $u_0$ is given by the following quantity:
\begin{eqnarray}
u_0= \frac{1-\Omega_{k_0}}{\Omega_{D_0}} -1, \label{u0}
\end{eqnarray}
while, using the result of Eq. (\ref{lgo17old}), we have that $u_0'$ can be written as follows:
\begin{eqnarray}
u_0'&=& \frac{2}{\beta}\left( \frac{1}{n^2\gamma_{n_0}} +\frac{\beta - \alpha}{\Omega_{D_0}} \right)  \left[  \Omega_{k_0}  - \left(u_0+1 \right)\left( 1-  \Omega_{D_0} \right) \right]  -u_0\left(u_0+1 \right) .\label{u0prime}
\end{eqnarray}
Moreover, we have that the present day value of $\gamma_{n}$ is given by:
\begin{eqnarray}
\gamma_{n_0} &=&   \frac{G_{cosmo}}{G_{grav}} = \frac{2}{3\lambda -1}, \label{lim1-1}
\end{eqnarray}
i.e. $\gamma_{n_0}$ is equivalent to the expression of $\gamma_{n}$, $\gamma_{n_0} =  \gamma_{n}  $.\\
Therefore, using the expression of $\gamma_{n_0} $ obtained in Eq. (\ref{lim1-1}), we can write $u_0'$ as folloes:
\begin{eqnarray}
u_0'&=& \frac{2}{\beta}\left( \frac{3\lambda-1}{2n^2} +\frac{\beta - \alpha}{\Omega_{D_0}} \right)  \left[  \Omega_{k_0}  - \left(u_0+1 \right)\left( 1-  \Omega_{D_0} \right) \right]  -u_0\left(u_0+1 \right) .\label{u0prime}
\end{eqnarray}
Inserting the results of Eqs. (\ref{lgo11}) and (\ref{lgo15}) in the general expression of the parameter $\omega_0$ given in Eq. (\ref{59}), we derive the following expression for $\omega_0$:
\begin{eqnarray}
\omega_0 &=& -\frac{2}{3\beta}\left[\left(\frac{1}{n^2\gamma_{n_0}} +  \frac{\beta - \alpha}{\Omega_{D_0}}\right) + \frac{\beta u_0}{2}\left( \frac{1-\Omega_{k_0}}{1-\Omega_{D_0} - \Omega_{k_0}}  \right)\right]. \label{59new}
\end{eqnarray}
We now want to calculate the terms involved in the expression of the parameter $\omega_1$.\\
Using the definition of the evolutionary form of the fractional energy density of DE $\Omega_D'$ given in Eq. (\ref{lgo11}), we have that:
\begin{eqnarray}
\frac{\Omega_D'}{\Omega_D}= \frac{2}{\beta}\left[\left(\frac{\Omega_D}{n^2\gamma_n} - \alpha + \beta\right) \frac{\left( 1-\Omega_D  \right)}{\Omega_D}  + \frac{ \beta u}{2} \right].   \label{lgo11newNAN}
\end{eqnarray}
Using the expression of $\Omega_D''$ obtained in Eq. (\ref{lgo19}), we have that the term $\left(\frac{\Omega_D''}{\Omega_D}\right)$ is equal to:
\begin{eqnarray}
\frac{\Omega_D''}{\Omega_D} = \frac{2}{\beta \Omega_D}\left\{ \Omega_D'\left[\frac{1-2\Omega_D}{n^2\gamma_n} +\alpha + \beta\left(\frac{ u}{2}-1\right)  \right] \right\} + u'. \label{iele1NON}
\end{eqnarray}
Adding the expressions of $\Omega_D''$ and $\Omega_k^{''}$ given, respectively, in Eqs. (\ref{lgo19}) and (\ref{lgo16}), we obtain:
\begin{eqnarray}
\frac{\Omega_D^{''}+\Omega_k^{''}}{1-\Omega_D-\Omega_k} &=& \frac{2}{\beta}\left\{\Omega_D' \left[\frac{1}{n^2\gamma_n} + \frac{\beta u}{2\left(  1-\Omega_D-\Omega_k \right)}   \right]    \right\} \nonumber \\
&& -\frac{2}{\beta}\frac{\Omega_D' + \Omega_k'}{1-\Omega_D-\Omega_k}\left( \frac{\Omega_D}{n^2\gamma_n} -\alpha + \beta \right) + \frac{ u'\Omega_D}{1-\Omega_D-\Omega_k}. \label{iele2}
\end{eqnarray}
Therefore, adding Eqs. (\ref{iele1NON}) and (\ref{iele2}), we can write:
\begin{eqnarray}
\frac{\Omega_D''}{\Omega_D} + \frac{\Omega_D^{''}+\Omega_k^{''}}{1-\Omega_D-\Omega_k} &=& \frac{2}{\beta \Omega_D}\left\{ \Omega_D'\left[\frac{1-2\Omega_D}{n^2\gamma_n} +\alpha + \beta\left(\frac{ u}{2}-1\right)  \right] \right\} \nonumber \\
&&+\frac{2}{\beta}\left\{\Omega_D' \left[\frac{1}{n^2\gamma_n} + \frac{\beta u}{2\left(  1-\Omega_D-\Omega_k \right)}   \right]    \right\}  \nonumber \\
&&-\frac{2}{\beta}\left(\frac{\Omega_D' + \Omega_k'}{1-\Omega_D-\Omega_k}\right)\left( \frac{\Omega_D}{n^2\gamma_n} -\alpha + \beta \right) + \frac{\left(1-\Omega_k\right)u'}{1-\Omega_D-\Omega_k}. \label{ieleNA2}
\end{eqnarray}
Finally, adding the expressions of the evolutionary form of the fractional energy density of DE $\Omega_D'$ and $\Omega_k^{'}$ given, respectively, in  Eqs. (\ref{lgo11}) and (\ref{lgo15}), we obtain:
\begin{eqnarray}
\frac{\Omega_D'+\Omega_k'}{1-\Omega_D-\Omega_k} = \frac{2}{\beta}\left[\left( \frac{\Omega_D}{n^2\gamma_n} -\alpha + \beta \right) +\frac{\beta u \Omega_D}{2\left( 1-\Omega_D-\Omega_k\right)}    \right].\label{iele3non}
\end{eqnarray}
Therefore, considering the results of above equations in Eq. (\ref{60}), the final expression of the parameter $\omega_1$ can be written as follows:
\begin{eqnarray}
\omega_1&=& \frac{4}{3\beta^2 \Omega_{D_0}}\left\{\left[\left(\frac{\Omega_{D_0}}{n^2\gamma_{n_0}} - \alpha + \beta\right) \left( 1-\Omega_{D_0}  \right)  + \frac{\Omega_{D_0} \beta u_0}{2} \right]\left[\frac{1-2\Omega_{D_0}}{n^2\gamma_{n_0}} +\alpha + \beta\left(\frac{ u_0}{2}-1\right)  \right] \right\} \nonumber \\
&&+\frac{4}{3\beta^2}\left\{\left[\left(\frac{\Omega_{D_0}}{n^2\gamma_{n_0}} - \alpha + \beta\right) \left( 1-\Omega_{D_0}  \right)  + \frac{\Omega_{D_0} \beta u_0}{2} \right] \left[\frac{1}{n^2\gamma_{n_0}} + \frac{\beta u_0}{2\left(  1-\Omega_{D_0}-\Omega_{k_0} \right)}   \right]    \right\}  \nonumber \\
&&-\frac{4}{3\beta^2}\left( \frac{\Omega_{D_0}}{n^2\gamma_{n_0}} -\alpha + \beta \right)\left[\left( \frac{\Omega_{D_0}}{n^2\gamma_{n_0}} -\alpha + \beta \right) +\frac{\beta u_0 \Omega_{D_0}}{2\left( 1-\Omega_{D_0}-\Omega_{k_0}\right)}    \right] \nonumber \\
&&+ \left\{  \frac{2}{\beta}\left( \frac{1}{n^2 \gamma_{n_0}} + \frac{\beta - \alpha}{\Omega_{D_0}}   \right) \left[\Omega_{k_0} - \left(u_0+1   \right)  \left(1-\Omega_{D_0}  \right)   \right] -u_0\left( u_0+1  \right)  \right\}\frac{1-\Omega_{k_0}}{3\left(1-\Omega_{D_0}-\Omega_{k_0}\right)}\nonumber \\
&&-\frac{4}{3\beta^2}\left[\left(\frac{\Omega_{D_0}}{n^2\gamma_{n_0}} - \alpha + \beta\right) \frac{\left( 1-\Omega_{D_0}  \right)}{\Omega_{D_0}}  + \frac{ \beta u_0}{2} \right]^2 \nonumber \\
&&+\frac{4}{3\beta^2}\left[\left( \frac{\Omega_{D_0}}{n^2\gamma_{n_0}} -\alpha + \beta \right) +\frac{\beta u_0 \Omega_{D_0}}{2\left( 1-\Omega_{D_0}-\Omega_{k_0}\right)}\right]^2 . \label{71}
\end{eqnarray}
We must emphasize here that  $\Omega_{D_0}$ and $\Omega_{k_0}$ represent, respectively, the present day values of the fractional energy densities for DE and curvature parameter $k$.\\
Using the expression of $\gamma_{n_0}$ obtained in Eq. (\ref{lim1-1}), we can finally write $\omega_0$ and $\omega_1$ as follows:
\begin{eqnarray}
\omega_0 &=& -\frac{2}{3\beta}\left[\left(\frac{3\lambda -1}{2n^2} +  \frac{\beta - \alpha}{\Omega_{D_0}}\right) + \frac{\beta u_0}{2}\left( \frac{1-\Omega_{k_0}}{1-\Omega_{D_0} - \Omega_{k_0}}  \right)\right], \label{59newfine}\\
\omega_1&=& \frac{4}{3\beta^2 \Omega_{D_0}}\left\{\left[\left(\frac{\left(3\lambda -1 \right)\Omega_{D_0}}{2n^2} - \alpha + \beta\right) \left( 1-\Omega_{D_0}  \right)  + \frac{\Omega_{D_0} \beta u_0}{2} \right]\times \right. \nonumber \\
&&\left.\left[\frac{\left(1-2\Omega_{D_0}\right)\left(3\lambda -1 \right)}{2n^2} +\alpha + \beta\left(\frac{ u_0}{2}-1\right)  \right] \right\} \nonumber \\
&&+\frac{4}{3\beta^2}\left\{\left[\left(\frac{\left(3\lambda -1 \right)\Omega_{D_0}}{2n^2} - \alpha + \beta\right) \left( 1-\Omega_{D_0}  \right)  + \frac{\Omega_{D_0} \beta u_0}{2} \right] \times \right. \nonumber \\
&&\left. \left[\frac{\left(3\lambda -1 \right)}{2n^2 } + \frac{\beta u_0}{2\left(  1-\Omega_{D_0}-\Omega_{k_0} \right)}   \right]    \right\}  \nonumber \\
&&-\frac{4}{3\beta^2}\left[ \frac{\left(3\lambda -1 \right)\Omega_{D_0}}{2n^2 } -\alpha + \beta \right]\left\{\left[ \frac{\left(3\lambda -1 \right)\Omega_{D_0}}{2n^2 } -\alpha + \beta \right] +\frac{\beta u_0 \Omega_{D_0}}{2\left( 1-\Omega_{D_0}-\Omega_{k_0}\right)}    \right\} \nonumber \\
&&+ \left\{  \frac{2}{\beta}\left[ \frac{\left(3\lambda -1 \right)}{2n^2  } + \frac{\beta - \alpha}{\Omega_{D_0}}   \right] \left[\Omega_{k_0} - \left(u_0+1   \right)  \left(1-\Omega_{D_0}  \right)   \right] -u_0\left( u_0+1  \right)  \right\}\left(\frac{1-\Omega_{k_0}}{3\left(1-\Omega_{D_0}-\Omega_{k_0}\right)}\right)\nonumber \\
&&-\frac{4}{3\beta^2}\left\{\left[\frac{\left(3\lambda -1 \right)\Omega_{D_0}}{2n^2 } - \alpha + \beta\right] \frac{\left( 1-\Omega_{D_0}  \right)}{\Omega_{D_0}}  + \frac{ \beta u_0}{2} \right\}^2 \nonumber \\
&&+\frac{4}{3\beta^2}\left\{\left[ \frac{\left(3\lambda -1 \right)\Omega_{D_0}}{2n^2 } -\alpha + \beta \right] +\frac{\beta u_0 \Omega_{D_0}}{2\left( 1-\Omega_{D_0}-\Omega_{k_0}\right)}\right\}^2 .   \label{71nino}
\end{eqnarray}
We now want to calculate the values of $\omega_0$ and $\omega_1$ for three different values of the running parameter $\lambda$, in particular $\lambda =1.02$ and
$\lambda = 0.98$ (which are values suggested by the work of Dutta and Saridakis) and $\lambda =1$ which leads to the Lorentz invariance case. Moreover, we will make the same examples also for the limiting case corresponding to the Ricci scalar curvature, i.e. when $\alpha =2$ and $\beta =1$.\\
Inserting in Eqs. (\ref{59newfine}) and (\ref{71nino}) the values of the parameters involved, we obtain, for $\lambda = 1.02$:
\begin{eqnarray}
\omega_0 &\approx & -1.81193, \\
\omega_1 &\approx & 0.416014,
\end{eqnarray}
therefore we obtain the following equation for the EoS parameter of DE $\omega_D$:
\begin{eqnarray}
\omega_D &\approx& -1.81193 + 0.416014z. \label{ceizze2}
\end{eqnarray}
In Figure \ref{omegad1}, we plot the behavior of  $\omega_D$ given in Eq. (\ref{ceizze2}). \\
At present time, i.e. for $z=0$, Eq. (\ref{ceizze2}) leads to $\omega_D \approx -1.81193 $ while the value $\omega_D = -1$ is obtained for a redshift of $z\approx 1.95$.\\
Moreover, inserting in Eqs. (\ref{59newfine}) and (\ref{71nino}) the values of the parameters involved, we obtain, for $\lambda = 0.98$:
\begin{eqnarray}
\omega_0 &=& -1.69188, \\
\omega_1 &=& 0.388546 ,
\end{eqnarray}
therefore we derive the following equation for the EoS parameter of DE $\omega_D$:
\begin{eqnarray}
\omega_D &\approx & -1.69188 + 0.388546 z. \label{ceizze1}
\end{eqnarray}
In Figure \ref{omegad2}, we plot the behavior of the EoS parameter of DE $\omega_D$ given in Eq. (\ref{ceizze1}). \\
At present time, i.e. for  $z=0$, Eq. (\ref{ceizze1}) leads to $\omega_D \approx -1.69188$ while the value $\omega_D = -1$ is obtained for a redshift of $z\approx 1.78$.\\
Finally, inserting in Eqs. (\ref{59newfine}) and (\ref{71nino}) the values of the parameters involved, we obtain, for $\lambda = 1$:
\begin{eqnarray}
\omega_0 &=& -1.75191, \\
\omega_1 &=& 0.40228,
\end{eqnarray}
therefore we derive the following equation for  the EoS parameter of DE $\omega_D$:
\begin{eqnarray}
\omega_D &\approx & -1.75191 + 0.40228   z. \label{ceizze3}
\end{eqnarray}
In Figure \ref{omegad3}, we plot the behavior of  the EoS parameter of DE  $\omega_D$ given in Eq. (\ref{ceizze3}). \\
At present time, i.e. for  $z=0$, Eq. (\ref{ceizze3}) leads to $\omega_D \approx -1.75191$ while the value $\omega_D = -1$ is obtained for a redshift of $z\approx 1.86$.\\

\begin{figure}[ht]
\centering\includegraphics[width=8cm]{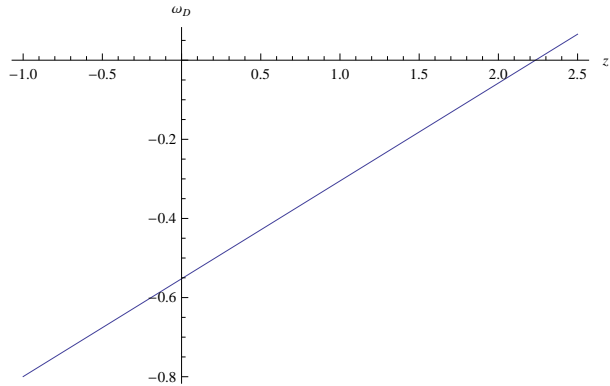}
\caption{Plot of  the EoS parameter of DE $\omega_D$ given in Eq. (\ref{ceizzericci1}) for $\lambda = 1.02$ in the limiting  case of Ricci scale. } \label{omegad4}
\end{figure}

\begin{figure}[ht]
\centering\includegraphics[width=8cm]{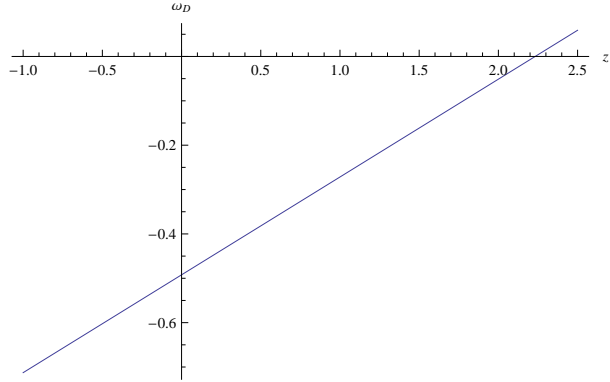}
\caption{Plot of  the EoS parameter of DE $\omega_D$ given in Eq. (\ref{ceizzericci2})  for $\lambda = 0.98$ in the limiting  case of Ricci scale. } \label{omegad5}
\end{figure}

\begin{figure}[ht]
\centering\includegraphics[width=8cm]{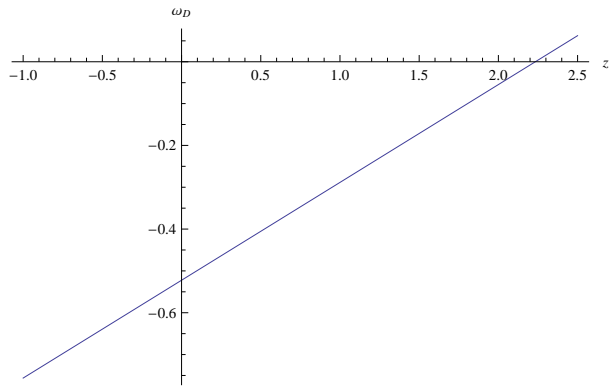}
\caption{Plot of  the EoS parameter of DE $\omega_D$ given in Eq. (\ref{ceizzericci3})  for $\lambda = 1.00$ in the limiting  case of Ricci scale. } \label{omegad6}
\end{figure}

We now want to obtain the expressions for $\omega_D$ for the Ricci scale, which is recovered for $\alpha =2$ and $\beta=1$.\\
Inserting in Eqs. (\ref{59newfine}) and (\ref{71nino}) the values of the parameters involved, we obtain, for $\lambda = 1.02$:
\begin{eqnarray}
\omega_0 &\approx & -0.552428, \\
\omega_1 &\approx &  0.247249,
\end{eqnarray}
therefore we obtain the following equation for the EoS parameter of DE $\omega_D$:
\begin{eqnarray}
\omega_D &\approx& -0.552428 + 0.247249 z. \label{ceizzericci1}
\end{eqnarray}
In Figure \ref{omegad4}, we plot the behavior of the EoS parameter of DE $\omega_D$ given in Eq. (\ref{ceizzericci1}). \\
At present time, i.e. for  $z=0$, Eq. (\ref{ceizzericci1}) leads to $\omega_D \approx  -0.552428$ while the value $\omega_D = -1$ is obtained for a redshift of $z\approx -1.81$.\\
Moreover, inserting in Eqs. (\ref{59newfine}) and (\ref{71nino}) the values of the parameters involved, we obtain, for $\lambda = 0.98$:
\begin{eqnarray}
\omega_0 &=&  -0.492207, \\
\omega_1 &=&0.220506,
\end{eqnarray}
therefore we derive the following equation for the EoS parameter of DE $\omega_D$:
\begin{eqnarray}
\omega_D &\approx & -0.492207 + 0.220506 z. \label{ceizzericci2}
\end{eqnarray}
In Figure \ref{omegad5}, we plot the behavior of the EoS parameter of DE  $\omega_D$ given in Eq. (\ref{ceizzericci2}). \\
At present time, i.e. for  $z=0$, Eq. (\ref{ceizzericci1}) leads to $\omega_D \approx -0.492207$ while the value $\omega_D = -1$ is obtained for a redshift of $z\approx -2.30$.\\
Finally, inserting in Eqs. (\ref{59newfine}) and (\ref{71nino}) the values of the parameters involved, we obtain, for $\lambda = 1$:
\begin{eqnarray}
\omega_0 &=&  -0.522318, \\
\omega_1 &=& 0.233877,
\end{eqnarray}
therefore we derive the following equation for the EoS parameter of DE $\omega_D$:
\begin{eqnarray}
\omega_D &\approx & -0.522318 + 0.233877 z. \label{ceizzericci3}
\end{eqnarray}
In Figure \ref{omegad6}, we plot the behavior of the EoS parameter of DE  $\omega_D$ given in Eq. (\ref{ceizzericci3}). \\
At present time, i.e. for  $z=0$, Eq. (\ref{ceizzericci1}) leads to $\omega_D \approx  -0.522318$ while the value $\omega_D = -1$ is obtained for a redshift of $z\approx -1.84$.\\

We now want ot obtain the present day  values of the deceleration parameter $q$ for the three different values of the running parameter $\lambda$ we are considering and for both sets of values of $\alpha$ and $\beta$ we are studying.\\
Using the general expression of $q$ obtained in Eq. (\ref{89-finale}) along with the considerations done in this Section, we have that the present day value of the deceleration parameter $q$ (indicated with $q_0)$ is given by:
\begin{eqnarray}
	q_0 &=& \frac{\left( \alpha - \beta  \right)n^2 \gamma_{n_0} - \Omega_{D_0}}{\beta n^2 \gamma_{n_0}} .  \label{89-finale-2old}
\end{eqnarray}
Using in Eq. (\ref{89-finale-2old})  the expression of $\gamma_{n0}$ given in Eq. (\ref{lim1-1}), we can write $q_0$ as follows:
\begin{eqnarray}
	q_0  &=&    -1 + \frac{\alpha}{\beta} - \frac{\left(3\lambda -1   \right)\Omega_{D_0}}{2\beta n^2} .  \label{89-finale-2}
\end{eqnarray}
We can now calculate the value of $q_0$ according to the values of the parameters involved.\\
We first consider the case with $\alpha =0.8824$ and $\beta =0.5016$.\\
We have that, for $\lambda = 1.02$, $q_0$ assumes the value of $-1.37734$, for $\lambda = 0.98$, it assumes the value of $-1.25288$ while for $\lambda = 1.00$ it assumes the value of $-1.31511$.\\
We now consider the limiting case of Ricci scale, which is recovered for $\alpha =2$ and $\beta =1$.\\
We obtain that, for $\lambda = 1.02$, $q_0$ assumes the value of $-0.0716745$, for $\lambda = 0.98$, it assumes the value of $-0.00924687$ while for $\lambda = 1.00$ it assumes the value of $-0.0404607$.\\
We can see, then, that for all the values of the running parameter $\lambda$ considered, the present day value of the deceleration parameter $q_0$ assumes a negative value for both set of values of $\alpha$ and $\beta$ we have chosen, which indicates an accelerated expansion of the Universe, result which is in agreement with the most recent cosmological observations. We also observe that for the Ricci scale case, we obtain values of $q_0$ which are closer to $q=0$ (which indicates the transition from decelerated to accelerated Universe) with respect to the case with $\alpha =0.8824$ and $\beta =0.5016$.

\subsection{Interacting Case}
\begin{figure}[ht]
\centering\includegraphics[width=8cm]{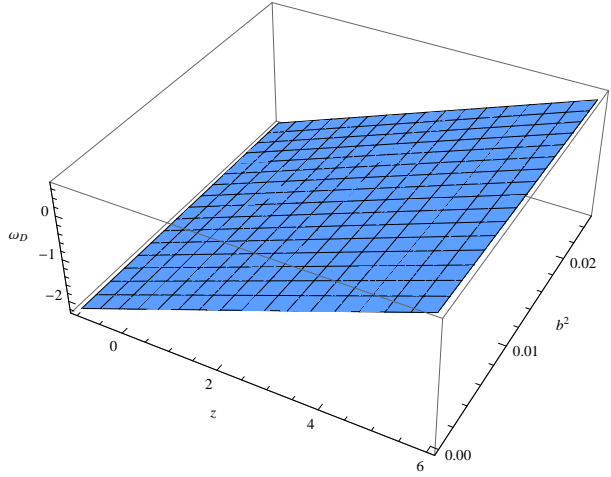}
\caption{Plot of  the EoS parameter of DE $\omega_D$ given in Eq. (\ref{ceizze2int}) for $\lambda = 1.02$. } \label{omegad7}
\end{figure}

\begin{figure}[ht]
\centering\includegraphics[width=8cm]{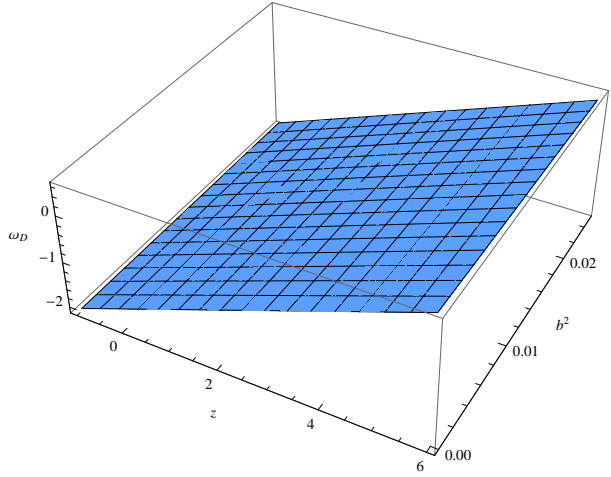}
\caption{Plot of  the EoS parameter of DE $\omega_D$ given in Eq. (\ref{ceizze1intlalla})  for $\lambda = 0.98$. } \label{omegad8}
\end{figure}

\begin{figure}[ht]
\centering\includegraphics[width=8cm]{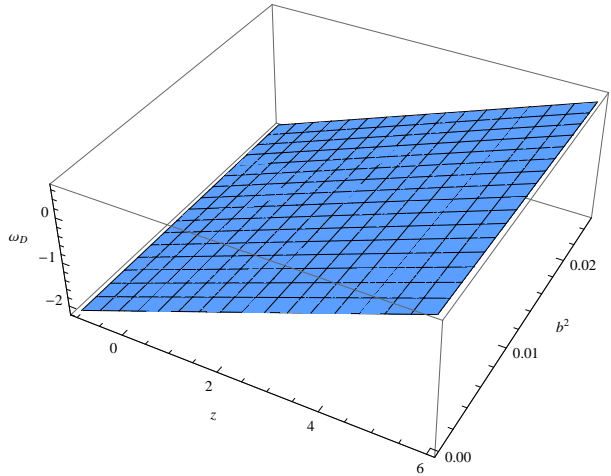}
\caption{Plot of  the EoS parameter of DE $\omega_D$ given in Eq. (\ref{ceizzeintint})  for $\lambda = 1.00$. } \label{omegad9}
\end{figure}

We now consider the interacting case, obtaining the same quantities of the previous subsection but with the contribution produced by the interaction between the two Dark Sectors.\\
Using the result of Eq. (\ref{58b}) along with the relation between all the fractional energy densities derived in Eq. (\ref{32}), we can write the following relation for $\rho_D$:
\begin{eqnarray}
\rho_D = \left( \frac{\rho_m}{\Omega_m}\right)\Omega_D =  \left[\frac{\rho_m}{\left(1 - \Omega_k - \Omega_D\right)}\right]\Omega_D. \label{76}
\end{eqnarray}
From the result of Eq. (\ref{76}), we can easily derive the following expression:
\begin{eqnarray}
\frac{d\ln{\rho_D}}{d\ln{a}} = \frac{\rho'_m}{\rho_m} - \frac{\Omega'_m}{\Omega_m} + \frac{\Omega'_D}{\Omega_D}. \label{77}
\end{eqnarray}
We also obtain, using the results of Eqs. (\ref{72}) and (\ref{49}), that the energy density of DE $\rho_D$ evolves according to the following relation:
\begin{eqnarray}
\frac{\rho_D}{\rho_{D_0}} = a^{-3\left(1+\omega_0 - \omega_1+b^2\right)}e^{3\omega_1 z}.
\label{78}
\end{eqnarray}
Using the Taylor expansion given in Eq. (\ref{52}) for small redshifts, Eq. (\ref{78}) leads to the following relation:
\begin{eqnarray}
\frac{\ln{\left(\rho_D / \rho_{D_0}\right)}}{\ln{a}} = -3\left(1+\omega_0+b^2\right) - \frac{3}{2} \omega_1 z. \label{79}
\end{eqnarray}
Comparing the results of Eqs. (\ref{78}) and (\ref{79}), we obtain that the parameters $\omega_0$ and $\omega_1$  for the interacting DE and DM can be written as follows:
\begin{eqnarray}
\omega_0 &=& \left. -\frac{1}{3}\frac{d\ln{\rho_D}}{d\ln{a}}\right|_0 - 1 - b^2, \label{80} \\
\omega_1 &=& \left. \frac{1}{3}\frac{d^2\ln{\rho_D}}{d \left(\ln{a}\right)^2}\right|_0. \label{81}
\end{eqnarray}
Inserting the result of Eq. (\ref{77}) in Eq. (\ref{80}) and considering the result of Eq. (\ref{73}), it is possible to write the parameter $\omega_0$ as follows:
\begin{eqnarray}
\omega_0 = -\frac{1}{3}\left[\frac{\Omega'_D}{\Omega_D}+\frac{\Omega'_D + \Omega'_k}{\left(1 - \Omega_k - \Omega_D\right)} \right]_0-b^2\left(\frac{1-\Omega_k}{1 - \Omega_k - \Omega_D}\right)_0. \label{82}
\end{eqnarray}
Moreover,  we also obtain the following relation for the parameter $\omega_1$:
\begin{eqnarray}
\omega_1 &=& \frac{1}{3}\left[\frac{3b^2\Omega'_D}{1 - \Omega_k - \Omega_D} + \frac{3b^2\Omega_D\left(\Omega'_D+\Omega'_k\right)}{\left(1 - \Omega_k - \Omega_D\right)^2}+\frac{\Omega''_D}{\Omega_D} \right. \nonumber\\
&-& \left. \frac{\Omega'^2_D}{\Omega^2_D}+\frac{\Omega''_D+\Omega''_k}{1 - \Omega_k - \Omega_D}+\frac{\left(\Omega'_D+\Omega'_k\right)^2}{\left(1 - \Omega_k - \Omega_D\right)^2}\right]_0. \label{83}
\end{eqnarray}
We now derive the explicit forms of the parameter $\omega_0$ and $\omega_1$ for the interacting case using the results of Eqs. (\ref{82}) and (\ref{83}). Therefore, we need to calculate the expressions of the quantities involved. \\
We have already obtained the expression of the evolutionary form of the fractional energy density of DE $\Omega_D'$ for the interacting case in Eq. (\ref{lgo13-1}) and it is given by the following relation:
\begin{eqnarray}
\Omega_D' &=& \frac{2}{\beta}\left[\left(\frac{\Omega_D}{n^2\gamma_n} - \alpha + \beta\right)\left(   1-\Omega_D  \right)  + \frac{\Omega_D \beta \left(u-3b^2\right)}{2} \right]. \label{paolone1}
\end{eqnarray}
Differentiating the result of Eq. (\ref{paolone1}) with respect to the variable $x$ leads to the following expression for $\Omega_D''$:
\begin{eqnarray}
\Omega_D'' = \frac{2}{\beta}\left\{ \Omega_D'\left[\frac{1-2\Omega_D}{n^2\gamma_n} +\alpha + \beta\left(\frac{u -2 -3b^2}{2}\right)  \right]
 + \frac{\Omega_D \beta u'}{2}    \right\}. \label{paolone2}
\end{eqnarray}
The expression of $\Omega_k'$ and $\Omega_k''$ are the same as those obtained for the non interacting case and given, respectively, in Eqs. (\ref{lgo15}) and (\ref{lgo16}).\\
Also in the interacting case, the expressions of $\gamma_{n}$ and $\gamma_{n}'$ are the same as the non-interacting case, given respectively in Eqs. (\ref{lim1}) and (\ref{lim2}). Instead, the expression of $u'$ is given by:
\begin{eqnarray}
u' &=& \frac{2}{\beta}\left( \frac{1}{n^2 \gamma_n} + \frac{\beta - \alpha}{\Omega_D}   \right) \left[\Omega_k - \left(u+1   \right)  \left(1-\Omega_D  \right)   \right] -\left(u -3b^2 \right)\left( u+1  \right), \label{lgo17int}
\end{eqnarray}
which implies that the final expression $u_0'$ is given by the following relation:
\begin{eqnarray}
u_0'&=& \frac{2}{\beta}\left( \frac{1}{n^2\gamma_{n_0}} +\frac{\beta - \alpha}{\Omega_{D_0}} \right)  \left[  \Omega_{k_0}  - \left(u_0+1 \right)\left( 1- \Omega_{D_0} \right) \right]  -\left( u_0 -3b^2\right)\left(u_0+1 \right). \label{lgo17-2-2old}
\end{eqnarray}
Using the expression of $\gamma_{n_0}$ obtained in Eq. (\ref{lim1-1}), we can write Eq. (\ref{lgo17-2-2old}) as follows:
\begin{eqnarray}
u_0'&=& \frac{2}{\beta}\left( \frac{3\lambda-1}{2n^2} +\frac{\beta - \alpha}{\Omega_{D_0}} \right)  \left[  \Omega_{k_0}  - \left(u_0+1 \right)\left( 1- \Omega_{D_0} \right) \right]  -\left( u_0 -3b^2\right)\left(u_0+1 \right). \label{lgo17-2-2}
\end{eqnarray}
Inserting  in Eq. (\ref{82}) the results obtained in Eqs. (\ref{paolone1}) and (\ref{paolone2}), we obtain the following expression for $\omega_0$:
\begin{eqnarray}
\omega_0 &=& -\frac{2}{3\beta}\left[\left(\frac{1}{n^2\gamma_{n_0}} +  \frac{\beta - \alpha}{\Omega_{D_0}}\right) + \frac{\beta u_0}{2}\left( \frac{1-\Omega_{k_0}}{1-\Omega_{D_0} - \Omega_{k_0}}  \right)\right] , \label{omega0}
\end{eqnarray}
which is the same result of the non interacting case.\\
In order to find the expression of the parameter $\omega_1$, we follow the same procedure of the non interacting case.\\
Using the definition of the evolutionary form of the fractional energy density of DE $\Omega_D'$ given in Eq. (\ref{paolone1}), we obtain that:
\begin{eqnarray}
\frac{\Omega_D'}{\Omega_D}= \frac{2}{\beta}\left[\left(\frac{\Omega_D}{n^2\gamma_n} - \alpha + \beta\right) \frac{\left( 1-\Omega_D  \right)}{\Omega_D}  + \frac{ \beta \left(u -3b^2\right)}{2} \right].   \label{lgo11new}
\end{eqnarray}
Moreover, using the expression of $\Omega_D''$ given in Eq. (\ref{paolone2}), we obtain that the term $\left(\frac{\Omega_D''}{\Omega_D}\right)$ is equal to:
\begin{eqnarray}
\frac{\Omega_D''}{\Omega_D} = \frac{2}{\beta \Omega_D}\left\{ \Omega_D'\left[\frac{1-2\Omega_D}{n^2\gamma_n} +\alpha + \beta\left(\frac{ u}{2}-1\right) -\frac{3}{2}\beta b^2 \right] \right\} + u'. \label{iele1dar}
\end{eqnarray}
Adding the expressions of $\Omega_k''$ and $\Omega_D^{''}$ given, respectively, in Eqs. (\ref{lgo16}) and (\ref{paolone2}), we obtain:
\begin{eqnarray}
\frac{\Omega_D^{''}+\Omega_k^{''}}{1-\Omega_D-\Omega_k} &=& \frac{2}{\beta}\left\{\Omega_D' \left[\frac{1}{n^2\gamma_n} + \frac{\beta \left(u-3b^2\right)}{2\left(  1-\Omega_D-\Omega_k \right)}   \right]    \right\} \nonumber \\
&& -\frac{2}{\beta}\frac{\Omega_D' + \Omega_k'}{1-\Omega_D-\Omega_k}\left( \frac{\Omega_D}{n^2\gamma_n} -\alpha + \beta \right) \frac{\Omega_D u'}{1-\Omega_D-\Omega_k}. \label{iele2dar}
\end{eqnarray}
Therefore, adding Eqs. (\ref{iele1dar}) and (\ref{iele2dar}), we can write:
\begin{eqnarray}
\frac{\Omega_D''}{\Omega_D} + \frac{\Omega_D^{''}+\Omega_k^{''}}{1-\Omega_D-\Omega_k} &=& \frac{2}{\beta \Omega_D}\left\{ \Omega_D'\left[\frac{1-2\Omega_D}{n^2\gamma_n} +\alpha + \beta\left(\frac{ u}{2}-1\right)  -\frac{3}{2}\beta b^2 \right] \right\} \nonumber \\
&&+\frac{2}{\beta}\left\{\Omega_D' \left[\frac{1}{n^2\gamma_n} + \frac{\beta \left(u -3b^2\right)}{2\left(  1-\Omega_D-\Omega_k \right)}   \right]    \right\}  \nonumber \\
&&-\frac{2}{\beta}\frac{\Omega_D' + \Omega_k'}{1-\Omega_D-\Omega_k}\left( \frac{\Omega_D}{n^2\gamma_n} -\alpha + \beta \right) + \frac{u'\left(1-\Omega_k\right)}{1-\Omega_D-\Omega_k}\label{iele2-2}.
\end{eqnarray}
Finally, adding the expressions of $\Omega_k'$ and $\Omega_D^{'}$, given, respectively, in Eqs. (\ref{lgo15}) and (\ref{paolone1}), we obtain:
\begin{eqnarray}
\frac{\Omega_D'+\Omega_k'}{1-\Omega_D-\Omega_k} = \frac{2}{\beta}\left[\left( \frac{\Omega_D}{n^2\gamma_n} -\alpha + \beta \right) +\frac{\beta \left(u-3b^2\right) \Omega_D}{2\left( 1-\Omega_D-\Omega_k\right)}    \right].\label{iele3}
\end{eqnarray}
Moreover, inserting the results of above equations in Eq. (\ref{83}), we obtain the following expression for $\omega_1$:
\begin{eqnarray}
\omega_1 &=&   \frac{2b^2}{\beta\left(1 - \Omega_{k_0} - \Omega_{D_0}\right)}\left[\left(\frac{\Omega_{D_0}}{n^2\gamma_{n_0}} - \alpha + \beta\right)\left(   1-\Omega_{D_0}  \right)  + \frac{\Omega_{D_0} \beta u_0}{2} - \frac{3}{2}\Omega_{D_0} \beta b^2 \right]\nonumber \\
&&  +\frac{2b^2\Omega_{D_0}}{\beta\left(1 - \Omega_{k_0} - \Omega_{D_0}\right)}\left[\left( \frac{\Omega_{D_0}}{n^2\gamma_{n_0}} -\alpha + \beta \right) +\frac{\beta \left(u_0-3b^2\right) \Omega_{D_0}}{2\left( 1-\Omega_{D_0}-\Omega_{k_0}\right)}    \right] \nonumber \\
&& -\frac{4}{3\beta^2}\left[\left(\frac{\Omega_{D_0}}{n^2\gamma_{n_0}} - \alpha + \beta\right) \frac{\left( 1-\Omega_{D_0}  \right)}{\Omega_{D_0}}  + \frac{ \beta \left(u_0 -3b^2\right)}{2} \right]^2 \nonumber \\
&& +\frac{4}{3\beta^2 \Omega_{D_0}}\left\{ \left[\left(\frac{\Omega_{D_0}}{n^2\gamma_{n_0}} - \alpha + \beta\right)\left(   1-\Omega_{D_0}  \right)  + \frac{\Omega_{D_0} \beta u_0}{2} - \frac{3}{2}\Omega_{D_0} \beta b^2 \right]\times  \right. \nonumber \\
&&\left. \left[\frac{1-2\Omega_{D_0}}{n^2\gamma_{n_0}} +\alpha + \beta\left(\frac{ u_0}{2}-1\right)  -\frac{3}{2}\beta b^2 \right] \right\} \nonumber \\
&&+\frac{4}{3\beta^2}\left\{\left[\left(\frac{\Omega_{D_0}}{n^2\gamma_{n_0}} - \alpha + \beta\right)\left(   1-\Omega_{D_0}  \right)  + \frac{\Omega_{D_0} \beta u_0}{2} - \frac{3}{2}\Omega_{D_0} \beta b^2 \right] \times \right. \nonumber \\
&&\left. \left[\frac{1}{n^2\gamma_{n_0}} + \frac{\beta \left(u_0 -3b^2\right)}{2\left(  1-\Omega_{D_0}-\Omega_{k_0} \right)}   \right]    \right\}  \nonumber \\
&&-\frac{4}{3\beta^2}\left[\left( \frac{\Omega_{D_0}}{n^2\gamma_{n_0}} -\alpha + \beta \right) +\frac{\beta \left(u_0-3b^2\right) \Omega_{D_0}}{2\left( 1-\Omega_{D_0}-\Omega_{k_0}\right)}    \right]\left( \frac{\Omega_{D_0}}{n^2\gamma_{n_0}} -\alpha + \beta \right) \nonumber \\
&&+\left\{ \frac{2}{\beta}\left( \frac{1}{n^2 \gamma_{n_0}} + \frac{\beta - \alpha}{\Omega_{D_0}}   \right) \left[\Omega_{k_0} - \left(u_0+1   \right)  \left(1-\Omega_{D_0}  \right)   \right] -\left(u_0 -3b^2 \right)\left( u_0+1  \right)   \right\}\times \nonumber \\
&&\frac{1-\Omega_{k_0}}{3\left(1-\Omega_{D_0}-\Omega_{k_0}\right)} \nonumber \\
&& +\frac{4}{3\beta^2}\left[\left( \frac{\Omega_{D_0}}{n^2\gamma_{n_0}} -\alpha + \beta \right) +\frac{\beta \left(u_0-3b^2\right) \Omega_{D_0}}{2\left( 1-\Omega_{D_0}-\Omega_{k_0}\right)}    \right]^2\label{omega1nino}.
\end{eqnarray}
Considering the expression of $\gamma_{n_0}$ given in Eq. (\ref{lim1-1}), we can rewrite Eqs. (\ref{omega0}) and (\ref{omega1nino}) as follows:
\begin{eqnarray}
\omega_0 &=& -\frac{2}{3\beta}\left\{\left[\frac{\left(3\lambda -1 \right)}{2n^2 } +  \frac{\beta - \alpha}{\Omega_{D_0}}\right] + \frac{\beta u_0}{2}\left( \frac{1-\Omega_{k_0}}{1-\Omega_{D_0} - \Omega_{k_0}}  \right)\right\} , \label{omega0-1}
\end{eqnarray}
\begin{eqnarray}
\omega_1 &=& \frac{2b^2}{\beta\left(1 - \Omega_{k_0} - \Omega_{D_0}\right)}\left\{\left[\frac{\left(3\lambda -1 \right)\Omega_{D_0}}{2n^2 } - \alpha + \beta\right]\left(   1-\Omega_{D_0}  \right)  + \frac{\Omega_{D_0} \beta u_0}{2} - \frac{3}{2}\Omega_{D_0} \beta b^2 \right\}\nonumber \\
&&  +\frac{2b^2\Omega_{D_0}}{\beta\left(1 - \Omega_{k_0} - \Omega_{D_0}\right)}\left\{\left[ \frac{\left(3\lambda -1 \right)\Omega_{D_0}}{2n^2 } -\alpha + \beta \right] +\frac{\beta \left(u_0-3b^2\right) \Omega_{D_0}}{2\left( 1-\Omega_{D_0}-\Omega_{k_0}\right)}    \right\} \nonumber \\
&& -\frac{4}{3\beta^2}\left\{\left[\frac{\left(3\lambda -1 \right)\Omega_{D_0}}{2n^2 } - \alpha + \beta\right] \frac{\left( 1-\Omega_{D_0}  \right)}{\Omega_{D_0}}  + \frac{ \beta \left(u_0 -3b^2\right)}{2} \right\}^2 \nonumber \\
&& +\frac{4}{3\beta^2 \Omega_{D_0}}\left\{ \left[\left(\frac{\left(3\lambda -1 \right)\Omega_{D_0}}{2n^2 } - \alpha + \beta\right)\left(   1-\Omega_{D_0}  \right)  + \frac{\Omega_{D_0} \beta u_0}{2} - \frac{3}{2}\Omega_{D_0} \beta b^2 \right]\times  \right. \nonumber \\
&&\left. \left[\frac{\left(1-2\Omega_{D_0}\right)\left(3\lambda -1 \right)}{2n^2 } +\alpha + \beta\left(\frac{ u_0}{2}-1\right)  -\frac{3}{2}\beta b^2 \right] \right\} \nonumber \\
&&+\frac{4}{3\beta^2}\left\{\left[\left(\frac{\left(3\lambda -1 \right)\Omega_{D_0}}{2n^2 } - \alpha + \beta\right)\left(   1-\Omega_{D_0}  \right)  + \frac{\Omega_{D_0} \beta u_0}{2} - \frac{3}{2}\Omega_{D_0} \beta b^2 \right] \times \right. \nonumber \\
&&\left. \left[\frac{\left(3\lambda -1 \right)}{2n^2 } + \frac{\beta \left(u_0 -3b^2\right)}{2\left(  1-\Omega_{D_0}-\Omega_{k_0} \right)}   \right]    \right\}  \nonumber \\
&&-\frac{4}{3\beta^2}\left\{\left[ \frac{\left(3\lambda -1 \right)\Omega_{D_0}}{2n^2 } -\alpha + \beta \right] +\frac{\beta \left(u_0-3b^2\right) \Omega_{D_0}}{2\left( 1-\Omega_{D_0}-\Omega_{k_0}\right)}    \right\}\left[ \frac{\left(3\lambda -1 \right)\Omega_{D_0}}{2n^2 } -\alpha + \beta \right] \nonumber \\
&&+\left\{ \frac{2}{\beta}\left[ \frac{\left(3\lambda -1 \right)}{2n^2  } + \frac{\beta - \alpha}{\Omega_{D_0}}   \right] \left[\Omega_{k_0} - \left(u_0+1   \right)  \left(1-\Omega_{D_0}  \right)   \right] -\left(u_0 -3b^2 \right)\left( u_0+1  \right)   \right\}\times \nonumber\\
&&\frac{1-\Omega_{k_0}}{3\left(1-\Omega_{D_0}-\Omega_{k_0}\right)} \nonumber \\
&& +\frac{4}{3\beta^2}\left\{\left[ \frac{\left(3\lambda -1 \right)\Omega_{D_0}}{2n^2 } -\alpha + \beta \right] +\frac{\beta \left(u_0-3b^2\right) \Omega_{D_0}}{2\left( 1-\Omega_{D_0}-\Omega_{k_0}\right)}    \right\}^2. \label{omega1nino-1}
\end{eqnarray}
Inserting in Eqs. (\ref{omega0-1}) and (\ref{omega1nino-1}) the values of the parameters involved, we obtain, for $\lambda = 1.02$:
\begin{eqnarray}
\omega_0 &\approx& -1.81193, \\
\omega_1 &\approx&\left(0.416014 -0.74713  b^2\right),
\end{eqnarray}
therefore we obtain the following equation for the EoS parameter of DE $\omega_D$ as function of the redshift $z$:
\begin{eqnarray}
\omega_D &\approx& -1.81193 + \left(0.416014 -0.74713  b^2\right) z\label{ceizze2int}.
\end{eqnarray}
In Figure \ref{omegad7}, we plot the behavior of  $\omega_D$ given in Eq. (\ref{ceizze2int}). \\
For $z=0$, Eq. (\ref{ceizze2int}) leads to $\omega_D \approx -1.81193 $ while the value $\omega_D = -1$ is obtained for $z \approx \frac{0.81193}{\left(0.416014 -0.74713  b^2\right)}$. \\
For $b^2=0.025$, we obtain:
\begin{eqnarray}
\omega_D &\approx& -1.81193 + 0.39733 z\label{ceizze2int2},
\end{eqnarray}
while $\omega_D = -1$ is obtained for $z \approx 2.04$.\\
Moreover, for $\lambda = 0.98$, we obtain:
\begin{eqnarray}
\omega_0 &\approx& -1.69188, \\
\omega_1 &\approx& \left(0.388546 -0.74713  b^2\right),
\end{eqnarray}
 therefore we obtain the following equation for the EoS parameter of DE $\omega_D$ as function of the redshift $z$:
\begin{eqnarray}
\omega_D &\approx& -1.69188 + \left(0.388546 -0.74713  b^2\right)  z.\label{ceizze1intlalla}
\end{eqnarray}
In Figure \ref{omegad8}, we plot the behavior of the EoS parameter of DE  $\omega_D$ given in Eq. (\ref{ceizze1intlalla}).\\
For $z=0$, Eq. (\ref{ceizze1intlalla}) leads to $\omega_D \approx 1.69188 $ while the value $\omega_D = -1$ is obtained for $z\approx \frac{0.69188}{\left(0.388546 -0.74713  b^2\right) }$.\\
 For $b^2=0.025$, we obtain:
\begin{eqnarray}
\omega_D &\approx& -1.69188 + 0.36987  z,\label{ceizze1int}
\end{eqnarray}
while $\omega_D = -1$ is obtained  for $z \approx 1.87$.\\
Finally, for $\lambda = 1.00$, we obtain the following values for $\omega_0$ and $\omega_1$:
\begin{eqnarray}
\omega_0 &\approx& -1.75191, \\
\omega_1 &\approx& \left(0.40228 -0.74713  b^2\right),
\end{eqnarray}
 therefore we obtain the following equation for the EoS parameter of DE $\omega_D$ as function of the redshift $z$:
\begin{eqnarray}
\omega_D &\approx& -1.75191 +\left(0.40228 -0.74713  b^2\right) z.\label{ceizzeintint}
\end{eqnarray}
In Figure \ref{omegad9}, we plot the behavior of the EoS parameter of DE $\omega_D$  given in Eq. (\ref{ceizzeintint}). \\
For $z=0$, Eq. (\ref{ceizze1int}) leads to $\omega_D \approx -1.75191 $ while the value $\omega_D = -1$ is obtained for $z \approx \frac{-1.75191}{\left(0.40228 -0.74713  b^2\right)}$.\\
 For $b^2=0.025$, we obtain:
\begin{eqnarray}
\omega_D &\approx& -1.75191 +0.38360 z,\label{ceizzeintint2}
\end{eqnarray}
while  $\omega_D = -1$  is obtained for for $z \approx 1.96$.\\

\begin{figure}[ht]
\centering\includegraphics[width=8cm]{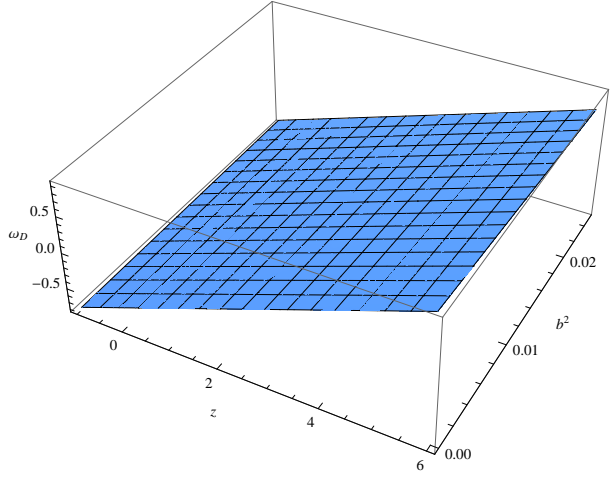}
\caption{Plot of  the EoS parameter of DE $\omega_D$ given in Eq. (\ref{ceizze2intricci}) for $\lambda = 1.02$ for the Ricci scale case. } \label{omegad10}
\end{figure}

\begin{figure}[ht]
\centering\includegraphics[width=8cm]{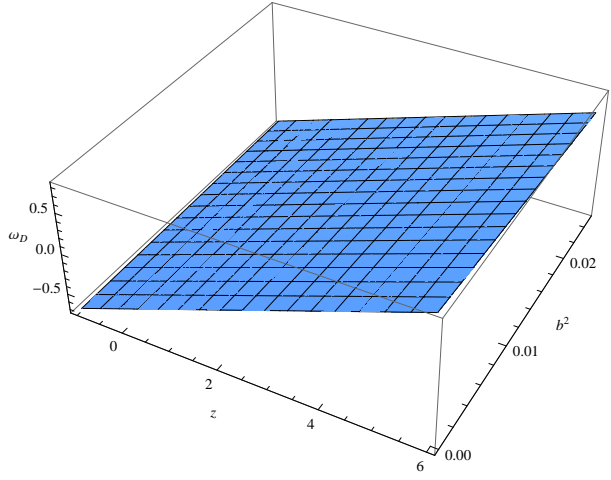}
\caption{Plot of  the EoS parameter of DE $\omega_D$ given in Eq. (\ref{ceizze1intricci})  for $\lambda = 0.98$ for the Ricci scale case. } \label{omegad11}
\end{figure}

\begin{figure}[ht]
\centering\includegraphics[width=8cm]{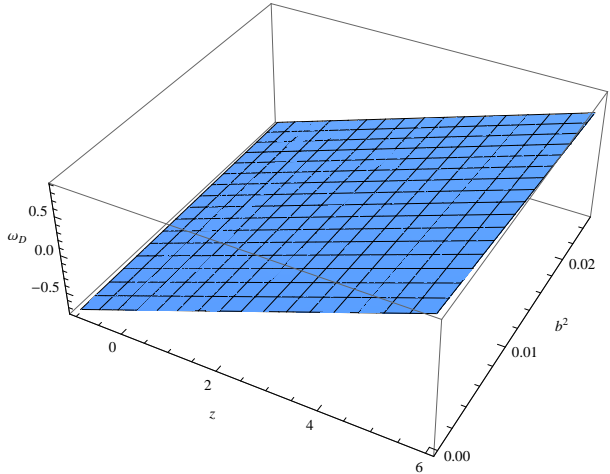}
\caption{Plot of  the EoS parameter of DE $\omega_D$ given in Eq. (\ref{ceizzeintintricci})  for $\lambda = 1.00$ for the Ricci scale case. } \label{omegad12}
\end{figure}

We now consider the limiting case corresponding to the Ricci scale, which is recovered for $\alpha =2$ and $\beta =1$.\\
Inserting in Eqs. (\ref{omega0-1}) and (\ref{omega1nino-1}) the values of the parameters involved, we obtain, for $\lambda = 1.02$:
\begin{eqnarray}
\omega_0 &\approx&  -0.552428, \\
\omega_1 &\approx& \left(0.247249 - 1.44407 b^2\right) z,
\end{eqnarray}
therefore we obtain the following equation for the EoS parameter of DE $\omega_D$ as function of the redshift $z$:
\begin{eqnarray}
\omega_D &\approx& -0.552428 + \left(0.247249 - 1.44407 b^2\right) z\label{ceizze2intricci}.
\end{eqnarray}
In Figure \ref{omegad10}, we plot the behavior of  $\omega_D$ given in Eq. (\ref{ceizze2intricci}). \\
For $z=0$, Eq. (\ref{ceizze2int}) leads to $\omega_D \approx  -0.552428 $ while the value $\omega_D = -1$ is obtained for $z \approx \frac{-0.447572 }{\left(0.247249 - 1.44407 b^2\right) }$.\\
 For $b^2=0.025$, we obtain:
\begin{eqnarray}
\omega_D &\approx& -0.552428 + 0.211115 z\label{ceizze2intricci2},
\end{eqnarray}
while  $\omega_D = -1$ is obtained for $z \approx -2.11$.\\
Moreover, for $\lambda = 0.98$, we obtain:
\begin{eqnarray}
\omega_0 &\approx&  -0.492207, \\
\omega_1 &\approx&  \left(0.220506 - 1.44407 b^2\right) z,
\end{eqnarray}
 therefore we obtain the following equation for the EoS parameter of DE $\omega_D$ as function of the redshift $z$:
\begin{eqnarray}
\omega_D &\approx& -0.492207 +   \left(0.220506 - 1.44407 b^2\right) z.\label{ceizze1intricci}
\end{eqnarray}
In Figure \ref{omegad11}, we plot the behavior of the EoS parameter of DE  $\omega_D$ given in Eq. (\ref{ceizze1intricci}). \\
For $z=0$, Eq. (\ref{ceizze1int}) leads to $\omega_D \approx  -0.492207 $ while the value $\omega_D = -1$ is obtained for $z\approx \frac{-0.507793}{\left(0.220506 - 1.44407 b^2\right)} $.\\
For $b^2=0.025$, we obtain:
\begin{eqnarray}
\omega_D &\approx& -0.492207 +   0.188958z.\label{ceizze1intricci2}
\end{eqnarray}
while  $\omega_D = -1$ is obtained for $z \approx  -2.69   $.\\
Finally, for $\lambda = 1.00$, we obtain the following values for $\omega_0$ and $\omega_1$:
\begin{eqnarray}
\omega_0 &\approx& -0.522318, \\
\omega_1 &\approx& \left(0.233877 - 1.44407 b^2\right) z ,
\end{eqnarray}
 therefore we obtain the following equation for the EoS parameter of DE $\omega_D$ as function of the redshift $z$:
\begin{eqnarray}
\omega_D &\approx& -0.522318 + \left(0.233877 - 1.44407 b^2\right) z.\label{ceizzeintintricci}
\end{eqnarray}
In Figure \ref{omegad12}, we plot the behavior of the EoS parameter of DE $\omega_D$ given in Eq. (\ref{ceizzeintintricci}). \\
For $z=0$, Eq. (\ref{ceizze1int}) leads to $\omega_D \approx -0.522318 $ while the value $\omega_D = -1$ is obtained for $z \approx \frac{-0.477682}{\left(0.233877 - 1.44407 b^2\right)}$. \\
For $b^2=0.025$, we obtain:
\begin{eqnarray}
\omega_D &\approx& -0.522318 + 0.197775 z.\label{ceizzeintintricci2}
\end{eqnarray}
while  $\omega_D = -1$ is obtained for $z \approx -2.42$.\\
We can clearly see in Eq. (\ref{omega1nino}) that the interaction parameter $b^2$ plays an important role in the final expression of the parameter $\omega_1$ for the interacting case. We can also observe that,
in the limiting case corresponding to $b^2=0$, the expression of the parameter $\omega_1$ given in Eq. (\ref{omega1nino-1}) leads to the expression for non-interacting case, given in Eq. (\ref{71nino}).

\section{Statefinder Diagnostic}
The study and the investigation of cosmological quantities like for example the Hubble parameter $H$, the EoS parameter of DE $\omega_D$ and the deceleration parameter $q$  have attracted a lot of
attention in present day cosmology. Since the different DE models usually lead to a positive Hubble parameter $H$ and a negative deceleration parameter $q$, i.e. to to $H > 0$ and $q < 0$, at the present epoch, the Hubble and the deceleration parameters $H$ and $q$ can not effectively discriminate
between the various DE models. We have, therefore, that a higher order of time derivative of the scale factor $a\left( t \right)$ is required in order to have a better and deeper comprehension and understanding of the DE model taken into account. Sahni et al. \citep{sahni}
and Alam et al. \citep{alam}, using the third time derivative of the scale factor $a(t)$, introduced the statefinder pair $\left\{r,s\right\}$ with the aim to remove the problems related to the values assumed by the Hubble parameter $H$ and the deceleration parameter $q$ at the present epoch. The general expressions of the statefinder parameters $r$ and $s$ are given, respectively, by the following relations:
\begin{eqnarray}
r &=& \frac{{ {...} \atop a}}{aH^3}, \label{r1}\\
s &=&   \frac{r -1}{3\left(q-1/2\right)}, \label{s1}
\end{eqnarray}
The statefinder parameters can be also written as functions of the total energy density $\rho$ and the total pressure $p$ of the model considered, defined as $\rho = \rho_D + \rho_m$ and $p = p_D$, as follows:
\begin{eqnarray}
r &=& 1 + \frac{9}{2}\left(\frac{\rho + p}{\rho}\right)\frac{\dot{p}}{\dot{\rho}} \nonumber \\
 &=& 1 + \frac{9}{2}\left(\frac{\rho_m + \rho_D + p_D}{\rho_m + \rho_D}\right)\frac{\dot{p}_D}{\dot{\rho}_m+\dot{\rho}_D} , \label{34p2}\\
s&=& \left(\frac{\rho + p}{p}\right)\frac{\dot{p}}{\dot{\rho}} \nonumber \\
&=&  \left(\frac{\rho_m +\rho_D + p_D}{p_D}\right)\frac{\dot{p}_D}{\dot{\rho}_m+\dot{\rho}_D}, \label{35p2}
\end{eqnarray}
which can be also written as follows:
\begin{eqnarray}
r  &=& 1 + \frac{9}{2}\left(\frac{\rho + p}{\rho}\right)\frac{p'}{\rho'}\nonumber \\
&=& 1 + \frac{9}{2}\left(\frac{\rho_m + \rho_D + p_D}{\rho_m + \rho_D}\right)\frac{p_D'}{\rho_m' + \rho_D'}, \label{34p2}\\
s&=&  \left(\frac{\rho + p}{p}\right)\frac{p'}{\rho'} \nonumber \\
&=& \left(\frac{\rho_m + \rho_D + p_D}{p_D}\right)\frac{p_D'}{\rho_m' + \rho_D'}. \label{35p2}
\end{eqnarray}
An alternative way to write the statefinder parameters $r$ and $s$ involves the Hubble parameter $H$ and its higher time derivatives as follows:
\begin{eqnarray}
r&=& 1 + 3\left(\frac{\dot{H}}{H^2}\right)+ \frac{\ddot{H}}{H^3}, \label{r2}\\
s&=& -\frac{3H\dot{H}+\ddot{H}}{3H\left( 2\dot{H}+3H^2  \right)}\nonumber \\
&=&  -\frac{3\dot{H}+\ddot{H}/H}{3\left( 2\dot{H}+3H^2  \right)}. \label{s2}
\end{eqnarray}
The most important features of the statefinder parameters which must be taken into account when a particular DE model is studied is that the point with coordinate corresponding to $\left\{r, s\right\} = \left\{1, 0\right\}$ in the $\left\{r, s\right\}$ plane  indicates the fixed point corresponding to the flat $\Lambda$CDM model. Departures of given DE models from this fixed point are good ways to establish the distance of these models from the flat $\Lambda$CDM model. \\
Moreover, we must underline here that, in the $\left \{r, s\right \}$ plane, the sector of positive $s$, i.e. $s > 0$, corresponds to  quintessence-like models of DE while the sector of negative $s$, i.e. $s < 0$, corresponds to  phantom-like models of DE. Furthermore, an evolution from phantom to quintessence or from quintessence to phantom is given by crossing of the fixed point $\left\{r, s\right\} = \left\{1, 0\right\}$ corresponding to the $\Lambda$CDM in the $\left \{r, s \right \}$ plane \citep{wu1}.\\
Braneworld, Cosmological Constant $\Lambda_{CC}$, Chaplygin gas and quintessence models were investigated by Alam et al. \citep{alam} using the statefinder diagnostic: they observed that the statefinder pair could differentiate between these different proposed models. An investigation on statefinder parameters for differentiating between DE and modified gravity was carried out in the paper of Wang et al. \citep{wang}. Statefinder diagnostics for the $f\left(T\right)$ modified gravity model has been studied in the paper of Wu $\&$ Yu \citep{wu1}. \\
Other authors have been studied the properties of various DE models from the viewpoint of statefinder diagnostic \citep{state1,state2,state3,state6}.\\
We now want to study the statefinder pair for the model considered in this paper, for this reason we need to derive the quantities useful in order to obtain the final expression of the pair $\left\{ r,s \right\}$. We underline here that for the statefinder parameter $r$ we will use the expression given in Eq. (\ref{r2}) since it will be easier to calculate the terms involved, in fact we have already obtained the expression of $\left(\frac{\dot{H}}{H^2}\right)$.\\

\subsection{Non Interacting Case}
We start studying the behavior of the statefinder parameters for the case corresponding to  the non interacting Dark Sectors in order to find if the model we are considering leads to a point close to the one of the $\Lambda$CDM model or it has a departure from it. \\
We have already derived the expression of $\left(\frac{\dot{H}}{H^2}\right)$ in Eq. (\ref{lgo6}). Differentiating the expression of $\left(\frac{\dot{H}}{H^2}\right)$ given in Eq. (\ref{lgo6}) with respect to the cosmic time $t$, we obtain:
\begin{eqnarray}
\frac{d}{dt} \left(  \frac{\dot{H}}{H^2} \right)= \frac{\ddot{H}}{H^2} -2\left(\frac{\dot{H}^2}{H^3}\right). \label{jorge}
\end{eqnarray}
Dividing the result of Eq. (\ref{jorge}) by the Hubble parameter $H$, after some algebraic calculations,  we can easily find the following expression for the term $\left(\frac{\ddot{H}}{H^3}\right)$:
\begin{eqnarray}
\frac{\ddot{H}}{H^3} &=&  2\left( \frac{\dot{H}}{H^2}  \right)^2 + \left( \frac{\dot{H}}{H^2}  \right)'. \label{}
\end{eqnarray}
Then, we have that the final expression of the statefinder parameter $r$ can be written as follows:
\begin{eqnarray}
r&=& 1 + 3\left(\frac{\dot{H}}{H^2}\right) + 2\left( \frac{\dot{H}}{H^2}  \right)^2 + \left( \frac{\dot{H}}{H^2}  \right)' . \label{}
\end{eqnarray}
Differentiating the expression of $\left( \frac{\dot{H}}{H^2}  \right)$ obtained in Eq. (\ref{lgo6}) with respect to the variable $x$, we find the following expression for $\left( \frac{\dot{H}}{H^2}  \right)'$:
\begin{eqnarray}
 \left( \frac{\dot{H}}{H^2}  \right)' = \frac{1}{\beta}\left(\frac{\Omega'_D}{n^2 \gamma_n}  \right). \label{}
\end{eqnarray}
Therefore, we conclude that $r$ can be written as follows:
\begin{eqnarray}
r&=& 1 + \frac{3}{\beta} \left(\frac{\Omega_D}{n^2\gamma_n}  -\alpha \right)  \nonumber \\
&& +\frac{2}{\beta^2} \left(\frac{\Omega_D}{n^2\gamma_n}  -\alpha \right)^2  + \frac{1}{\beta}\left(\frac{\Omega'_D}{n^2 \gamma_n}  \right). \label{rfinalenino}
\end{eqnarray}
Inserting in Eq. (\ref{rfinalenino}) the expression of the evolutionary form of the fractional energy density of DE $\Omega_D'$ obtained in Eq. (\ref{lgo11}), we obtain the following expression for the statefinder parameter $r$:
\begin{eqnarray}
r&=& 1 + \frac{3}{\beta} \left(\frac{\Omega_D}{n^2\gamma_n}  -\alpha \right)  +\frac{2}{\beta^2} \left(\frac{\Omega_D}{n^2\gamma_n}  -\alpha \right)^2 \nonumber \\
&& + \frac{1}{\beta n^2 \gamma_n}\left\{\frac{2}{\beta}\left[\left(\frac{\Omega_D}{n^2\gamma_n} - \alpha + \beta\right) \left( 1-\Omega_D  \right)  + \frac{\Omega_D \beta u}{2} \right]  \right\}. \label{nataliar}
\end{eqnarray}
Considering the present day values of the quantities involved in the final expression of the statefinder parameter $r$ given in Eq. (\ref{nataliar}) and using the expression of $\gamma_{n_0}$ obtained in Eq. (\ref{lgo4}), we have that the expression of the present day value of the statefinder parameter $r$ can be also written as follows:
\begin{eqnarray}
r_0&=& 1 + \frac{3}{\beta} \left[\frac{\left(3\lambda -1  \right)\Omega_{D_0}}{2n^2}  -\alpha \right]  +\frac{2}{\beta^2} \left[\frac{\left(3\lambda -1   \right)\Omega_{D_0}}{2n^2}  -\alpha \right]^2\nonumber \\
&&  + \frac{\left(3\lambda -1\right)}{\beta^2n^2}\left\{\left[\frac{\left( 3\lambda -1 \right)\Omega_{D_0}}{2n^2} - \alpha + \beta\right] \left( 1-\Omega_{D_0}  \right)  + \frac{\Omega_{D_0} \beta u_0}{2} \right\}   . \label{rzerofinalenino}
\end{eqnarray}
The present day value of the statefinder parameter $s$ can be obtained from Eq. (\ref{s1}) and it is given by:
\begin{eqnarray}
s_0 =   \frac{r_0 -1}{3\left(q_0-1/2\right)}, \label{s11}
\end{eqnarray}
where $r_0$ is given in Eq. (\ref{rzerofinalenino}) while $q_0$ is given in Eq. (\ref{89-finale-2}).\\
Inserting in Eq. (\ref{rzerofinalenino}) the values of the parameters involved, we find that $r_0\approx 6.00853$ for $\lambda =1.02$,  $r_0 \approx 5.0452$ for $\lambda =0.98$ and $r_0 \approx 5.51566$ for $\lambda =1.00$.\\
Using the values of $r_0$ derived above along with the values of $q_0$ obtained in previous Section, we can easily obtain that $s_0 \approx -0.889295$   for $\lambda =1.02$,  $s_0\approx -0.769247$ for $\lambda =0.98$ and $s_0 \approx -0.829271$ for $\lambda =1.00$.\\
Therefore we obtain the following pairs of values: $\left\{ r_0,s_0 \right\} \approx  \left\{ 6.00853, -0.889295\right\}$ for $\lambda =1.02$, $\left\{ r_0,s_0 \right\} \approx  \left\{ 5.0452,-0.769247 \right\}$ for $\lambda =0.98$ and $\left\{ r_0,s_0 \right\} \approx  \left\{ 5.51566, -0.829271\right\}$ for $\lambda =1.00$.\\
We can observe that the values of the statefinder pair $\left\{ r,s \right\}$ for all cases of the running parameter $\lambda$ taken into account considerably differs from the values corresponding to the $\Lambda$CDM model. Moreover, since we have $s<0$ for all cases considered, we can conclude that we are dealing with a phantom-like model.\\
We now consider the limiting case corresponding to the Ricci scale, which is recovered for $\alpha =2$ and $\beta =1$.\\
Inserting in Eq. (\ref{rzerofinalenino})  the values of the parameters involved, we find that $r_0\approx 0.489371$ for $\lambda =1.02$,  $r_0 \approx 0.453289$ for $\lambda =0.98$ and $r_0 \approx 0.46851 $ for $\lambda =1.00$.\\
Using the values of $r_0$ derived above along with the values of $q_0$ obtained in previous Section, we can easily obtain that $s_0 \approx 0.297739$   for $\lambda =1.02$,  $s_0\approx 0.357856$ for $\lambda =0.98$ and $s_0 \approx 0.3278$ for $\lambda =1.00$.\\
Therefore we obtain the following pairs of values: $\left\{ r_0,s_0 \right\} \approx  \left\{ 0.489371,0.297739 \right\}$ for $\lambda =1.02$, $\left\{ r_0,s_0 \right\} \approx  \left\{ 0.453289, 0.357856\right\}$ for $\lambda =0.98$ and $\left\{ r_0,s_0 \right\} \approx  \left\{ 0.46851, 0.3278\right\}$ for $\lambda =1.00$.\\
We can clearly observe that, in this case, we obtain values which are closer to the point corresponding to the $\Lambda$CDM model if compared with the result of the other set of values of $\alpha$ and $\beta$. Moreover, since we obtained $s>0$ for all cases considered, we have that for the limiting case of Ricci scale we deal with a quintessence-like model.

\subsection{Interacting Case}
We now consider the case corresponding to presence of interaction between the Dark Sectors.\\
Following the same procedure of the non interacting case, we have that the expression of the statefinder parameter $r$ for the interacting case can be written as follows:
\begin{eqnarray}
r_0&=& 1 + \frac{3}{\beta} \left[\frac{\left(3\lambda -1  \right)\Omega_{D_0}}{2n^2}  -\alpha \right]  \nonumber \\
&& +\frac{2}{\beta^2} \left[\frac{\left(3\lambda -1   \right)\Omega_{D_0}}{2n^2}  -\alpha \right]^2\nonumber \\
&&  + \frac{\left(3\lambda -1\right)}{\beta^2n^2}\left\{\left[\frac{\left( 3\lambda -1 \right)\Omega_{D_0}}{2n^2} - \alpha + \beta\right] \left( 1-\Omega_{D_0}  \right)  + \frac{\Omega_{D_0} \beta u_0}{2}\left( u_0 -3b^2  \right)\right\}. \label{rzerofinaleninoint}
\end{eqnarray}
The present day value of the statefinder parameter $s$ can be obtained from Eq. (\ref{s1}) and it is given by:
\begin{eqnarray}
s_0 =   \frac{r_0 -1}{3\left(q_0-1/2\right)}, \label{s11int}
\end{eqnarray}
where $r_0$ has been obtained in Eq. (\ref{rzerofinaleninoint})  while $q_0$ has been obtained in Eq. (\ref{89-finale-2}).\\
Inserting in Eq. (\ref{rzerofinaleninoint})  the values of the parameters involved, we find that $r_0\approx 6.00853 - 6.40954 b^2$ for $\lambda =1.02$,   $r_0\approx 5.0452 - 6.03617 b^2$ for $\lambda =0.98$ and $r_0 \approx 5.51566 - 6.22285 b^2$ for $\lambda = 1.00$.\\
Using the expression of $r_0$ derived above along with the expression of $q_0$ obtained in previous Section, we can easily obtain that $s_0 \approx -0.889295 + 1.13805 b^2$   for $\lambda =1.02$,  $s_0\approx -0.769247 + 1.14785 b^2$ for $\lambda =0.98$ while  $s_0\approx -0.829271 + 1.14278 b^2$ for $\lambda =1.00$.\\
Therefore, we obtain $\left\{ r_0,s_0 \right\} \approx \left\{ 6.00853 - 6.40954 b^2, -0.889295 + 1.13805 b^2  \right\}$ for $\lambda = 1.02$,  $\left\{ r_0,s_0 \right\} \approx \left\{ 5.0452 - 6.03617 b^2, -0.769247 + 1.14785 b^2\right\}$ for $\lambda = 0.98$ and $\left\{ r_0,s_0 \right\} \approx \left\{ 5.51566 - 6.22285 b^2,-0.829271 + 1.14278 b^2 \right\}$ for $\lambda = 1.00$.\\
In the limiting case of $b^2 = 0.025$, we obtain $\left\{ r_0,s_0 \right\} \approx \left\{ 5.84829,  -0.86084 \right\}$ for $\lambda = 1.02$,  $\left\{ r_0,s_0 \right\} \approx \left\{ 4.89429, -0.74055 \right\}$ for $\lambda = 0.98$ and $\left\{ r_0,s_0 \right\} \approx \left\{ 5.36009,-0.80070\right\}$ for $\lambda = 1.00$. Moreover, in the limiting case of $b^2=0$, we recover the same results of the non interacting case.  \\
We obtain, then, that also in the interacting case, the values of the pair $\left\{ r,s \right\}$ differ (even if they are closer with respect to the non interacting case) from the $\Lambda$CDM model for all the cases of the running parameter $\lambda$ considered. Moreover, since we obtain $s<0$ for all the cases of the running parameter $\lambda$ considered, we conclude that we deal with a phantom-like model for the set of values of the parameters considered. \\
We now consider the limiting case corresponding to the Ricci scale, then for $\alpha =2$ and $\beta=1$.\\
Inserting in Eq. (\ref{rzerofinaleninoint})  the values of the parameters involved, we find that $r_0\approx 0.489371 - 3.21502 b^2$ for $\lambda =1.02$,  $r_0\approx 0.453289 - 3.02774 b^2$ for $\lambda =0.98$ while   $r_0\approx 0.46851 - 3.12138 b^2$ for $\lambda =1.00$.\\
Using the expressions of $r_0$ derived above along with the expressions of $q_0$ obtained in previous Section, we can easily obtain that $s_0 \approx 0.297739 + 1.87462 b^2$   for $\lambda =1.02$, $s_0\approx 0.357856 + 1.98184 b^2 b^2$ for $\lambda =0.98$ while $s_0\approx 0.3278 + 1.92514 b^2$ for $\lambda =1.00$.\\
Therefore we obtain the following pairs of values: $\left\{ r_0,s_0 \right\} \approx  \left\{ 0.489371 - 3.21502 b^2,0.297739 + 1.87462 b^2 \right\}$ for $\lambda =1.02$, $\left\{ r_0,s_0 \right\} \approx  \left\{ 0.453289 - 3.02774 b^2,0.357856 + 1.98184 b^2 \right\}$ for $\lambda =0.98$ and $\left\{ r_0,s_0 \right\} \approx  \left\{ 0.46851 - 3.12138 b^2,0.3278 + 1.92514 b^2 \right\}$ for $\lambda =1.00$.\\
In the limiting case of $b^2 = 0.025$, we obtain $\left\{ r_0,s_0 \right\} \approx \left\{0.40900 ,0.34460   \right\}$ for $\lambda = 1.02$,  $\left\{ r_0,s_0 \right\} \approx \left\{ 0.37760,0.40740 \right\}$ for $\lambda = 0.98$ and $\left\{ r_0,s_0 \right\} \approx \left\{0.390476, 0.37593\right\}$ for $\lambda = 1.00$. Therefore, for $b^2=0.025$, the results obtained lead to a values of the statefinder parameters which are  more distant with respect to the point  $\left\{ r_0,s_0 \right\} = \left\{1,0   \right\}$ corresponding to the $\Lambda$CDM model.  Moreover, since we obtain that $s>0$ for all the cases of the running parameter $\lambda$ considered, we conclude we deal with a quintessence-like model for the set of values of the parameters taken into account. Furthermore, in the limiting case of $b^2=0$, we recover the same results of the non interacting case.  \\

\section{Cosmographic Parameters}
In this Section, we want to obtain some important cosmological information about the PLECHDE model with Granda-Oliveros cut-off we are considering using the properties of the cosmographic parameters.\\
Standard candles (like SNe Ia) represent powerful instruments in present day cosmology since they can be used in order to reconstruct the Hubble diagram, i.e. the redshift-distance relation up to high redshifts $z$. It is quite common to constrain a parameterized model against available cosmological data in order to check the validity of the model considered and in order to constraint the free parameters of the model. However, it is known that this type of approach is highly model-dependent, for this reason there are still doubts in scientific community on the validity and reliability of the constraints on the derived cosmological quantities obtained with this method.\\
In order to avoid this kind of problem, it is possible to consider the cosmography, i.e. it is possible to expand the scale factor $a\left( t \right)$ in Taylor series with respect to the cosmic time $t$. This type of expansion produces a distance-redshift relation  which is based  only on the assumption of the FLRW metric, therefore it is fully model independent because it is independent on the particular form of the solution of the cosmic equations. Cosmography can be considered as a milestone in the study of the main properties of Universe dynamics, which any theoretical model studied and considered has to take into account and also to satisfy.\\
 It is useful to introduce the following four quantities \citep{cosmo1,cosmo2}:
\begin{eqnarray}
q &=& -\left(\frac{\ddot{a}}{a}\right)H^{-2}= -\frac{\ddot{a}a}{\dot{a}^2} = -\frac{a^{\left(2\right)}a}{\dot{a}^2}, \label{par1} \\
j &=& \left(\frac{1}{a}\frac{d^3a}{dt^3}\right)H^{-3} = \frac{a^{\left(3\right)}a^2}{\dot{a}^3}, \label{par2} \\
s &=& \left(\frac{1}{a}\frac{d^4a}{dt^4}\right)H^{-4}= -\frac{a^{\left(4\right)}a^3}{\dot{a}^4}, \label{par3} \\
l &=& \left(\frac{1}{a}\frac{d^5a}{dt^5}\right)H^{-5}= \frac{a^{\left(5\right)}a^4}{\dot{a}^5}, \label{par4}
\end{eqnarray}
where the number in parenthesis indicates the order of the derivative with respect to the cosmic time $t$ while the numbers without parenthesis indicates the power of the relevant quantity.\\
In general, we have that the $i$-th parameter $x^i$ can be obtained thanks to the following general expression:
\begin{eqnarray}
x^{i} &=& \left( -1  \right)^{i+1}\left(\frac{1}{H^{i}}\right)\frac{a^{\left(i\right)}}{a}\nonumber \\
&=& \left( -1  \right)^{i+1} \frac{a^{\left(i\right)}a^{i-1}}{\dot{a}^{i+1}}.
\end{eqnarray}
The quantities given in Eqs. (\ref{par1}), (\ref{par2}), (\ref{par3}) and (\ref{par4}) are known, respectively, as deceleration, jerk, snap and lerk parameters. We must underline that we have already derived the expressions of the deceleration parameter $q$.\\
In order to avoid confusion with the statefinder parameter $s$, we will denote since now the snap parameter with $s_{cosmo}$.\\
Making some algebraic calculations, we can obtain the following useful relations between the time derivatives of the Hubble parameter and the cosmographic parameters $q$, $j$, $s_{cosmo}$ and $l$:
\begin{eqnarray}
\frac{d H}{dt}&=&\dot{H} = -H^2 \left(1 + q \right), \label{ddd1} \\
\frac{d^2H}{dt^2}&=&\ddot{H}= H^3 \left(j+3q+2   \right), \label{ddd2} \\
\frac{d^3H}{dt^3}&=& \dot{\ddot{H}} =H^4 \left[s_{cosmo} - 4j -3q\left(q+4\right) -6   \right], \label{ddd3} \\
\frac{d^4H}{dt^4}&=& \ddot{\ddot{H}} =H^5 \left[l-5s_{cosmo} +10\left( q+2  \right)j +30\left( q+2 \right)q +24   \right]. \label{ddd4}
\end{eqnarray}
The present-day values of these cosmographic parameters, denoted with the subscript 0 (which indicates the value of the parameter for $z=0$ or equivalently for $t=0$) can be used in order to characterize the evolutionary status of the present day Universe. For example, a negative value of  $q_0$ indicates an accelerated expansion of the Universe (as it is suggested by recent cosmological measurements), while the value of $j_0$ allows to discriminate among different accelerating models.  \\
More information on the calculations made in order to obtain the previous equations  can be found in the paper of Capozziello et al. \citep{capozziello}.\\
Some constraints about the present day values of the snap parameter $s_{cosmo}$ and of the lerk parameter $l$ have been recently obtained. For example, Capozziello $\&$ Izzo \citep{values0} have found that $s_{cosmo,0} = 8.32 \pm 12.16$, while John \citep{values0-1,values0-2} has derived that  $s_{cosmo,0} = 36.5 \pm 52.9$ and  $l_0= 142.7 \pm 320$. As we can clearly see, the errors associated with the values derived   in these two works for the snap and lerk cosmographic parameters $s_{cosmo}$ and $l$ are of the order of 200$\%$, for future more precise comparisons between cosmological constraints of $s_{cosmo}$ and $l$ and the values obtained from theoretical models, it will be useful to have better constraints with more accurate errors.\\
Using the definitions of the cosmographic parameters given in Eqs. (\ref{par1}), (\ref{par2}), (\ref{par3}) and (\ref{par4}), we can easily obtain the fifth order Taylor expansion of the scale factor $a\left( t \right)$ as follows:
\begin{eqnarray}
\frac{a\left(t\right)}{a\left(t_0\right)} &=& 1+H_0 \left( t -t_0 \right) - \left(\frac{q_0}{2}\right)H_0^2 \left( t -t_0 \right)^2 +  \left(\frac{j_0}{3!}\right)H_0^3 \left( t -t_0 \right)^3  \nonumber \\
 &&+\left(\frac{s_0}{4!}\right)H_0^4 \left( t -t_0 \right)^4   + \left(\frac{l_0}{5!}\right)H_0^5 \left( t -t_0 \right)^5 + O \left[ \left( t-t_0  \right)^6 \right], \label{exp}
\end{eqnarray}
where $t_0$ represents the present day age of the Universe, which is given by $t\approx 1/H_0$. It must be here underlined that Eq. (\ref{exp}) is also the fifth order expansion of $\left(1 + z\right)^{-1}$, since, from the definition of redshift,  we obtain:
\begin{eqnarray}
z = \frac{a\left( t_0 \right)}{a\left( t \right)} -1. \label{}
\end{eqnarray}
The jerk parameter $j$ is also another name of the statefinder parameter $r$ we have studied in the previous Section  and it represents a natural next step beyond the Hubble parameter $H$ and the deceleration parameter $q$.\\
The snap parameter $s_{cosmo}$, which involves the fourth time derivative of the scale factor $a\left(t \right)$, is also
sometimes called kerk parameter and it has been well discussed in the works of Dabrowski \citep{dabro}, Dunajski $\&$ Gibbons \citep{duna} and Arabsalmania $\&$ Sahni \citep{arab}.
Another useful relation which can be used in order to find the expression of $s_{cosmo}$ involves the deceleration and the jerk parameters and it is given by:
\begin{eqnarray}
s_{cosmo} &=& \frac{\dot{j}}{H}-j\left( 2+3q  \right)\nonumber \\
& =& j' -j\left( 2+3q  \right). \label{scosmogeneral}
\end{eqnarray}
The lerk parameter $l$ involves the fifth time derivative of scale factor $a\left(  t \right)$. More information about the lerk parameter can be found in the paper of Dabrowski \citep{dabro}. An useful relation between the lerk parameter $l$ and the deceleration and snap parameters is given by:
\begin{eqnarray}
l= s_{cosmo}' -\left( 3+4q  \right)s_{cosmo}. \label{lcosmo}
\end{eqnarray}
Using the definition of the snap parameter $s_{cosmo}$ given in Eq. (\ref{scosmogeneral}), we can rewrite Eq. (\ref{lcosmo}) as follows:
\begin{eqnarray}
l=  s_{cosmo}'-\left( 3+4q  \right)\left[ j'-j\left( 2+3q  \right)  \right].\label{lcosmo2}
\end{eqnarray}
Moreover, from the definition of the snap parameter $s_{cosmo}$ given in Eq. (\ref{scosmogeneral}), we obtain the following expression for the derivative of the snap parameter $s_{cosmo}$ with respect to the variable $x$:
\begin{eqnarray}
s_{cosmo}' = j'' - j'\left( 2+3q  \right) -3jq'.\label{sprimo}
\end{eqnarray}
Therefore, $l$ can be also written as follows:
\begin{eqnarray}
l&=&   j'' - j' \left( 5+7q  \right) -j\left( 3q' - 3q -2  \right). \label{lcosmo3}
\end{eqnarray}

\subsection{Non Interacting Case}
We now want to derive the final expressions for $s_{cosmo}$ and $l$ for the non interacting case.\\
We start calculating the final expression of the snap parameter $s_{cosmo}$. \\
Differentiating with respect to the variable $x$ the expression of the statefinder parameter $r$ given in Eq. (\ref{rfinalenino}), we find that:
\begin{eqnarray}
j' &\equiv& r' = \left(\frac{\Omega'_{D}}{n^2\gamma_{n}} \right) \left[\frac{3}{\beta}  +\frac{4}{\beta^2} \left(\frac{\Omega_{D}}{n^2\gamma_{n}}  -\alpha \right)  \right] + \frac{1}{\beta}\left(\frac{\Omega''_{D}}{n^2 \gamma_{n}}    \right) . \label{jprime}
\end{eqnarray}
Using the expression of $\Omega_D''$ for the non interacting case, we can write $j'$ as follows:
\begin{eqnarray}
j'  &=& \frac{1}{\beta n^2\gamma_n}  \left[ \Omega_D' \left( \frac{2}{\beta n^2 \gamma_n} - \frac{2\alpha}{\beta} + u -2  \right)  + \Omega_D u'\right]. \label{jprime2}
\end{eqnarray}
Using in Eq. (\ref{jprime2}) the expressions of the evolutionary form of the fractional energy density of DE $\Omega_D'$ and $u'$ for the non interacting case,  we obtain:
\begin{eqnarray}
j' &=& \frac{1}{\beta n^2\gamma_n}  \left\{ \frac{2}{\beta}\left[\left(\frac{\Omega_D}{n^2\gamma_n} - \alpha + \beta\right) \left( 1-\Omega_D  \right)  + \frac{\Omega_D \beta u}{2} \right] \left( \frac{2}{\beta n^2 \gamma_n} - \frac{2\alpha}{\beta} + u -2  \right) \right. \nonumber \\
&&\left.+  \frac{2}{\beta}\left( \frac{\Omega_D}{n^2 \gamma_n} + \beta - \alpha   \right) \left[\Omega_k - \left(u+1   \right)  \left(1-\Omega_D  \right)   \right] -u\Omega_D\left( u+1  \right)\right\}. \label{jprime3}
\end{eqnarray}

Therefore, we can conclude that the cosmographic parameter $s_{cosmo,non}$ for the non interacting case can be written as follows:
\begin{eqnarray}
s_{cosmo,non} &=& \frac{1}{\beta n^2\gamma_n}  \left\{ \frac{2}{\beta}\left[\left(\frac{\Omega_D}{n^2\gamma_n} - \alpha + \beta\right) \left( 1-\Omega_D  \right)  + \frac{\Omega_D \beta u}{2} \right]\times \right. \nonumber \\
 &&\left. \left( \frac{2}{\beta n^2 \gamma_n} - \frac{2\alpha}{\beta} + u -2  \right) \right. \nonumber \\
&&\left.+  \frac{2}{\beta}\left( \frac{\Omega_D}{n^2 \gamma_n} + \beta - \alpha   \right) \left[\Omega_k - \left(u+1   \right)  \left(1-\Omega_D  \right)   \right] -u\Omega_D\left( u+1  \right)\right\}\nonumber \\
&& -j\left( 2+3q  \right). \label{scosmogeneralfinalnon}
\end{eqnarray}
considering the present day values of the parameter involved along with the expression of $\gamma_{n0}$ given in Eq. (\ref{lim1-1}), we can write:
\begin{eqnarray}
s_{cosmo,non0} &=& \left(\frac{3\lambda -1}{2\beta n^2}\right)  \left\{ \frac{2}{\beta}\left[\left(\frac{\left( 3\lambda -1  \right)\Omega_{D_0}}{2n^2} - \alpha + \beta\right) \left( 1-\Omega_{D_0}  \right)  + \frac{\Omega_{D_0} \beta u_0}{2} \right] \times \right. \nonumber \\
&&\left. \left( \frac{3\lambda -1}{\beta n^2 } - \frac{2\alpha}{\beta} + u_0 -2  \right) \right. \nonumber \\
&&\left.+  \frac{2}{\beta}\left[ \frac{\left( 3\lambda -1  \right)\Omega_{D_0}}{2n^2 } + \beta - \alpha   \right] \times \right. \nonumber \\
&& \left.\left[\Omega_{k_0} - \left(u_0+1   \right)  \left(1-\Omega_{D_0}  \right)   \right] -u_0\Omega_{D_0}\left( u_0+1  \right)\right\}\nonumber \\
&& -j_0\left( 2+3q_0  \right). \label{scosmogeneralfinalnon0}
\end{eqnarray}
Inserting in Eq. (\ref{scosmogeneralfinalnon0}) the values of the parameters involved, we obtain, for $\lambda = 1.02$, that $s_{cosmo,non0} \approx 17.1515$, for $\lambda = 0.98$ we obtain $s_{cosmo,non0}\approx 11.9$ while for $\lambda = 1.00$ we obtain $s_{cosmo,non0}\approx 14.3768$. We obtain, therefore, values of $s_{cosmo,non0}$ which are between the errors obtained in the works of   Capozziello $\&$ Izzo \citep{values0} and John \citep{values0-1,values0-2}. \\
We now consider the case corresponding to the Ricci scale, which is recovered in the limiting case of $\alpha =2$ and $\beta =1$. \\
Inserting in Eq. (\ref{scosmogeneralfinalnon0}) the values of the parameters involved, we obtain, for $\lambda = 1.02$, that $s_{cosmo,non0} \approx -0.230849$, for $\lambda = 0.98$ we obtain $s_{cosmo,non0}\approx -0.15236$ while for $\lambda = 1.00$ we obtain $s_{cosmo,non0}\approx -0.18851$.  Also for the limiting case of the Ricci scale,  we obtain  values of $s_{cosmo,non0}$ which are between the errors obtained in the works of   Capozziello $\&$ Izzo \citep{values0} and John \citep{values0-1,values0-2}.\\
We now want to obtain the final expression for the lerk parameter $l$, therefore, in order to use the general expression given in Eq. (\ref{lcosmo3}), we need to calculate the expressions of $j''$ and $q'$.\\
Considering the expression of $r'$ given in Eq. (\ref{jprime}), we obtain the following expression for $j''$:
\begin{eqnarray}
j'' &=& r''=  \frac{1}{\beta n^2\gamma_n}  \left[ \Omega_D'' \left( \frac{2}{\beta n^2 \gamma_n} - \frac{2\alpha}{\beta} + u -2  \right)+2 \Omega_D' u'  + \Omega_D u'' \right] . \label{jsecond}
\end{eqnarray}
We have already obtained the expressions of the evolutionary form of the fractional energy density of DE $\Omega_D'$, $\Omega_D''$ and $u'$ for the non interacting case. We now need to find the expression of $u''$ for the non interacting case.\\
 Using the general expression of $u'$, we find the following relation for $u''$:
\begin{eqnarray}
u'' &=& \frac{2}{\beta}\left( \frac{1}{n^2 \gamma_n} + \frac{\beta - \alpha}{\Omega_D}   \right) \left[\Omega'_k - u'     \left(1-\Omega_D  \right)  + \Omega_D'\left(1+u \right)  \right] -u\left( u+1  \right) \nonumber \\
&&-\frac{2}{\beta}\left[ \frac{\left(\beta - \alpha\right)\Omega_D'}{\Omega_D^2}   \right] \left[\Omega_k - \left(u+1   \right)  \left(1-\Omega_D  \right)   \right] \nonumber \\
&&-u'\left( 2u+1  \right). \label{usecond}
\end{eqnarray}
Inserting the expressions of $\Omega_k'$, $\Omega_D'$ and $u'$ in Eq. (\ref{usecond}), we obtain:
\begin{eqnarray}
u'' &=& -\frac{4\Omega_k}{\beta^2}\left( \frac{1}{n^2 \gamma_n} + \frac{\beta - \alpha}{\Omega_D}   \right)\left( \frac{\Omega_D}{n^2\gamma_n}  -\alpha + \beta  \right)  \nonumber \\
&&-\frac{2}{\beta}\left( \frac{1}{n^2 \gamma_n} + \frac{\beta - \alpha}{\Omega_D}   \right)    \left(1-\Omega_D  \right) \times \nonumber \\
&&\left\{ \frac{2}{\beta}\left( \frac{1}{n^2 \gamma_n} + \frac{\beta - \alpha}{\Omega_D}   \right) \left[\Omega_k - \left(u+1   \right)  \left(1-\Omega_D  \right)   \right] -u\left( u+1  \right)  \right\}\nonumber \\
&&+\frac{4}{\beta^2}\left( \frac{1}{n^2 \gamma_n} + \frac{\beta - \alpha}{\Omega_D}   \right) \left(1+u \right) \left[\left(\frac{\Omega_D}{n^2\gamma_n} - \alpha + \beta\right) \left( 1-\Omega_D  \right)  + \frac{\Omega_D \beta u}{2} \right]\nonumber \\
&&-\frac{4}{\beta^2}\left( \frac{\beta - \alpha}{\Omega_D^2}   \right) \left[\Omega_k - \left(u+1   \right)  \left(1-\Omega_D  \right) \right]\left[\left(\frac{\Omega_D}{n^2\gamma_n} - \alpha + \beta\right) \left( 1-\Omega_D  \right)  + \frac{\Omega_D \beta u}{2} \right] \nonumber \\
&&-\left\{  \frac{2}{\beta}\left( \frac{1}{n^2 \gamma_n} + \frac{\beta - \alpha}{\Omega_D}   \right) \left[\Omega_k - \left(u+1   \right)  \left(1-\Omega_D  \right)   \right] -u\left( u+1  \right)\right\}\left( 2u+1  \right). \label{usecond2}
\end{eqnarray}
The final expression of $j''$ can be obtained inserting in Eq. (\ref{jsecond}) the expressions of $u'$, $\Omega_D'$ and $u''$ for the non interacting case.\\
Finally, differentiating the expression of $q$ given in Eq. (\ref{89-finale-2}) with respect to the variable $x$, we have that $q'$ is given by:
\begin{eqnarray}
	q'= -   \frac{1}{\beta}\left(  \frac{\Omega_D'}{n^2\gamma_n}   \right). \label{qfinale}
\end{eqnarray}
Therefore, for the non interacting case, using in Eq. (\ref{qfinale}) the expression of the evolutionary form of the fractional energy density of DE $\Omega_D'$ for the non interacting case, we obtain the following expression for $q'$:
\begin{eqnarray}
	q'= -   \frac{2}{\beta^2}\left(  \frac{1}{n^2\gamma_n}   \right)\left[\left(\frac{\Omega_D}{n^2\gamma_n} - \alpha + \beta\right) \left( 1-\Omega_D  \right)  + \frac{\Omega_D \beta u}{2} \right].  \label{}
\end{eqnarray}
We can now calculate the present day values of the lerk parameter $l$ taking into account all the results of the above equations. \\
For $\lambda = 1.02$, we obtain that $l_{non,0}\approx 80.5574$,  for $\lambda = 0.98$ we derive that $l_{non,0}\approx 57.667$ while for $\lambda = 1.00$ we obtain $l_{non,0} \approx 68.3621$.   We obtain, therefore, values of $l_{non0}$ which are between the errors obtained in the work of  John \citep{values0-1,values0-2}.\\
We now consider the case corresponding to the Ricci scale, which is recovered for $\alpha =2$ and $\beta =1$. \\
Inserting in Eq. (\ref{scosmogeneralfinalnon0}) the values of the parameters involved, we obtain, for $\lambda = 1.02$, that $l_{non,0} \approx -10.0602$, for $\lambda = 0.98$ we obtain $l_{non,0}\approx -11.2141$ while for $\lambda = 1.00$ we obtain $l_{non,0}\approx -10.6489$. Also for the limiting case of the Ricci scale,  we obtain,  values of $l_{non0}$ which are between the errors obtained in the work of John \citep{values0-1,values0-2}. \\

\subsection{Interacting Case}
\begin{figure}[ht]
\centering\includegraphics[width=8cm]{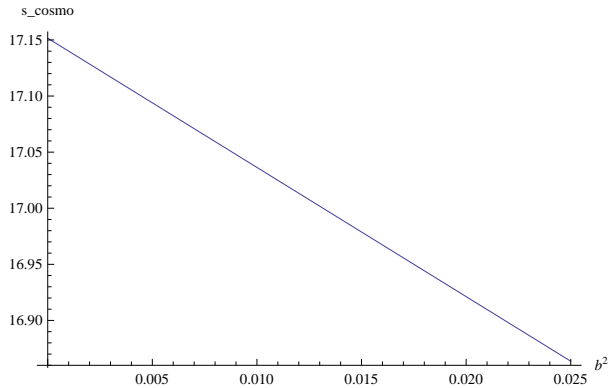}
\caption{Plot of  $s_{cosmo,int0} $ obtained in Eq. (\ref{debbymary1})  for $\lambda = 1.02$. } \label{celeste1}
\end{figure}

\begin{figure}[ht]
\centering\includegraphics[width=8cm]{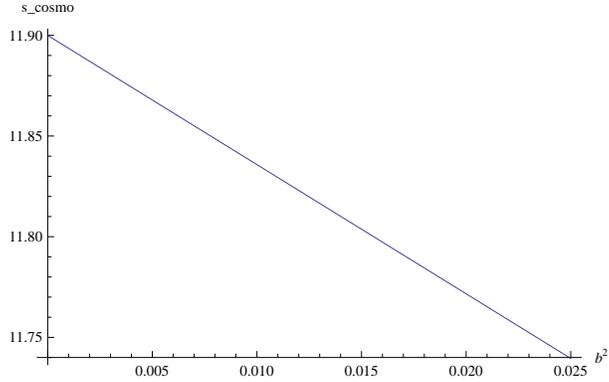}
\caption{Plot of  $s_{cosmo,int0} $ obtained in Eq. (\ref{debbymary2})   for $\lambda = 0.98$. } \label{celeste2}
\end{figure}

\begin{figure}[ht]
\centering\includegraphics[width=8cm]{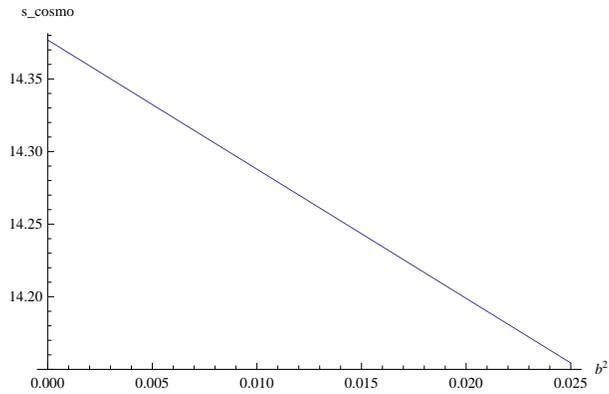}
\caption{Plot of  $s_{cosmo,int0} $ obtained in Eq. (\ref{debbymary3}) for $\lambda = 1.00$. } \label{celeste3}
\end{figure}

\begin{figure}[ht]
\centering\includegraphics[width=8cm]{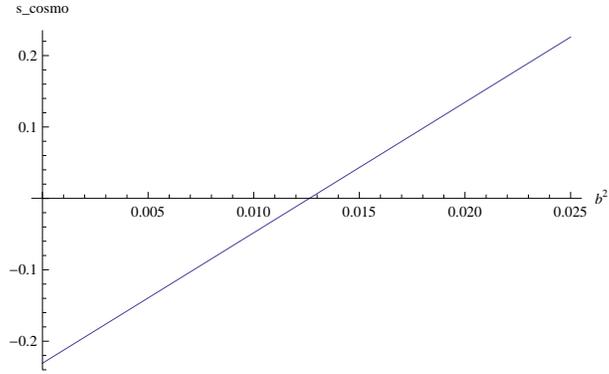}
\caption{Plot of  $s_{cosmo,int0} $ obtained in Eq. (\ref{debbymary4})  for $\lambda = 1.02$. } \label{celeste4}
\end{figure}

\begin{figure}[ht]
\centering\includegraphics[width=8cm]{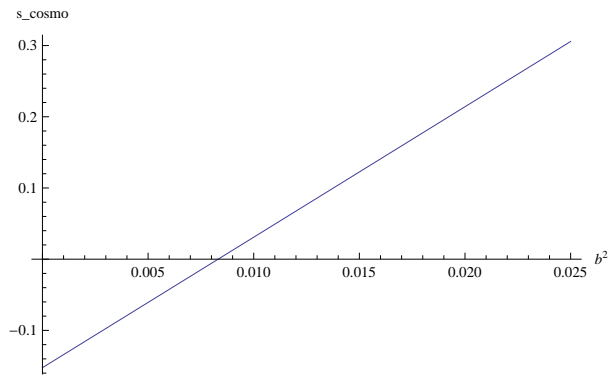}
\caption{Plot of   $s_{cosmo,int0} $ obtained in Eq. (\ref{debbymary5}) for $\lambda = 0.98$. } \label{celeste5}
\end{figure}

\begin{figure}[ht]
\centering\includegraphics[width=8cm]{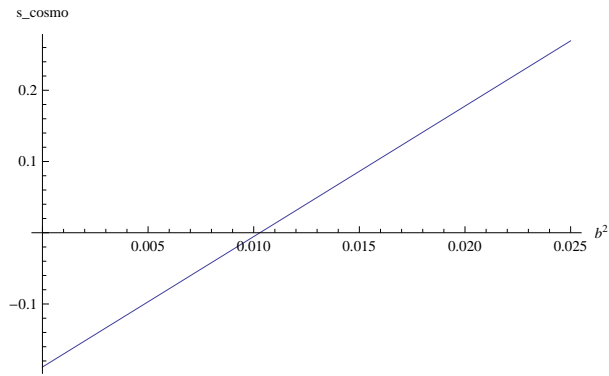}
\caption{Plot of  $s_{cosmo,int0} $  obtained in Eq. (\ref{debbymary6})  for $\lambda = 1.00$. } \label{celeste6}
\end{figure}

\begin{figure}[ht]
\centering\includegraphics[width=8cm]{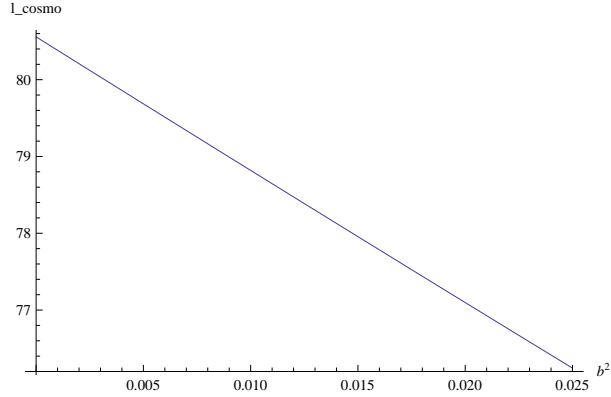}
\caption{Plot of $l_{int,0}$ given in Eq. (\ref{driutti1})  for $\lambda = 1.02$. } \label{celeste7}
\end{figure}

\begin{figure}[ht]
\centering\includegraphics[width=8cm]{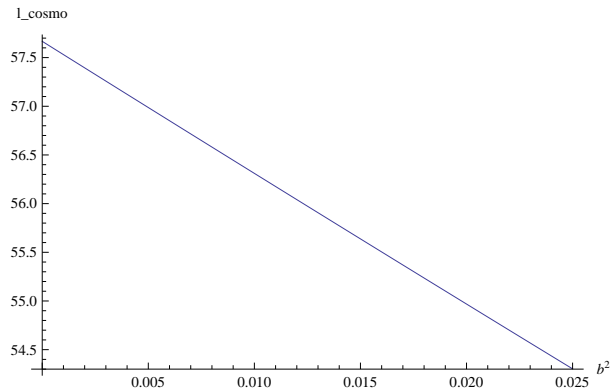}
\caption{Plot of $l_{int,0}$  given in Eq. (\ref{driutti2}) for $\lambda=0.98$. } \label{celeste8}
\end{figure}

\begin{figure}[ht]
\centering\includegraphics[width=8cm]{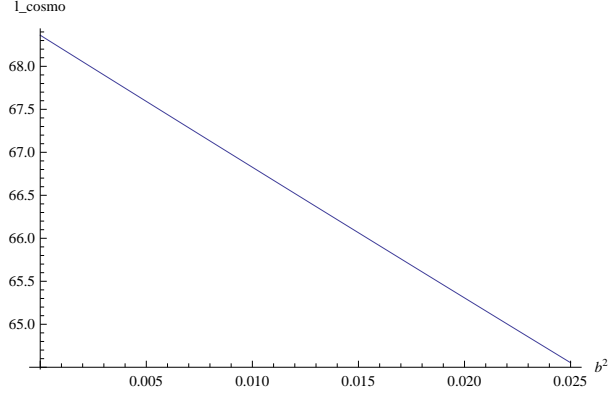}
\caption{Plot of $l_{int,0}$  given in Eq. (\ref{driutti3}) for $\lambda =1.00$. } \label{celeste9}
\end{figure}

\begin{figure}[ht]
\centering\includegraphics[width=8cm]{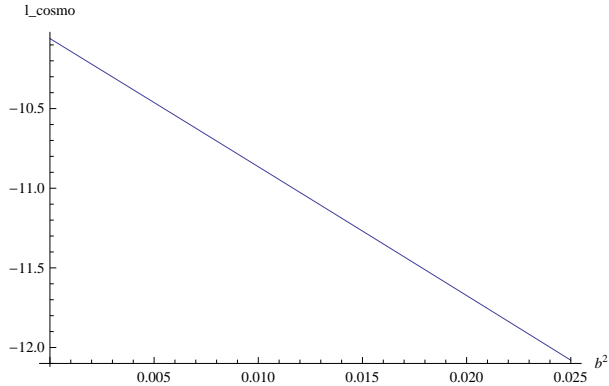}
\caption{Plot of  $l_{int,0}$ given in Eq. (\ref{driutti4}) for $\lambda = 1.02$ for the limiting case corresponding to the Ricci scale. } \label{celeste10}
\end{figure}

\begin{figure}[ht]
\centering\includegraphics[width=8cm]{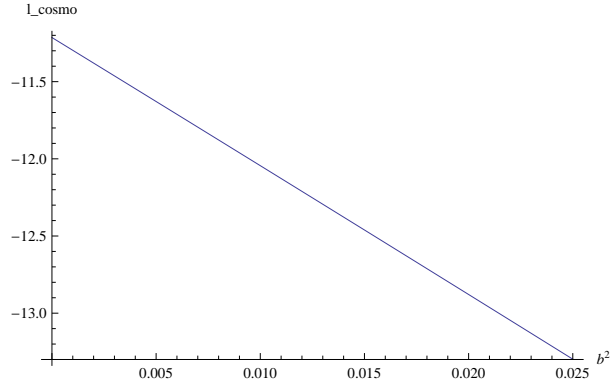}
\caption{Plot of  $l_{int,0}$ given in Eq. (\ref{driutti5}) for $\lambda = 0.98$ for the limiting case corresponding to the Ricci scale. } \label{celeste11}
\end{figure}

\begin{figure}[ht]
\centering\includegraphics[width=8cm]{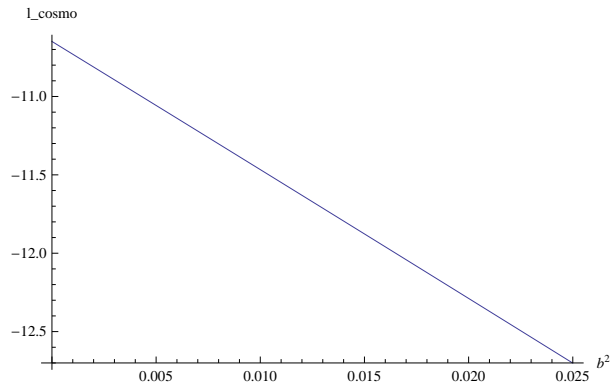}
\caption{Plot of  $l_{int,0}$ given in Eq. (\ref{driutti6}) for $\lambda = 1.00$ for the limiting case corresponding to the Ricci scale. } \label{celeste12}
\end{figure}

We now consider the interacting case, following the same procedure of the non interacting case in order to find the final expressions of $s_{cosmo}$ and $l$.\\
We start calculating $s_{cosmo}$. We still start from the following equation for $j'$:
\begin{eqnarray}
j'  &=& \frac{1}{\beta n^2\gamma_n}  \left[ \Omega_D' \left( \frac{2}{\beta n^2 \gamma_n} - \frac{2\alpha}{\beta} + u -2  \right)  + \Omega_D u'\right]. \label{jprime2int}
\end{eqnarray}
Instead, using in Eq. (\ref{jprime2int}) the expressions of the evolutionary form of the fractional energy density of DE $\Omega_D'$ and $u'$ for the  interacting case, we obtain:
\begin{eqnarray}
j' &=& \frac{1}{\beta n^2\gamma_n}  \left\{\frac{2}{\beta}\left[\left(\frac{\Omega_D}{n^2\gamma_n} - \alpha + \beta\right)\left(   1-\Omega_D  \right)  + \frac{\Omega_D \beta }{2}\left( u-3b^2 \right)  \right] \times \right. \nonumber \\
&&\left. \left( \frac{2}{\beta n^2 \gamma_n} - \frac{2\alpha}{\beta} + u -2  \right) \right. \nonumber \\
&&\left.+ \frac{2}{\beta}\left( \frac{\Omega_D}{n^2 \gamma_n} + \beta - \alpha   \right) \left[\Omega_k - \left(u+1   \right)  \left(1-\Omega_D  \right)   \right] -\Omega_D\left(u-3b^2\right)\left( u+1  \right)\right\}. \label{}
\end{eqnarray}
Therefore, we can conclude that the cosmographic parameter $s_{cosmo,int}$ for the interacting case can be written as follows:
\begin{eqnarray}
s_{cosmo,int} &=&\frac{1}{\beta n^2\gamma_n}  \left\{\frac{2}{\beta}\left[\left(\frac{\Omega_D}{n^2\gamma_n} - \alpha + \beta\right)\left(   1-\Omega_D  \right)  + \frac{\Omega_D \beta }{2}\left( u-3b^2 \right)  \right] \times \right. \nonumber \\
&&\left. \left( \frac{2}{\beta n^2 \gamma_n} - \frac{2\alpha}{\beta} + u -2  \right) \right. \nonumber \\
&&\left.+ \frac{2}{\beta}\left( \frac{\Omega_D}{n^2 \gamma_n} + \beta - \alpha   \right) \left[\Omega_k - \left(u+1   \right)  \left(1-\Omega_D  \right)   \right] -\Omega_D\left(u-3b^2\right)\left( u+1  \right)\right\}\nonumber \\
&& -j\left( 2+3q  \right). \label{scosmogeneralfinalint}
\end{eqnarray}
Considering the present day values of the quantities involved, we can write the present day value of $s_{cosmo,int}$ as follows:
\begin{eqnarray}
s_{cosmo,int0} &=&\left(\frac{3\lambda-1}{2\beta n^2}\right)  \left\{\frac{2}{\beta}\left[\left(\frac{\left(3\lambda -1 \right)\Omega_{D_0}}{2n^2} - \alpha + \beta\right)\left(   1-\Omega_{D_0}  \right)  + \frac{\Omega_{D_0} \beta }{2}\left( u_0-3b^2 \right)  \right]\times \right. \nonumber \\
&&\left. \left( \frac{3\lambda -1}{\beta n^2 } - \frac{2\alpha}{\beta} + u_0 -2  \right) \right. \nonumber \\
&&\left.+ \frac{2}{\beta}\left[ \frac{\left(3\lambda -1  \right)\Omega_{D_0}}{2n^2 } + \beta - \alpha   \right]\times \right. \nonumber \\
&&\left. \left[\Omega_{k_0} - \left(u_0+1   \right)  \left(1-\Omega_{D_0}  \right)   \right] -\Omega_{D_0}\left(u_0-3b^2\right)\left( u_0+1  \right)\right\}\nonumber \\
&& -j_0\left( 2+3q_0  \right). \label{scosmogeneralfinalint0}
\end{eqnarray}
Inserting in Eq. (\ref{scosmogeneralfinalint0}) the values of the parameters involved, we obtain, for $\lambda = 1.02$, that:
 \begin{eqnarray}
 s_{cosmo,int0} \approx 17.1515 - 11.5155 b^2. \label{debbymary1}
 \end{eqnarray}
In Figure \ref{celeste1},  we plot the behavior of $s_{cosmo,int0} $ for the case with $\lambda = 1.02$.\\
Instead, for $\lambda = 0.98$ we obtain:
 \begin{eqnarray}
 s_{cosmo,int0} \approx 11.9 - 6.41689 b^2.\label{debbymary2}
 \end{eqnarray}
In Figure \ref{celeste2},    we plot the behavior of $s_{cosmo,int0} $ for the case with $\lambda = 0.98$.\\
Finally, for $\lambda = 1.00$, we obtain:
\begin{eqnarray}
s_{cosmo,int0} \approx 14.3768 - 8.89771 b^2.\label{debbymary3}
\end{eqnarray}
In Figure \ref{celeste3},    we plot the behavior of $s_{cosmo,int0} $ for the case with $\lambda = 1.00$.\\
In the limiting case of $b^2=0$, the results of the interacting case lead to the same results of the non interacting case. Instead, for $b^2=0.025$, the results of the interacting case reduce to $s_{cosmo,int0}\approx  16.86366$ for $\lambda = 1.02$, $s_{cosmo,int0}\approx  11.7396$ for $\lambda = 0.98$ and $s_{cosmo,int0}\approx 14.1544 $ for $\lambda = 1.00$. \\
We now consider the limiting case corresponding to the Ricci scale, i.e. for $\alpha =2$ and $\beta =1$.\\
Inserting in Eq. (\ref{scosmogeneralfinalint0}) the values of the parameters involved, we obtain, for $\lambda = 1.02$:
\begin{eqnarray}
s_{cosmo,int0} \approx  -0.230849 + 18.273 b^2.\label{debbymary4}
\end{eqnarray}
In Figure (\ref{celeste4}),   we plot the behavior of $s_{cosmo,int0} $ for the case with $\lambda = 1.02$.\\
Instead, for $\lambda = 0.98$, we obtain:
\begin{eqnarray}
s_{cosmo,int0} \approx -0.15236 + 18.3226 b^2.\label{debbymary5}
\end{eqnarray}
In Figure (\ref{celeste5}),    we plot the behavior of $s_{cosmo,int0} $ for the case with $\lambda = 0.98$.\\
Finally, for $\lambda = 1.00$,  we obtain:
\begin{eqnarray}
s_{cosmo,int0} \approx  -0.18851 + 18.315 b^2. \label{debbymary6}
\end{eqnarray}
In Figure (\ref{celeste6}),    we plot the behavior of $s_{cosmo,int0} $ for the case with $\lambda = 1.00$.\\
In the limiting case of $b^2=0$, the results of the interacting case lead to the same results of the non interacting case. Instead, for $b^2=0.025$, the results of the interacting case reduce to $s_{cosmo,int0}\approx  0.22598$ for $\lambda = 1.02$, $s_{cosmo,int0}\approx 0.305705 $ for $\lambda = 0.98$ and $s_{cosmo,int0}\approx  0.269365$ for $\lambda = 1.00$.\\
We now want to find the final expression of the lerk parameter $l$ for the interacting case.\\
We follow also in this case the same procedure of the non interacting case.\\
For the interacting case, we have that $u''$ is given by the following relation:
\begin{eqnarray}
u'' &=& \frac{2}{\beta}\left( \frac{1}{n^2 \gamma_n} + \frac{\beta - \alpha}{\Omega_D}   \right) \left[\Omega'_k - u'     \left(1-\Omega_D  \right)  + \Omega_D'\left(1+u \right)  \right] \nonumber \\
&&-\frac{2}{\beta}\left[ \frac{\left(\beta - \alpha\right)\Omega_D'}{\Omega_D^2}   \right] \left[\Omega_k - \left(u+1   \right)  \left(1-\Omega_D  \right)   \right] \nonumber \\
&&-u'\left( 2u-3b^2+1  \right). \label{usecondint}
\end{eqnarray}
Inserting the expressions of $\Omega_k'$, $\Omega_D'$ and $u'$ in Eq. (\ref{usecondint}), we obtain:
\begin{eqnarray}
u'' &=& -\frac{4\Omega_k}{\beta^2}\left( \frac{1}{n^2 \gamma_n} + \frac{\beta - \alpha}{\Omega_D}   \right) \left( \frac{\Omega_D}{n^2\gamma_n}  -\alpha + \beta  \right) \nonumber \\
&&-  \frac{2}{\beta}\left( \frac{1}{n^2 \gamma_n} + \frac{\beta - \alpha}{\Omega_D}   \right)  \left(1-\Omega_D  \right) \times \nonumber \\
&&\left\{ \frac{2}{\beta}\left( \frac{1}{n^2 \gamma_n} + \frac{\beta - \alpha}{\Omega_D}   \right) \left[\Omega_k - \left(u+1   \right)  \left(1-\Omega_D  \right)   \right] -\left(u-3b^2\right)\left( u+1  \right)   \right\} \nonumber \\
&&+ \frac{4}{\beta^2}\left( \frac{1}{n^2 \gamma_n} + \frac{\beta - \alpha}{\Omega_D}   \right) \left(1+u \right) \left[\left(\frac{\Omega_D}{n^2\gamma_n} - \alpha + \beta\right)\left(   1-\Omega_D  \right)  + \frac{\Omega_D \beta }{2}\left( u-3b^2 \right)  \right] \nonumber \\
&&-\frac{4}{\beta^2}\left( \frac{\beta - \alpha}{\Omega_D^2}   \right) \left[\Omega_k - \left(u+1   \right)  \left(1-\Omega_D  \right)   \right]\left[\left(\frac{\Omega_D}{n^2\gamma_n} - \alpha + \beta\right)\left(   1-\Omega_D  \right)  + \frac{\Omega_D \beta }{2}\left( u-3b^2 \right)  \right] \nonumber \\
&&-\left( 2u-3b^2+1  \right)\left\{ \frac{2}{\beta}\left( \frac{1}{n^2 \gamma_n} + \frac{\beta - \alpha}{\Omega_D}   \right) \left[\Omega_k - \left(u+1   \right)  \left(1-\Omega_D  \right)   \right] \right. \nonumber \\
&&\left.-\left(u-3b^2\right)\left( u+1  \right)   \right\} \label{usecondint2}.
\end{eqnarray}
The final expression of $j''$ can be obtained inserting in Eq. (\ref{jsecond}) the expressions of $u'$, $\Omega_D'$ and $u''$ for the interacting case.\\
For the interacting case, using in Eq. (\ref{qfinale}) the expression of the evolutionary form of the fractional energy density of DE $\Omega_D'$ for the interacting case, we obtain the following expression for $q'$:
\begin{eqnarray}
	q'= -   \frac{2}{\beta^2}\left(  \frac{1}{n^2\gamma_n}   \right)\left[\left(\frac{\Omega_D}{n^2\gamma_n} - \alpha + \beta\right)\left(   1-\Omega_D  \right)  + \frac{\Omega_D \beta }{2}\left( u-3b^2 \right)  \right].  \label{}
\end{eqnarray}
We can now calculate the present day values of $l$ taking into account all the results of the above equations. \\
For the interacting case, we obtain, for $\lambda = 1.02$, that the present day value of the lerk parameter is given by the following relation:
\begin{eqnarray}
 l_{int,0} \approx     80.5574 +  \left(-174.752 + 88.9182 b^2\right)b^2. \label{driutti1}
 \end{eqnarray}
 In Figure \ref{celeste7},    we plot the behavior of $l_{int,0}$ obtained in Eq. (\ref{driutti1}).\\
For $\lambda = 0.98$ we derive that the present day value of the lerk parameter is given by the following relation:
\begin{eqnarray}
  l_{int,0} \approx  57.667 + \left(-136.334 + 70.4551b^2 \right)b^2. \label{driutti2}
\end{eqnarray}
In Figure \ref{celeste8},    we plot the behavior of $l_{int,0}$ obtained in Eq. (\ref{driutti2}).\\
Finally, for $\lambda =1.00$, we obtain:
\begin{eqnarray}
  l_{int,0} \approx  68.3621 + \left(-154.359 + 79.4812 b^2\right)b^2. \label{driutti3}
\end{eqnarray}
In Figure   \ref{celeste9},  we plot the behavior of $l_{int,0}$  obtained in Eq. (\ref{driutti3}).\\
In the limiting case of $b^2=0$, the results of the interacting case lead to the same results of the non interacting case. Instead, for $b^2=00.25$, we obtain, for $\lambda = 1.02$, that $l_{int,0} \approx  76.2442$, for $\lambda = 0.98$, we obtain $l_{int,0} \approx 54.3027$ while for $\lambda = 1.00$ we obtain $l_{int,0} \approx 64.5520 $.\\
We now consider the limiting case of Ricci scale, which is obtained in the limiting case of $\alpha =2$ and $\beta =1$.\\
For the interacting case, we obtain, for $\lambda = 1.02$, that the present day value of the lerk parameter is given by the following relation:
\begin{eqnarray}
 l_{int,0} \approx   -10.0602 - \left(80.2631 + 20.5776 b^2\right) b^2 . \label{driutti4}
 \end{eqnarray}
 In Figure \ref{celeste10},    we plot the behavior of  $l_{int,0}$ obtained in Eq. (\ref{driutti4}).\\
For $\lambda = 0.98$ we derive that the present day value of the lerk parameter is given by the following relation:
\begin{eqnarray}
  l_{int,0}\approx  -11.2141 - \left(82.7856 + 22.721 b^2\right) b^2. \label{driutti5}
\end{eqnarray}
In Figure  \ref{celeste11},   we plot the behavior of  $l_{int,0}$ obtained in Eq. (\ref{driutti5}).\\
Finally, for $\lambda =1.00$, we obtain:
\begin{eqnarray}
  l_{int,0}\approx  -10.6489 - \left(81.5413 + 21.701 b^2\right) b^2. \label{driutti6}
\end{eqnarray}
In Figure \ref{celeste12},    we plot the behavior of  $l_{int,0}$ obtained in Eq. (\ref{driutti6}).\\
In the limiting case of $b^2=0$, the results of the interacting case lead to the same results of the non interacting case. Instead, for $b^2=00.25$, we obtain, for $\lambda = 1.02$, that $l_{int,0} \approx  -12.0796$, for $\lambda = 0.98$ we obtain $l_{int,0} \approx -13.2979$ while for $\lambda = 1.00$ we obtain $l_{int,0} \approx  -12.7010$.\\
 We obtain, therefore, values of $s_{cosmo,int0}$ and $l_{int0}$ which are between the errors obtained in the works of   Capozziello $\&$ Izzo \citep{values0} and John \citep{values0-1,values0-2} for both sets of values of $\alpha$ and $\beta$ we considered and for all the values of $\lambda$ taken into account.

\section{Squared speed of the sound $v_s^2$}
We now consider an important quantity which is used in order check the stability of any DE model studied, named as squared speed of sound $v_s^2$, which is generally  defined as follows \citep{myung}:
\begin{eqnarray}
v_s^2 = \frac{\dot{p}}{\dot{\rho}} =  \frac{p'}{\rho'}   , \label{genvs}
\end{eqnarray}
where, as before, $p= p_D$ and $\rho = \rho_D + \rho_m$ are, respectively, the total pressure and the total energy density of the DE model taken into account. We must also remember that we are considering pressureless DM. Inserting in Eq. (\ref{genvs}) the expressions of $p$ and $\rho$, we can write:
\begin{eqnarray}
v_s^2 &=&  \frac{\dot{p}_D}{\dot{\rho}_D + \dot{\rho}_m}\nonumber \\
 &=&  \frac{p'_D}{\rho'_D + \rho'_m}. \label{genvs2}
\end{eqnarray}
The sign of the squared speed of the sound $v_{s}^{2}$ has a fundamental role when the stability of a background
evolution is wanted to be studied since it discriminates between instable and stable models. A negative value of $v_s^2$ indicates a classical instability of a given perturbation in General Relativity (GR) \citep{myung,kim} while a positive value of $v_s^2$ indicates a stable model. Myung \citep{myung} observed that $v_{s}^{2}$ for the HDE model is always negative if the future event horizon is taken as IR cut-off of the system, while for Chaplygin gas and tachyon models it is observed to be non negative. Kim et al. \citep{kim} observed that the squared speed of the sound $v_{s}^{2}$ for the Agegraphic DE (ADE) model is always negative leading to an instability of the perfect fluid for this particular model. In a recent paper, Sharif $\&$ Jawad \citep{sharif} have obtained that the interacting new HDE model leads to a negative value of $v_{s}^{2}$. Jawad et al. \citep{jawad} observed that $v_s^2$ remains negative for the HDE model based reconstructed $f\left(G\right)$ model with the choice of the scale factor in the power law form. In the recent work of Pasqua et al. \citep{miofrt}, the Authors observed that the interacting Modified Holographic Ricci DE (MHRDE) model in the  framework of $f\left(R,T\right) = \mu R + \nu T$ modified gravity model (with $\mu$ and $\nu$ being two constant parameters) is classically stable.
Pasqua et al. \citep{miovs2} showed  that the New ADE (NADE) model  based on the Generalized Uncertainty Principle (GUP) with power-law form of the scale factor $a\left(t\right)$ is classically instable.\\
\subsection{Non Interacting Case}
We can start studying the behavior of the squared speed of the sound $v_s^2$  for the non interacting case.\\
We have already obtained the expression of $\rho_D'$ in Eq. (\ref{lgo10evoother}), we now need to find the expressions of $p_D'$ and $\rho_m'$.
From the continuity equation for DE given in Eq. (\ref{38}), using the general definition of the EoS parameter of DE $\omega_D$, we can write the following expression of the pressure of DE $p_D$:
\begin{eqnarray}
p_D = -\rho_D - \frac{\rho_D'}{3}  . \label{pressurevs}
\end{eqnarray}
Therefore, we obtain that the derivative of the pressure of DE $p_D$ with respect to the variable $x$ is given by:
\begin{eqnarray}
p'_D = -\rho'_D - \frac{\rho_D''}{3}  . \label{pressurevs}
\end{eqnarray}
We must now calculate the expression of $\rho_D''$. We have that the expression of $\rho'_D$ obtained in Eq. (\ref{lgo10evoother}) is given by:
\begin{eqnarray}
\rho'_D =  \frac{2\rho_D}{ \beta \Omega_D}\left[\frac{\Omega_D}{n^2\gamma_n} -\alpha + \beta  + \beta \Omega_D \left( \frac{u-2}{2} \right) \right].  \label{lgo10evootherderi}
\end{eqnarray}
Differentiating the expression of $\rho_D'$ given in Eq. (\ref{lgo10evootherderi}) with respect to the variable $x$, we obtain that:
\begin{eqnarray}
\rho''_D &=&  =\frac{2\rho_D'}{ \beta \Omega_D}\left[\frac{\Omega_D}{n^2\gamma_n} -\alpha + \beta  + \beta \Omega_D \left( \frac{u-2}{2} \right) \right] \nonumber \\
&&+\frac{2\rho_D\Omega_D'}{ \beta \Omega_D^2}\left(\beta - \alpha  \right) +\rho_D u'.  \label{rhosecond}
\end{eqnarray}
The expressions of the EoS parameter of DE $\omega_D$ and $u'$ have been derived in Eqs. (\ref{lgo11}) and (\ref{lgo17}). Using the result of Eq. (\ref{rhosecond}) in Eq. (\ref{pressurevs}), we can write:
\begin{eqnarray}
p'_D &=& - \frac{2\rho_D'}{ 3\beta \Omega_D}\left[\frac{\Omega_D}{n^2\gamma_n} -\alpha + \beta  + \beta \Omega_D \left( \frac{u-2}{2} \right) + \frac{3\beta \Omega_D}{2}\right] \nonumber \\
&&+\frac{2\rho_D\Omega_D'}{ 3\beta \Omega_D^2}\left( \alpha - \beta  \right) -\frac{\rho_D u'}{3}  . \label{pressurevs2pao}
\end{eqnarray}
Inserting in Eq. (\ref{pressurevs2pao}) the expressions of the evolutionary form of the fractional energy density of DE $\Omega_D'$, $\rho_D'$ and $u'$ for the non interacting case, we obtain:
\begin{eqnarray}
p'_D &=& - \frac{4\rho_D}{ 3\beta^2 \Omega_D^2}\left[\frac{\Omega_D}{n^2\gamma_n} -\alpha + \beta  + \beta \Omega_D \left( \frac{u-2}{2} \right) + \frac{3\beta \Omega_D}{2}\right]\times \nonumber \\
&&\left[\frac{\Omega_D}{n^2\gamma_n} -\alpha + \beta  + \beta \Omega_D \left( \frac{u-2}{2} \right) \right] \nonumber \\
&& +\frac{4\rho_D}{ 3\beta^2 \Omega_D^2}\left(\alpha - \beta \right)\times \nonumber \\
&&\left[\left(\frac{\Omega_D}{n^2\gamma_n} - \alpha + \beta\right) \left( 1-\Omega_D  \right)  + \frac{\Omega_D \beta u}{2} \right] \nonumber \\
&&-\frac{\rho_D}{3}\left\{  \frac{2}{\beta}\left( \frac{1}{n^2 \gamma_n} + \frac{\beta - \alpha}{\Omega_D}   \right) \left[\Omega_k - \left(u+1   \right)  \left(1-\Omega_D  \right)   \right] -u\left( u+1  \right)  \right\}  . \label{pressurevs3}
\end{eqnarray}
We must now find an expression for $\rho_m'$.\\
 From the continuity equation for DM given in Eq. (\ref{39}), we obtain:
\begin{eqnarray}
\rho_m'= -3\rho_m= -3u\rho_D  , \label{rhomprime}
\end{eqnarray}
where we used the general definition of $u$.\\
Therefore, we have that:
\begin{eqnarray}
\rho_D' +\rho_m'= \rho_D' -3u\rho_D  . \label{rhosum}
\end{eqnarray}
Inserting in Eq. (\ref{rhosum}) the expression of $\rho_D'$ given in Eq. (\ref{lgo10evootherderi}), we obtain:
\begin{eqnarray}
\rho_D' +\rho_m'= \frac{2\rho_D}{ \beta \Omega_D}\left[\frac{\Omega_D}{n^2\gamma_n} -\alpha + \beta  + \beta \Omega_D \left( \frac{u-2}{2} \right) \right] -3u\rho_D  . \label{rhosumnew}
\end{eqnarray}
Finally, we can write $v_s^2$ as follows:
\begin{eqnarray}
v_s^2 = \frac{A_1}{B_1}, \label{}
\end{eqnarray}
where $A_1$ and $B_1$ are defined as follows:
\begin{eqnarray}
A_1 &=& - \frac{4}{ 3\beta^2 \Omega_D^2}\left[\frac{\Omega_D}{n^2\gamma_n} -\alpha + \beta  + \beta \Omega_D \left( \frac{u-2}{2} \right) + \frac{3\beta \Omega_D}{2}\right]\times \nonumber \\
&&\left[\frac{\Omega_D}{n^2\gamma_n} -\alpha + \beta  + \beta \Omega_D \left( \frac{u-2}{2} \right) \right] \nonumber \\
&&+\frac{4}{ 3\beta^2 \Omega_D^2}\left(\beta - \alpha \right)\times \nonumber \\
&&\left[\left(\frac{\Omega_D}{n^2\gamma_n} - \alpha + \beta\right) \left( 1-\Omega_D  \right)  + \frac{\Omega_D \beta u}{2} \right] \nonumber \\
&&-\frac{1}{3}\left\{  \frac{2}{\beta}\left( \frac{1}{n^2 \gamma_n} + \frac{\beta - \alpha}{\Omega_D}   \right) \left[\Omega_k - \left(u+1   \right)  \left(1-\Omega_D  \right)   \right] -u\left( u+1  \right)  \right\}, \label{a1}\\
B_1 &=&  \frac{2}{ \beta \Omega_D}\left[\frac{\Omega_D}{n^2\gamma_n} -\alpha + \beta  + \beta \Omega_D \left( \frac{u-2}{2} \right) -\frac{3\beta \Omega_D u}{2} \right]. \label{b1}
\end{eqnarray}
Using the present day values of the parameters involved along with the expression of $\gamma_{n_0}$ given in Eq. (\ref{lim1-1}), we can write the terms $A_1$ and $B_1$ as follows:
\begin{eqnarray}
A_1 &=& - \frac{4}{ 3\beta^2 \Omega_{D_0}^2}\left[\frac{\Omega_{D_0}\left( 3\lambda -1  \right)}{2n^2} -\alpha + \beta  + \beta \Omega_{D_0} \left( \frac{u_0-2}{2} \right) + \frac{3\beta \Omega_{D_0}}{2}\right]\times \nonumber \\
&&\left[\frac{\Omega_{D_0}\left( 3\lambda -1  \right)}{2n^2} -\alpha + \beta  + \beta \Omega_{D_0} \left( \frac{u_0-2}{2} \right) \right] \nonumber \\
&&+ \frac{4\left(\beta - \alpha \right)}{ 3\beta^2 \Omega_{D_0}^2}\times \nonumber \\
&&\left\{\left[\frac{\Omega_{D_0}\left( 3\lambda -1  \right)}{2n^2} - \alpha + \beta\right] \left( 1-\Omega_{D_0}  \right)  + \frac{\Omega_{D_0} \beta u_0}{2} \right\} \nonumber \\
&&-\frac{1}{3}\left\{  \frac{2}{\beta}\left[ \frac{\left( 3\lambda -1  \right)}{2n^2 } + \frac{\beta - \alpha}{\Omega_{D_0}}   \right] \left[\Omega_{k_0} - \left(u_0+1   \right)  \left(1-\Omega_{D_0}  \right)   \right] -u_0\left( u_0+1  \right)  \right\}, \label{a11}\\
B_1   &=& \frac{2}{ \beta \Omega_{D_0}}\left[\frac{\Omega_{D_0}\left( 3\lambda -1  \right)}{2n^2} -\alpha + \beta  + \beta \Omega_{D_0} \left( \frac{u_0-2}{2} \right) -\frac{3\beta \Omega_{D_0} u_0}{2} \right]. \label{b11}
\end{eqnarray}
Inserting in Eqs. (\ref{a11}) and (\ref{b11}) the values of the parameters involved, we obtain, for $\lambda = 1.02$:
\begin{eqnarray}
A_1 &\approx& -6.96235,\\
B_1 &\approx& 1.08621,
\end{eqnarray}
which lead to the following value of  $v_s^2$:
\begin{eqnarray}
v_s^2 &\approx& -6.40974. \label{vs1}
\end{eqnarray}
Inserting in Eqs. (\ref{a11}) and (\ref{b11}) the values of the parameters involved, we obtain, for $\lambda = 0.98$:
\begin{eqnarray}
A_1 &\approx& -5.87013,\\
B_1 &\approx& 0.726044,
\end{eqnarray}
which lead to the following value of $v_s^2$:
\begin{eqnarray}
v_s^2 &\approx& -8.08509. \label{vs1}
\end{eqnarray}
Inserting in Eqs. (\ref{a11}) and (\ref{b11}) the values of the parameters involved, we obtain, for $\lambda = 1.00$:
\begin{eqnarray}
A_1 &\approx& -6.40543,\\
B_1 &\approx& 0.906129,
\end{eqnarray}
which lead to the following value of $v_s^2$:
\begin{eqnarray}
v_s^2 &\approx& -7.069. \label{vs1}
\end{eqnarray}
The negative values obtained for the squared speed of the sound $v_s^2$ for all cases of the running parameter $\lambda$ considered indicate that the model we are studying is unstable for the set of values considered.\\
We now consider the limiting case corresponding to the Ricci scale, i.e. for $\alpha =2$ and $\beta =1$.\\
Inserting in Eqs. (\ref{a11}) and (\ref{b11}) the values of the parameters involved, we obtain, for $\lambda = 1.02$:
\begin{eqnarray}
A_1 &\approx& -0.0630672,\\
B_1 &\approx& -2.6923,
\end{eqnarray}
which lead to the following value of $v_s^2$:
\begin{eqnarray}
v_s^2 &\approx& 0.023425. \label{vs1}
\end{eqnarray}
Inserting in Eqs. (\ref{a11}) and (\ref{b11}) the values of the parameters involved, we obtain, for $\lambda = 0.98$:
\begin{eqnarray}
A_1 &\approx& 0.0794059,\\
B_1 &\approx& -2.87297,
\end{eqnarray}
which lead to the following value of $v_s^2$:
\begin{eqnarray}
v_s^2 &\approx& -0.027639. \label{vs1}
\end{eqnarray}
Inserting in Eqs. (\ref{a11}) and (\ref{b11}) the values of the parameters involved, we obtain, for $\lambda = 1.00$:
\begin{eqnarray}
A_1 &\approx& 0.0108893,\\
B_1 &\approx& -2.78263,
\end{eqnarray}
which lead to the following value of $v_s^2$:
\begin{eqnarray}
v_s^2 &\approx& -0.00391329. \label{vs1}
\end{eqnarray}
We observe that the squared soeed of the sound $v_s^2$ can assume both positive and negative values according to the value of $\lambda$ considered.\\
Therefore, we can conclude that we can obtain both a stable or an unstable model according to the value of the running parameter $\lambda$ considered.

\subsection{Interacting Case}
\begin{figure}[ht]
\centering\includegraphics[width=8cm]{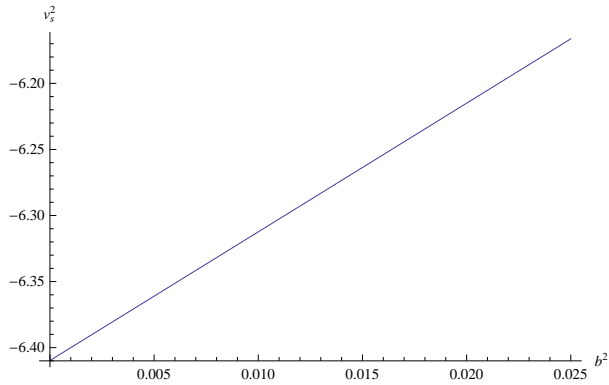}
\caption{Plot of  the squared speed of the sound $v_s^2$  given in Eq. (\ref{schi1}) for $\lambda = 1.02$. } \label{schila1}
\end{figure}

\begin{figure}[ht]
\centering\includegraphics[width=8cm]{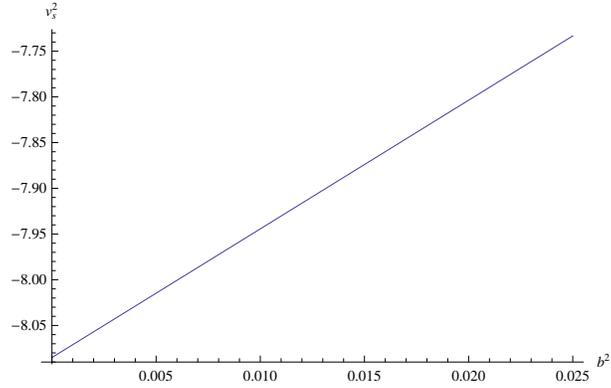}
\caption{Plot of the squared speed of the sound $v_s^2$  given in Eq. (\ref{schi2}) for $\lambda = 9.98$. } \label{schila2}
\end{figure}

\begin{figure}[ht]
\centering\includegraphics[width=8cm]{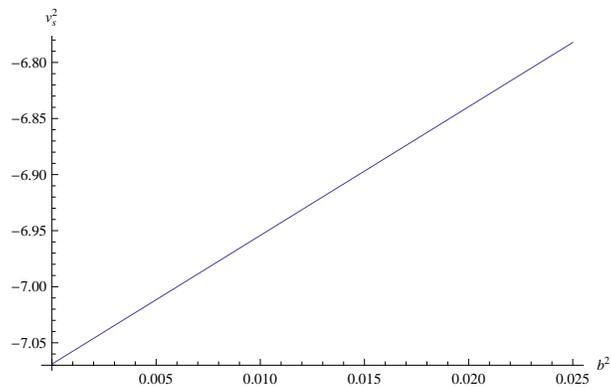}
\caption{Plot of  the squared speed of the sound $v_s^2$  given in Eq. (\ref{schi3}) for $\lambda = 1.00$. } \label{schila3}
\end{figure}

We now want to study the behavior of the squared speed of the sound $v_s^2$ in the case of interacting Dark Sectors.\\
From the continuity equation for DE given in Eq. (\ref{72}), using the general definition of the EoS parameter of DE $\omega_D$, we can write the following expression of the pressure of DE $p_D$:
\begin{eqnarray}
p_D &=& -\rho_D - \frac{\rho_D'}{3}  -\frac{Q}{3H} \nonumber \\
 &=&  -\rho_D - \frac{\rho_D'}{3}  -b^2 \rho_D\nonumber \\
  &=& -\rho_D\left( 1+b^2 \right) - \frac{\rho_D'}{3}. \label{pressurevsint}
\end{eqnarray}
Differentiating Eq. (\ref{pressurevsint}) with respect to the variable $x$, we obtain the following expression for $p'_D$:
\begin{eqnarray}
p'_D = -\rho'_D\left( 1+b^2 \right) - \frac{\rho_D''}{3}. \label{pressurevsintnew}
\end{eqnarray}
We have already obtained the expression of $\rho'_D$ in Eq. (\ref{lgo12evoint2}) and it is given by:
\begin{eqnarray}
\rho'_D =\frac{2\rho_D}{ \beta\Omega_D}\left[\frac{\Omega_D}{n^2\gamma_n} -\alpha + \beta  + \beta \Omega_D \left( \frac{u-2-3b^2}{2} \right) \right].  \label{lgo12evovs2}
\end{eqnarray}
We must now calculate the expression of $\rho_D''$.\\
Differentiating the expression of $\rho_D'$ given in Eq. (\ref{lgo12evovs2}) with respect to the variable $x$,we obtain that:
\begin{eqnarray}
\rho''_D &=&  =\frac{2\rho_D'}{ \beta \Omega_D}\left[\frac{\Omega_D}{n^2\gamma_n} -\alpha + \beta  + \beta \Omega_D \left( \frac{u-2-3b^2}{2} \right) \right] \nonumber \\
&&+\frac{2\rho_D\Omega_D'}{ \beta \Omega_D^2}\left(\beta-\alpha \right) +\rho_D u'.  \label{rhosecond}
\end{eqnarray}
Using the result of Eq. (\ref{rhosecond}) in Eq. (\ref{pressurevsintnew}), we can write:
\begin{eqnarray}
p'_D &=& - \frac{2\rho_D'}{ 3\beta \Omega_D}\left[\frac{\Omega_D}{n^2\gamma_n} -\alpha + \beta  + \beta \Omega_D \left( \frac{u-2}{2} \right) + \frac{3\beta \Omega_D}{2}\right] \nonumber \\
&&+\frac{2\rho_D\Omega_D'}{ 3\beta \Omega_D^2}\left(\alpha - \beta  \right) -\frac{\rho_D u'}{3}  , \label{pressurevs2}
\end{eqnarray}
which is the same general expression obtained for the non interacting case.\\
Inserting  in Eq. (\ref{pressurevs2}) the expressions of $\Omega_D'$, $\rho_D'$ and $u'$ obtained for the interacting case, we obtain:
\begin{eqnarray}
p'_D &=& - \frac{4\rho_D}{ 3\beta^2 \Omega_D^2}\left[\frac{\Omega_D}{n^2\gamma_n} -\alpha + \beta  + \beta \Omega_D \left( \frac{u-2}{2} \right) + \frac{3\beta \Omega_D}{2}\right]\times  \nonumber \\
&&\left[\frac{\Omega_D}{n^2\gamma_n} -\alpha + \beta  + \beta \Omega_D \left( \frac{u-2-3b^2}{2} \right) \right] \nonumber \\
&&+\frac{4\rho_D}{ 3\beta^2 \Omega_D^2}\left(\alpha - \beta \right) \times \nonumber \\
&&\left[\left(\frac{\Omega_D}{n^2\gamma_n} - \alpha + \beta\right)\left(   1-\Omega_D  \right)  + \frac{\Omega_D \beta u}{2} - \frac{3}{2}\Omega_D \beta b^2 \right] \nonumber \\
&&-\frac{\rho_D}{3}\left\{ \frac{2}{\beta}\left( \frac{1}{n^2 \gamma_n} + \frac{\beta - \alpha}{\Omega_D}   \right) \left[\Omega_k - \left(u+1   \right)  \left(1-\Omega_D  \right)   \right] -\left(u-3b^2\right)\left( u+1  \right)  \right\}  . \label{pressurevs3}
\end{eqnarray}
We must now find an expression for $\rho_m'$.\\
From the continuity equation for DM given in Eq. (\ref{73}), we obtain:
\begin{eqnarray}
\rho_m'&=& -3\rho_m + \frac{Q}{H}\nonumber \\
&=& -3u\rho_D  +3b^2\rho_D\nonumber \\
 &=& 3\rho_D \left( b^2 - u  \right), \label{rhomprime}
\end{eqnarray}
where we used the general definition of $u$.\\
Therefore, we have that:
\begin{eqnarray}
\rho_D' +\rho_m'= \rho_D' + 3\rho_D \left( b^2 - u  \right) . \label{rhosum2}
\end{eqnarray}
Inserting in Eq. (\ref{rhosum2}) the expression of $\rho_D'$ given in Eq. (\ref{lgo12evovs2}), we obtain:
\begin{eqnarray}
\rho_D' +\rho_m'= \frac{2\rho_D}{ \beta\Omega_D}\left[\frac{\Omega_D}{n^2\gamma_n} -\alpha + \beta  + \beta \Omega_D \left( \frac{u-2-3b^2}{2} \right) \right] + 3\rho_D \left( b^2 - u  \right)  . \label{}
\end{eqnarray}
Finally, we can write $v_s^2$ as follows:
\begin{eqnarray}
v_s^2 = \frac{A_2}{B_2}, \label{}
\end{eqnarray}
where $A_2$ and $B_2$ are defined as follows:
\begin{eqnarray}
A_2 &=&- \frac{4}{ 3\beta^2 \Omega_D^2}\left[\frac{\Omega_D}{n^2\gamma_n} -\alpha + \beta  + \beta \Omega_D \left( \frac{u-2}{2} \right) + \frac{3\beta \Omega_D}{2}\right]\times  \nonumber \\
&&\left[\frac{\Omega_D}{n^2\gamma_n} -\alpha + \beta  + \beta \Omega_D \left( \frac{u-2-3b^2}{2} \right) \right] \nonumber \\
&&+\frac{4\left( \alpha - \beta  \right)}{ 3\beta^2 \Omega_D^2} \times \nonumber \\
&&\left[\left(\frac{\Omega_D}{n^2\gamma_n} - \alpha + \beta\right)\left(   1-\Omega_D  \right)  + \frac{\Omega_D \beta u}{2} - \frac{3}{2}\Omega_D \beta b^2 \right] \nonumber \\
&&-\frac{1}{3}\left\{ \frac{2}{\beta}\left( \frac{1}{n^2 \gamma_n} + \frac{\beta - \alpha}{\Omega_D}   \right) \left[\Omega_k - \left(u+1   \right)  \left(1-\Omega_D  \right)   \right] -\left(u-3b^2\right)\left( u+1  \right)  \right\}, \label{a2} \\
B_2 &=& \frac{2}{ \beta\Omega_D}\left[\frac{\Omega_D}{n^2\gamma_n} -\alpha + \beta  + \beta \Omega_D \left( \frac{u-2-3b^2}{2} \right) \right]  + 3 \left( b^2 - u  \right).\label{b2}
\end{eqnarray}
Using the present day values of the parameters involved along with the expression of $\gamma_{n_0}$ given in Eq. (\ref{lim1-1}), we can write $A_2$ and $B_2$ as follows:
\begin{eqnarray}
A_2 &=&- \frac{4}{ 3\beta^2 \Omega_{D_0}^2}\left[\frac{\Omega_{D_0}\left( 3\lambda -1  \right)}{2n^2} -\alpha + \beta  + \beta \Omega_{D_0} \left( \frac{u_0-2}{2} \right) + \frac{3\beta \Omega_{D_0}}{2}\right]\times  \nonumber \\
&&\left[\frac{\Omega_{D_0}\left( 3\lambda -1  \right)}{2n^2} -\alpha + \beta  + \beta \Omega_{D_0} \left( \frac{u_0-2-3b^2}{2} \right) \right] \nonumber \\
&&+\frac{4\left( \alpha - \beta  \right)}{ 3\beta^2 \Omega_{D_0}^2} \times \nonumber \\
&&\left\{\left[\frac{\Omega_{D_0}\left( 3\lambda -1  \right)}{2n^2} - \alpha + \beta\right]\left(   1-\Omega_{D_0}  \right)  + \frac{\Omega_{D_0} \beta u_0}{2} - \frac{3}{2}\Omega_{D_0} \beta b^2 \right\} \nonumber \\
&&-\frac{1}{3}\left\{ \frac{2}{\beta}\left[ \frac{\left( 3\lambda -1  \right)}{2n^2 } + \frac{\beta - \alpha}{\Omega_{D_0}}   \right]\times \right. \nonumber \\
 &&\left.\left[\Omega_{k_0} - \left(u_0+1   \right)  \left(1-\Omega_{D_0}  \right)   \right] -\left(u_0-3b^2\right)\left( u_0+1  \right)  \right\}, \label{a22} \\
B_2 &=& \frac{2}{ \beta\Omega_{D_0}}\left[\frac{\Omega_{D_0}\left( 3\lambda -1  \right)}{2n^2} -\alpha + \beta  + \beta \Omega_{D_0} \left( \frac{u_0-2-3b^2}{2} \right) \right]  + 3 \left( b^2 - u_0  \right)\label{b22}.
\end{eqnarray}
Inserting in Eqs. (\ref{a22}) and (\ref{b22}) the values of the parameters involved, we obtain, for $\lambda = 1.02$:
\begin{eqnarray}
A_2 &\approx& -6.96235 + 10.5769 b^2,\\
B_2 &\approx&1.08621,
\end{eqnarray}
which lead to the following value of $v_s^2$:
\begin{eqnarray}
v_s^2 &\approx& -6.40974 + 9.73741 b^2. \label{schi1}
\end{eqnarray}
In Figure \ref{schila1},    we plot the behavior of $v_s^2 $ obtained in Eq. (\ref{schi1}).\\
In the limiting case of $b^2 =0$ we recover the same result of the non interacting case, while for $b^2=0.025$ we obtain $A_2 \approx -6.69793$ and $v_s^2 \approx -6.1663$.\\
Inserting in Eqs. (\ref{a22}) and (\ref{b22}) the values of the parameters involved, we obtain, for $\lambda = 0.98$:
\begin{eqnarray}
A_2 &\approx& -5.87013 + 10.2167 b^2,\\
B_2&\approx& 0.726044,
\end{eqnarray}
which lead to the following value of $v_s^2$:
\begin{eqnarray}
v_s^2 &\approx& -8.08509 + 14.0718 b^2. \label{schi2}
\end{eqnarray}
In Figure \ref{schila2},    we plot the behavior of $v_s^2$ obtained in Eq. (\ref{schi2}).\\
In the limiting case of $b^2 =0$ we recover the same result of the non interacting case, while for $b^2=0.025$ we obtain $A_2 \approx -5.6147$ and $v_s^2 \approx -7.7339$.\\
Inserting in Eqs. (\ref{a22}) and (\ref{b22}) the values of the parameters involved, we obtain, for $\lambda = 1.00$:
\begin{eqnarray}
A_2 &\approx& -6.40543 + 10.3968 b^2,\\
B_2 &\approx& 0.906129,
\end{eqnarray}
which lead to the following value of $v_s^2$:
\begin{eqnarray}
v_s^2 &\approx& -7.069 + 11.4739 b^2. \label{schi3}
\end{eqnarray}
In Figure  \ref{schila3},   we plot the behavior of $v_s^2 $ obtained in Eq. (\ref{schi3}).\\
In the limiting case of $b^2 =0$ we recover the same result of the non interacting case, while for $b^2=0.025$ we obtain $A_2 \approx -6.14551$ and $v_s^2 \approx -6.7821$.\\
We can observe that the squared speed of the sound $v_s^2$ assumes a negative value for all the cases of the running parameter $\lambda$ considered for $b^2$ in the range $\left[0, 0.025  \right]$, therefore we deal with a model which is unstable for the set of values considered.\\

\begin{figure}[ht]
\centering\includegraphics[width=8cm]{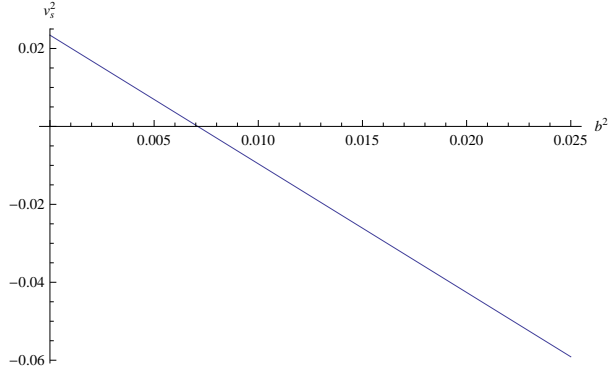}
\caption{Plot of  the squared speed of the sound $v_s^2$  given in Eq. (\ref{schi4}) for $\lambda = 1.02$ for the limiting case of Ricci scale. } \label{schila4}
\end{figure}

\begin{figure}[ht]
\centering\includegraphics[width=8cm]{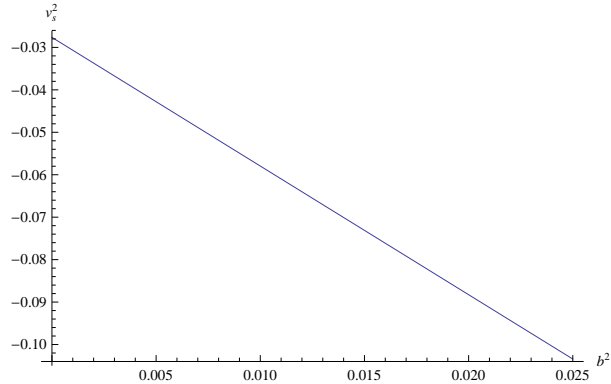}
\caption{Plot of  the squared speed of the sound $v_s^2$  given in Eq. (\ref{schi5}) for $\lambda = 0.98$ for the limiting case of Ricci scale. } \label{schila5}
\end{figure}

\begin{figure}[ht]
\centering\includegraphics[width=8cm]{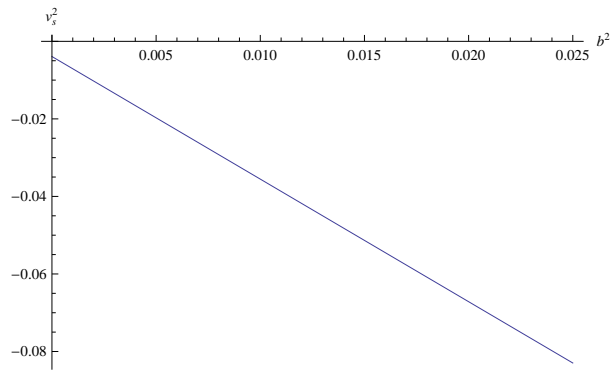}
\caption{Plot of  the squared speed of the sound $v_s^2$  given in Eq. (\ref{schi6}) for $\lambda = 1.00$ for the limiting case of Ricci scale. } \label{schila6}
\end{figure}

We now consider the limiting case corresponding to the Ricci scale, i.e. in the limiting case of $\alpha =2$ and $\beta =1$.\\
Inserting in Eqs. (\ref{a22}) and (\ref{b22}) the values of the parameters involved, we obtain, for $\lambda = 1.02$:
\begin{eqnarray}
A_2 &\approx& -0.0630672 + 8.88923 b^2,\\
B_2 &\approx& -2.6923,
\end{eqnarray}
which lead to the following value of $v_s^2$:
\begin{eqnarray}
v_s^2 &\approx& 0.023425 - 3.30172 b^2. \label{schi4}
\end{eqnarray}
In Figure \ref{schila4},    we plot the behavior of $v_s^2 $ obtained in Eq. (\ref{schi4}).\\
In the limiting case of $b^2 =0$ we recover the same result of the non interacting case, while for $b^2=0.025$ we obtain $A_2 \approx 0.159164$ and $v_s^2 \approx  -0.0591181$.\\
Inserting in Eqs. (\ref{a22}) and (\ref{b22}) the values of the parameters involved, we obtain, for $\lambda = 0.98$:
\begin{eqnarray}
A_2 &\approx& 0.0794059 + 8.70857 b^2,\\
B_2 &\approx& -2.87297,
\end{eqnarray}
which lead to the following value of $v_s^2$:
\begin{eqnarray}
v_s^2 &\approx& -0.027639 - 3.03121 b^2. \label{schi5}
\end{eqnarray}
In Figure \ref{schila5},    we plot the behavior of $v_s^2$ obtained in Eq. (\ref{schi5}).\\
In the limiting case of $b^2 =0$ we recover the same result of the non interacting case, while for $b^2=0.025$ we obtain $A_2\approx 0.29712$ and $v_s^2 \approx -0.103419$.\\
Inserting in Eqs. (\ref{a22}) and (\ref{b22}) the values of the parameters involved, we obtain, for $\lambda = 1.00$:
\begin{eqnarray}
A_2 &\approx& 0.0108893 + 8.7989 b^2,\\
B_2 &\approx& -2.78263,
\end{eqnarray}
which lead to the following value of $v_s^2$:
\begin{eqnarray}
v_s^2 &\approx& -0.00391329 - 3.16208 b^2. \label{schi6}
\end{eqnarray}
In Figure \ref{schila6},    we plot the behavior of $v_s^2 $ obtained in Eq. (\ref{schi6}).\\
In the limiting case of $b^2 =0$ we recover the same result of the non interacting case, while for $b^2=0.025$ we obtain $A_2 \approx 0.230862$ and $v_s^2 \approx -0.0829653$.\\
We can conclude that, for $\lambda = 1.02$, we can obtain a model which can be stable or unstable according to the value of $b^2$. Instead, for $\lambda =1.00$ and $\lambda = 0.98$, we obtain an unstable model for all the range of values of $b^2$.\\

\section{Conclusions}
In this work, we studied the Power Law Entropy Corrected versions of the HDE (PLECHDE) model with infrared (IR) cut-off the one recently suggested and studied by Granda and Oliveros, which contains two terms, one proportional to the Hubble parameter squared $H^2$ and one proportional to the first derivative with respect of the cosmic time $t$ of the Hubble parameter $H$, i.e.  $\dot{H}$. Moreover, this model is characterized by two constant parameters indicated with $\alpha$ and $\beta$. In the limiting case of $\alpha=2$ and $\beta=1$, we obtain that the Granda-Oliveros cut-off becomes proportional to the average radius of the Ricci scalar curvature.  We have investigated this model in a FLRW Universe in the framework of Ho\v{r}ava-Lifshitz gravity for both non-interacting and interacting DE and DM. We must underline that we have considered three different values of the running parameter $\lambda$ (which is one of the parameter characterizing the Ho\v{r}ava-Lifshitz gravity): in particular, following the results of Dutta $\&$ Saridakis \citep{duttasari}, we have considered $\lambda =1.02$ and $\lambda = 0.98$. We must also underline that Lorentz invariance is restored for $\lambda =1$. We also considered the case with $\lambda =1$ in order to obtain results when the Lorentz invariance is restored. Moreover, we also studied the limiting case corresponding to the Ricci scale for all the values of the running parameter $\lambda$ taken into account. \\
Using a low redshift expansion of the EoS parameter of DE as $\omega_D \left( z \right) = \omega_0 + \omega_1 z$, we calculated the expressions of the parameters $\omega_0$ and $\omega_1$ as functions of the DE and curvature fractional energy density parameters and of the interaction parameter $b^2$ for the interacting case. We found that the parameter $\omega_0$ assumes the same expression for both non interacting and interacting Dark Sectors. Instead, $\omega_1$ has a clear dependence on $b^2$ for the interacting case. We also calculated the value of the redshift which lead to $\omega_D = -1$ and the present day values of the EoS parameter of DE $\omega_D$.\\
We also derived an expression for deceleration parameter $q$, which was found to be function of the fractional energy density of DE $\Omega_D$ and of the parameter $n$, the running parameter $\lambda$ and the two constants $\alpha$ and $\beta$ characterizing the GO cut-off. The values obtained for the deceleration parameter $q$ for the present day values of the parameters involved indicate that the model considered leads to an accelerated Universe since $q$ assumes negative values, which is in agreement with the most recent cosmological observations.\\
Studying the statefinder parameters $r$ and $s$, for $\alpha = 8824$ and $\beta = 0.5016$,    we found that the PLECHDE model with GO cut-off considered in this work leads to points that are far from the point corresponding to the $\Lambda$CDM model for both non interacting and interacting Dark Sectors; moreover, since we obtained $s<0$, we derive that we deal with a phantom-like model. Instead, for the limiting case corresponding to the Ricci scale (recovered for $\alpha=2$ and $\beta=1$), we obtain points with are closer to  the point corresponding to the $\Lambda$CDM model (with a departure  which is a  bit more evident for the interacting case); moreover, since we obtained $s>0$, we derive that we deal with a quintessence-like model for this case. \\
We have also derived the expressions of the cosmographic parameters $s_{cosmo}$ and $l$, also known as snap and lerk parameters. We must remember that the cosmographic parameters are useful in order to characterize the main properties of a particular DE model. Being dependent on the higher time derivatives of the scale factor $a\left(r\right)$, the cosmographic parameter $s_{cosmo}$ and $l$ can give more cosmological details respect to the Hubble and and deceleration parameters $H$ and $q$. We have derived that the present day values of $s_{cosmo}$ and $l$ are  between the errors of the values found in some recent papers for both non interacting and interacting DE and DM and for both sets of values of $\alpha$ and $\beta$ considered in this paper.  \\
We also studied the behavior of the squared speed of the sound $v_s^2$ in order to check the stability of the model we are studying. We obtained that, for both non interacting and interacting Dark Sectors, the model we are considering in unstable since we obtained a negative value of $v_s^2$ for the case with $\alpha = 8824$ and $\beta = 0.5016$. Instead, for the limiting case corresponding to the Ricci scale, we can obtain a model which is stalbe or unstable depending on the value of the running parameter $\lambda$,  on the absence or presence of interaction and (in the case an interaction between the Dark Sectors really exists) on the strength of the interaction. \\
Some of the parameters involved in the equations used and derived in this work have not a precise value yet or they have a considerable errors (like it happens for the snap and lerk  cosmographic parameters),    so it is difficult to obtain exact constraints and comparisons using these parameters. For this reason,  future precision cosmological missions could help to obtain better constraints in order to also allow the accurate determination of the values of the parameter derived in this work.

\end{document}